\documentclass[twocolumn,pra,superscriptaddress,aps,10pt]{revtex4-2}
\usepackage{graphicx}          
\usepackage{amsfonts,amsmath,amssymb}    
\usepackage{dsfont}                      
\usepackage[
colorlinks=true,
linkcolor=blue,
citecolor=blue,
urlcolor=blue]{hyperref}                
\usepackage{comment}  
\usepackage[caption=false]{subfig}                  
\usepackage{appendix}
\usepackage{enumerate}

\newcommand{\argp}[1]{\left( #1 \right)}
\newcommand{\args}[1]{\left[ #1 \right]}
\newcommand{\argc}[1]{\left\{ #1 \right\}}
\newcommand{\ket}[1]{\left\vert #1 \right\rangle}
\newcommand{\bra}[1]{\left\langle #1\right\vert}

\newcommand{\braket}[2]{\left\langle #1 \vert #2 \right\rangle}

\begin{document}

\title{Sharing quantum nonlocality and teleportation over long distance using optical hybrid states}

\author{Subhankar Bera}
\affiliation{S. N. Bose National Centre for Basic Sciences, Sector III, Saltlake, Kolkata 700106, India}

\author{Soumyakanti Bose}
\affiliation{NextQuantum, Department of Physics $\&$ Astronomy, Seoul National University, Gwanak-ro 1, Gwanak-gu, Seoul 08826, Korea}
\email{soumyakanti.bose09@gmail.com}

\author{Hyunseok Jeong}
\affiliation{NextQuantum, Department of Physics $\&$ Astronomy, Seoul National University, Gwanak-ro 1, Gwanak-gu, Seoul 08826, Korea} 

\author{A. S. Majumdar}
\affiliation{S. N. Bose National Centre for Basic Sciences, Sector III, Saltlake, Kolkata 700106, India}

\begin{abstract} 
We analyze sharing Bell-type nonlocal correlation between two distant parties with optical hybrid states comprising a single photon polarization state and a multiphoton coherent state. 
By deploying entanglement swapping over the coherent state parts at the middle station, we show that the optical hybrid states can efficiently generate a polarization-entangled state that violates Clauser-Horne-Shimony-Holt (CHSH) Bell-inequality well over a metropolitan distance.
We further assess the quality of the shared entangled state in the information processing task of quantum teleportation of an unknown polarization qubit.
Our results with realistic devices, embedding detection inefficiency and transmission losses, indicate the viability of faithful quantum teleportation over large distances, consistent with the quality of the shared correlation.
\end{abstract}

\maketitle
\section{Introduction}

Bell nonlocality \cite{Bell_Bell1964, Bell_CHSH1969, BellBook_Aspect2004}, which enables correlations between two parties that surpass the limitations imposed by local hidden variable models, lies at the heart of many state-of-the-art applications. These include secure communication \cite{DIQKD_Zhang2022, DIQKD_Liu2022, DIQKD_Victor2023, DIQKD_Lewis2024, DIQKD_Tan2024}, efficient quantum computation \cite{QCBell_Daniel2022, QCBell_Jelena2023, QCBell_Kalai2023}, and foundational tasks such as self-testing \cite{BellSelfTEst_Andrea2017, BellSelfTest_Goswami2018, BellSelfTest_Zhang2019, BellSelfTest_Zhang2019_2} and device-independent certification \cite{BellDICert_Peter2016, BellDICert_Acin2016, BellDICert_Acin2016_2, BellDICert_Joseph2018, BellDICert_Ole2018}. Although Bell nonlocality has been extensively studied both theoretically \cite{BellRev_Ghadimi2018, BellRevBook_Scarani2019} and experimentally across various platforms, the practical realization of such correlations over large distances using ground-based optical fiber networks \cite{BellExp_Forgues2015, BellExp_Prabhakar2015, BellExp_Juan2016, BellExp_Martin2016, BellExp_Gonzalo2017, BellExp_Igor2018, BellExp_Oliver2018, BellExp_Zhang2018, BellExp_Zhong2019, BellFibOpt_Tim2022} remains a significant challenge. This limitation continues to be a major bottleneck in the development of a sustainable and efficient ground-based quantum internet \cite{QuantInt_Wehner2018, QuantInt_Koji2023}.

Bell nonlocality and teleportation are predominantly studied in two broad categories of physical systems: discrete-variable (DV) systems, such as spin-like particles \cite{BellRev_Ghadimi2018, BellRevBook_Scarani2019}, and continuous-variable (CV) systems characterized by Gaussian optical states \cite{BellGauss_Wiseman2007, BellGauss_Buono2014, Qureshi2020, BellGauss_Lantz2021, BellGauss_Bohr2023}. 
While each system offers specific advantages and drawbacks \cite{DVCVCompare_Xu2015, DVCVCompare_Pirandola2015, DVCVCompare_Diamanti2016}, recent progress in producing weak coherent pulses \cite{WCP_Costanzo2017, WCP_Jain2020, WCP_Wiseman2023, WCP_Duranti2024} have enhanced the prospects of DV-based information processing. 
Nonetheless, identifying an optimal physical platform for quantum information processing—especially within linear-optics-based telecommunication frameworks—remains an open and active area of research.

On the other hand, a distinct class of physical systems—optical hybrid states—combines both discrete-variable (DV) and continuous-variable (CV) components \cite{HOSDescrip_Jeong2005, HOSDescript_Li2006, HOSDescrip_Bing2011, HOSDescript_Le2021}. These states exhibit intrinsic correlations \cite{HOSBell_Kwon2013, HOSBell_Kwon2014, HOSBell_Moradi2024} that are evident under both particle-like and wave-like measurement schemes \cite{HOSBellMeas_Daniel2011}. Optical hybrid states have proven instrumental in various quantum information protocols, including quantum teleportation \cite{HOSTeleport_Park2012, HOSTeleport_Lee2013, HOSTeleport_Kim2016, HOSTelport_Ulanov2017, HOSTelport_Sychev2018, HOSTelport_Francisco2020, HOSTeleport_Bose2022, HOSTeleport_He2022, HOSTeleport_Kirdi2023, HOSQT_Liu2024} and fault-tolerant quantum computation \cite{HOSQComp_Omkar2020, HOSQComp_Omkar2021, HOSQComp_Tom2023, HOSQComp_Lee2024}. They also offer a compelling alternative for distributing quantum correlations, bridging the gap between DV-only and CV-only approaches \cite{HOSSwap_Lim2016, HOSSwap_Parker2017, HOSEnt_Bose2024}.

Although optical hybrid states have been successfully generated across various experimental platforms \cite{HOSTelport_Sychev2018, HOSExp_Jeong2014, HOSEXxp_Morin2014, HOSExp_Jeannic2018, HOSExp_Huang2019, HOSExp_Hacker2019, HOSExp_Giovanni2020, HOSExp_Gouzien2020, HOSExp_Wen2021, HOSExp_Li2021}, their capacity to share stronger-than-entanglement correlations over long distances remains a promising area for further investigation.

In this article, within the current state-of-the-art architecture \cite{BellSwap_Zopf2019, BellSwap_Tsujimoto2020, BellSwap_Huang2022, BellSwap_Anders2023}, we analyze entanglement-swapping protocol with optical hybrid states, in the context of sharing highly correlated DV Bell pairs between two distant laboratories.
It is worth noting that, despite apparent similarities in the basic setup and the overall structure of the final DV-entangled state, the present work is distinct from the previous attempt reported in \cite{HOSSwap_Lim2016}. 
That earlier study was based on single-photon hybrid states, which are known to yield higher teleportation fidelity and enhanced robustness against photon loss \cite{HOSTeleport_Kim2016}, resulting in a single-photon entangled state between two spatial modes.
In contrast, the current work investigates polarization-based hybrid states, which offer advantages for fault-tolerant encoding \cite{HOSTeleport_Lee2013, HOSQComp_Omkar2020} and entanglement purification \cite{HOSEntPurify_Sheng2013}. 
Consequently, the final DV-entangled state achieved here is a two-photon entangled state, where each photon occupies a distinct polarization mode.
Moreover, the recent utilization of polarization-entangled states in Bell tests \cite{BellPolarization_Sheng2015, BellPolarization_Matthias2023, BellPolarization_Mishra2024} and quantum communication protocols \cite{PolarizationQuantNet_Elena2024, PolarizationQKD_Matteo2024} underscores the contemporary relevance of our approach.

We further evaluate the quality of the shared Bell-CHSH correlations through the lens of quantum teleportation \cite{Teleport_Bennett1993}, a foundational quantum information processing task that underpins the vision of distributed quantum computing \cite{QCDist_Caleffi2024}. 
In addition to various schemes proposed to witness quantum teleportation using entangled states \cite{EntWit_Ganguly2011} and in the presence of noise \cite{QT_Adhikari2010}, Bell-CHSH nonlocality has been shown to play a crucial role in surpassing classical teleportation limits \cite{BellTeleport_Gisin1996, BellTeleport_Horodecki1996}.

Our numerical results, demonstrating the feasibility of quantum teleportation in fiber-optic infrastructures \cite{FibOptTp_Hiroki2015, FibOptTp_Huo2018, FibOptTp_Zhao2022, BellFibOpt_Tim2022, FibOptTp_Shen2023, FibOptTp_Rivera2023}, offer an operational characterization of the proposed scheme. 
These findings underscore its effectiveness in realizing Bell-CHSH violations over long distances, reinforcing its potential for real-world quantum network implementations.

To further strengthen our study, we begin by analyzing the performance of the proposed scheme under transmission losses, assuming ideal detectors. 
Our results show that optical hybrid states exhibit notable robustness against transmission losses, enabling significant Bell-CHSH violations over large laboratory separations (up to $~250$ km). 
This robustness is further supported by near-perfect or high-fidelity quantum teleportation across the same distances.

We observe that while the magnitude of Bell-CHSH violation decreases with increasing lab separation and coherent amplitude, the success probability exhibits a non-monotonic trend. 
Specifically, increasing the coherent amplitude raises the average photon number, thereby enhancing the success probability. 
However, a higher photon number also increases the state’s susceptibility to transmission losses. 
As a result, the success probability reaches a maximum at an optimal value of the coherent amplitude ($\alpha$).

To evaluate the impact of detector inefficiencies alongside transmission losses, we next consider Bell-CHSH violation and teleportation performance with non-ideal detectors. 
The results indicate a sharp decline in performance as detection efficiency decreases, scaling approximately with the square of the detection efficiency. 
For example, with $5\%$ and $10\%$ detection inefficiencies, both the Bell-CHSH violation and teleportation fidelity drop to approximately $90\%$ and $81\%$, respectively. 
Nevertheless, transmission distances of up to $~200$ km remain achievable, even with imperfect detectors.

These findings highlight the practical viability of optical hybrid states for distributing quantum correlations within fiber-optics-based architectures. 
Moreover, they advocate for this hybrid approach as a promising alternative to traditional DV-only and CV-only systems, consistent with earlier observations \cite{HOSEnt_Bose2024}.

The paper is organized as follows.
In Sec. \ref{sec:protocol} we first describe our protocol for sharing Bell-type nonlocal correlation between two distant parties.
Sec. \ref{sec:Bell_Teleport_descript} contains mathematical descriptions and analytical results on the success of generating the DV Bell pair, corresponding Bell-CHSH violation and teleportation of an unknown qubit with the shared state.
In Sec.~\ref{sec:Results} we provide our simulation results on Bell nonlocality and teleportation with both ideal and non-ideal detectors. 
Finally, in Sec.~\ref{sec:conclusion} we conclude our results with discussion and future prospects.

\section{Protocol}
\label{sec:protocol}

\begin{figure*}
    \centering
    \includegraphics[width=1.0\textwidth]{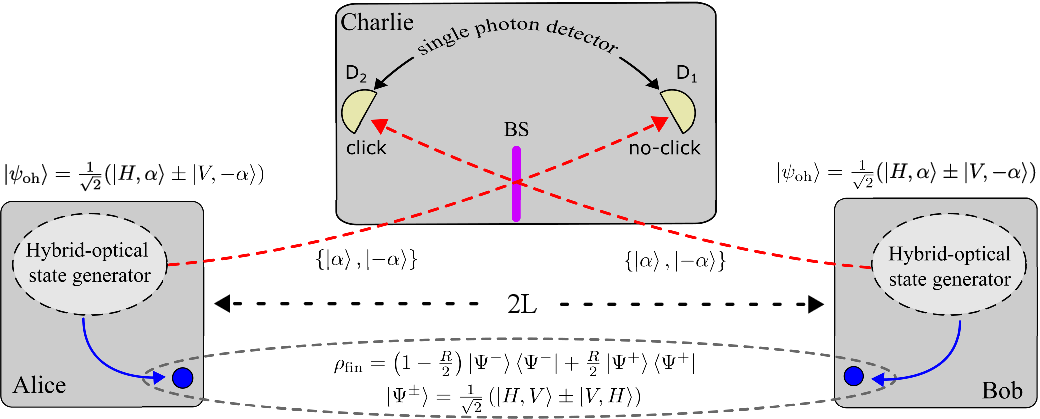}
    \caption{Schematic for sharing distant DV Bell-pair using optical hybrid states.
    Two parties, say alice and Bob, send the coherent states $\lbrace \ket\alpha, \ket{-\alpha} \rbrace$ to a third party in the middle, say Charlie. 
    Subsequently, Charlie mixes the incoming signals through a balanced beam splitter followed by photon measurement by two on-off detectors. 
    Upon receiving the information about which detectors clicks, Alice and Bob post-select the overall state to the desired form.}
    \label{fig:protocol}
\end{figure*}

In this paper, we consider a particular type of optical hybrid state that represents entanglement between a polarization states ($H$/$V$) and the coherent state ($\ket\alpha$) as \cite{HOSBell_Kwon2014, HOSTelport_Sychev2018} 
\begin{equation}
    \ket{\psi_\text{oh}} = \frac{1}{\sqrt{2}}\argp{ \ket{H,\alpha} + \ket{V,-\alpha}},
    \label{eq:hos}
\end{equation}
where $H$ ($V$) stands for the horizontal (vertical) polarization.
For the sake of simplicity, we have considered real $\alpha$.
Our protocol for sharing correlation with the hybrid states \eqref{eq:hos}, schematically shown in Fig. \ref{fig:protocol}, proceeds as described below:
\begin{itemize}
    \item {\bf Step 1 - channel transmission:} Alice and Bob first prepare their individual optical hybrid states and send the coherent state signal ($\argc{\ket\alpha,\ket{-\alpha}}$) to a third party, say Charlie.
    We consider that the transmission channels (optical fiber) to be lossy described by the transmission coefficient $T$ ($0\leq T\leq 1$) such that $T=1$ corresponds to ideal lossless channel and $T=0$ stands for complete loss.
    For a standard optical cable, channel transmittance is given by $T=10^{-l L_\text{ab}/10}$, where $L_\text{ab}$ is the channel length (in km) and $l=0.2$ dB/km is the average photon loss per km.
    Here we consider a symmetric setup, i.e., Charlie sits midway between Alice and Bob.
    As a consequence, for the coherent signals from Alice and Bob to travel a distance of $L$, the lab separation becomes $L_\text{ab}=2L$.
    One may also consider a more general transmission channel with additional thermal noise.
    However, the loss-only channels closely mimic these general channels as shown earlier \cite{HOSEnt_Bose2024}.

    \item {\bf Step 2 - swapping measurement:} Charlie then mixes the two incoming signals in a balanced ($50:50$) beam splitter followed by detection through two single-photon detectors.
    The individual detectors at Charlie's laboratory are described by the measurement setting $\mathcal{M} = \lbrace \Pi_1, \Pi_{\neg 1} \rbrace$, where $\Pi_1 = \ket 1\bra 1$ is the projection along the photon-number state $\ket 1$ and $\Pi_{\neg 1} = \mathds{1} - \Pi_1$.
    This step is considered successful only if one of the detectors click while the other remain dormant (no-click event).
    For the sake of simplicity (without loss of generality) let us consider that the detector on the left (Fig. \ref{fig:protocol}) clicks which correspondss to the measurement operator $\mathcal{M}_\text{succ} = \Pi_1^\text{left} \otimes \Pi_{\neg 1}^\text{right}$.
    To model the detectors realistically, we also consider that the efficiency of the detectors is given by $\eta_0$ ($0\leq \eta_0\leq 1$) where $\eta_0 = 1$ represents  perfect detector.
    Once the click event successfully takes place, Charlie declares it.

    \item {\bf Step 3 - sharing final state:} After the Step 2 completes successfully, Charlie declares which of the detectors have clicked.
    Based on the information announced by Charlie, Alice and Bob can then post-select their state.
    
\end{itemize}
    
Let us now define the $4$-Bell states in the polarization basis as   
\begin{align}
    \ket{\Psi^\pm} &= \frac{1}{\sqrt{2}}\argp{\ket{H,V} \pm \ket{V,H}}
    \nonumber 
    \\
    \ket{\Phi^\pm} &= \frac{1}{\sqrt{2}}\argp{\ket{H,H} \pm \ket{V,V}}.
    \label{eq:bell_projector}
\end{align}

Considering that the detector on the left side (as shown in the Fig. \ref{fig:protocol}) clicks, the final shared state between Alice and Bob is given by \eqref{append_eq:dvstate_final_bellbasis}
\begin{equation}
    \rho_\text{fin} = \argp{1-R}\ket{\Psi^-}\bra{\Psi^-} + R\ket{\Psi^+}\bra{\Psi^+}
    \label{eq:finalstate}
\end{equation}
with probability 
\begin{equation}
    \text{Pr} = T\eta_0\alpha^2 e^{-2T\eta_0\alpha^2},
    \label{eq:probability}
\end{equation}
where $R = \frac{1 - e^{-4(1-T\eta_0)\alpha^2}}{2}$ is the overall effective loss factor such that $0\leq R\leq \frac{1}{2}$.

\section{Bell-nonlocality and Teleportation}
\label{sec:Bell_Teleport_descript}

In this section, we first analyze the Bell nonlocal character of the shared DV state \eqref{eq:finalstate} by using polarization-based measurements. 
To characterize the operational utility of the shared nonlocality we further analyze an information processing task, to be precise teleportation of an unknown input qubit state.
Here, we provide the mathematical expressions for various quantities related to Bell-CHSH violation as well as derive the fidelity of teleportation of an unknown polarization qubit state using the shared DV state.
It must be noted that for the sake of generality we consider all the detectors to be imperfect  with the efficiency $\eta_0$ ($0\leq \eta_0\leq 1$).

\subsection{Bell-violation}
\label{subsec:bell_descript}

In the polarization basis ($\{ \ket H,\ket V \}$), the binary ($2$-outcome) operator  $\Pi = \ket H\bra H - \ket V\bra V$  yields either of $\pm 1$ based on the state of the polarization.
The influence of noise (check Supplementary material) leads to the noisy binary measurement $\Pi \rightarrow \Pi(\eta_0) = \eta_0 \argp{\ket H\bra H - \ket V\bra V}$.
Now, a unitary rotation between the polarization degrees of freedom could be implemented through a polarization-beam-splitter (PBS) with phase components defined by the unitary matrix
\begin{align}
    U(\vec\zeta) = U(\zeta,\theta) = 
    \begin{pmatrix}
        \sqrt{\zeta} &\sqrt{1-\zeta}e^{\mathsf{i}\theta}\\
        -\sqrt{1-\zeta}e^{-\mathsf{i}\theta} &\sqrt{\zeta}
    \end{pmatrix},
\end{align}
where $0\leq \zeta \leq 1$ and $0\leq \theta \leq 2\pi$ and $\vec\zeta = \argc{\zeta,\theta}$.
This leads to the polarization-rotated-binary-operator (PRBO) as \eqref{append_eq:bell_operator_rotated}
\begin{align}
    \hat{O}(\vec\zeta) &= U(\vec\zeta) \Pi(\eta_0) U^\dagger(\vec\zeta)
    \nonumber 
    \\
    &= N_{hh}(\vec\zeta) \ket H\bra H + N_{vv}(\vec\zeta) \ket V\bra V - N_{hv}(\vec\zeta)\ket H\bra V 
    \nonumber 
    \\
    &~~~~- N_{hv}^*(\vec\zeta)\ket V\bra H,
    \label{eq:binary_measurement_rotated}
\end{align}
where $N_{hh}(\vec\zeta) = -\eta_0(1-2\zeta)$, $N_{vv}(\vec\zeta) = \eta_0(1-2\zeta)$ and $N_{hv}(\vec\zeta) = 2e^{\mathsf{i}\theta}\eta_0\sqrt{\zeta(1-\zeta)}$.

Accordingly, one can define the joint binary-outcome measurement as $\hat{O}_{a_1,b_1}(\vec\zeta,\vec\xi) = \hat{O}_{a_1}(\vec\zeta) \otimes \hat{O}_{b_1}(\vec\xi)$ and its expectation value as \eqref{append_eq:expectation_joint_measure}
\begin{align}
    &\mathcal{E}\argp{\vec\zeta,\vec\xi} = \text{Tr}\args{ \rho_{a_1b_1} \hat{O}_{a_1,b_1}(\vec\zeta,\vec\xi) } 
    \nonumber 
    \\
    &= -\eta_0^2 \left[
    (1-2\zeta)(1-2\xi) + 4e^{-4(1-T\eta_0)\alpha^2} \cos{2(\theta-\phi)}
    \right.
    \nonumber 
    \\
    &~~~~\left.\times \sqrt{\zeta(1-\zeta)\xi(1-\xi)}
    \right],
    \label{eq:expression_jointmeasure}
\end{align}
where $\vec\zeta = \argc{\zeta,\theta}$ and $\vec\xi = \argc{\xi,\phi}$.
As a consequence, by varying over $\argc{\vec\zeta,\vec\xi}$ one can recast the Bell-function as
\begin{align}
    \mathcal{B} = \mathcal{E}\argp{\zeta_1,\xi_1} + \mathcal{E}\argp{\zeta_1,\xi_2} + \mathcal{E}\argp{\zeta_2,\xi_2} - \mathcal{E}\argp{\zeta_2,\xi_1}
    \label{eq:bell_function}
\end{align}
which implies CHSH Bell nonlocality for $\mathcal{B} > 2$ \cite{Bell_CHSH1969}.
To obtain the optimal measurement setting for Bell nonlocality one needs to optimize the Bell-function ($\mathcal{B}$) over the set of $\left\{ \zeta,\theta,\xi,\phi \right\}$ where $0\leq \left\{ \zeta,\xi \right\} \leq 1$ and $0\leq \left\{ \theta,\phi \right\} \leq 2\pi$.
It should also be noted the optimal setting also varies with the distance between the laboratories ($L_\text{ab}$).

Nonetheless, it may be noted that for a general mixed entangled channel of the form \eqref{eq:finalstate}, the optimal Bell-violation, that could be obtained by optimizing over all measurement settings, could be written as \cite{BellTeleport_Horodecki1996} $\mathcal{B}_\text{opt} = \text{max}_{i>j}~ 2\sqrt{\lambda_i^2+\lambda_j^2}$ where $\lambda_i$'s are the eigenvalues of the correlation matrix $T_\text{cor}$ for which elements are defined by $t_{ij} = \text{Tr} [(\sigma_i \otimes \sigma_j)\rho]$. 
In a simple and straightforward calculation it can be shown that 
\begin{equation}
    T_{\text{cor}} =
    \begin{pmatrix}
    2R - 1 & 0 & 0 \\
    0 & 2R - 1 & 0 \\
    0 & 0 & -1
    \end{pmatrix}
\end{equation} 
such that the eigenvalues, in the descending order, are $\lambda_1=1$, $\lambda_2=2R - 1$, $\lambda_3=2R - 1$. 
This immediately leads to the result
\begin{equation}
    \mathcal{B}_\text{opt} = 2\sqrt{1+(1-2R)^2} = 2\sqrt{1 + e^{-8(1-T\eta_0)\alpha^2}},
    \label{eq:bellmax}
\end{equation}
such that the state \eqref{eq:finalstate} is always Bell-CHSH nonlocal ($\mathcal{B}_\text{opt}>2$) for all values of $0\leq R <\frac{1}{2}$.

\subsection{Teleportation of unknown qubit input state}
\label{subsec:teleport_descript}

Let us now consider the case of teleporting an unknown input pure-state $\ket{\psi_\text{in}} = \sqrt{p} \ket H + \sqrt{1-p}e^{\mathsf{i}\theta} \ket V$ using the shared resource \eqref{eq:finalstate}.
In the ideal case, Alice's measurements are given by the $4$ Bell-state projectors \eqref{eq:bell_projector} $\Pi_{\Psi}^\pm = \ket{\Psi^\pm}\bra{\Psi^\pm}$ and $\Pi_{\Phi}^\pm = \ket{\Phi^\pm}\bra{\Phi^\pm}$ over the input mode (we denote it by ``in") and mode $a_1$.
However, in the presence of inefficient detectors (check Supplementary material), these ideal Bell-state measurements change  to $\Pi_\Psi^\pm (\eta_0) = \eta_0^2 \Pi_{\Psi}^\pm$ and $\Pi_\Phi^\pm (\eta_0) = \eta_0^2 \Pi_{\Phi}^\pm$.

For an input pure state $\ket{\psi_\text{in}}$ and the output mixed state $\rho^\text{out}$, the fidelity of teleportation is given by $F = \text{Tr}\argp{\ket{\psi_\text{in}}\bra{\psi_\text{in}} \rho^\text{out}}$.
This, given the probabilistic nature of the Bell-state measurements $P_{\Psi/\Phi}^\pm(\eta_0)$, leads to the average fidelity of teleportation for the chosen measurement settings as 
\begin{equation}
    F_\text{meas} = \sum_{\lambda,\pm} P_\lambda^\pm(\eta_0) F_\lambda^\pm(\eta_0),
    \label{eq:avfid_teleport_bellmeasure_expression}
\end{equation}
where $F_\lambda^\pm(\eta_0) = \bra{\psi_\text{in}} U_\lambda^\pm \rho_{b_1,\lambda}^\pm(\eta_0) \argp{U_\lambda^\pm}^\dagger \ket{\psi_\text{in}}$ such that $U_\lambda^\pm$ is the suitable unitary operator to be applied on output state in mode $b_1$, i.e., $\rho_{b_1,\lambda}^\pm(\eta_0)$, the corresponding to the projection operator $\Pi_\lambda^\pm(\eta_0)$ ($\lambda = \Psi, \Phi$).

Here, we are interested in the fidelity of teleportation averaged over all possible input states, i.e.,
\begin{equation}
    F_\text{av} = \frac{1}{2\pi}\int_0^1 dp \int_0^{2\pi} d\theta F_\text{meas}.
    \label{eq:avfid_teleport_input_expression}
\end{equation}
Accordingly, the quantum teleportation for the unknown input state can be characterized as $F_\text{av} > 2/3$.
The average fidelity of teleportation with noisy/imperfect detectors can be shown to be \eqref{append_eq:avfid_teleport_input}
\begin{equation}
    F_\text{av} = \eta_0^2 \argp{ 1 - \frac{2R}{3} } = \eta_0^2 \frac{2 + e^{-4(1-T\eta_0)\alpha^2}}{3}.
    \label{eq:avfid_teleport}
\end{equation}

This may further be noted that with perfect detectors, i.e., $\eta_0=1$, \eqref{eq:avfid_teleport} represents the fidelity of teleportation for the mixed quantum channel \eqref{eq:finalstate} for which the criterion for quantum teleportation ($F_\text{av}>2/3$) becomes $R<1/2$.
This condition matches exactly with the condition of Bell violation as observed from \eqref{eq:bellmax}.

We evaluate the Bell-CHSH violation and quantum teleportation of the input polarization qubit, for the shared DV entangled state, in the next section.

\section{Simulation Results}
\label{sec:Results}

\subsection{Probability of generating DV entangled-pair against transmission losses}
\label{subsec:prob_transloss}
Here, we analyze the efficacy of our scheme by considering its robustness against the transmission losses first. 
To facilitate that we consider the detectors to be ideal, i.e., $\eta_0=1$. 
Next, we consider the impact of detection in-efficiency on the performance of our scheme by considering non-ideal detectors.

\begin{figure}[htpb]
    \centering
    \includegraphics[width=\columnwidth]{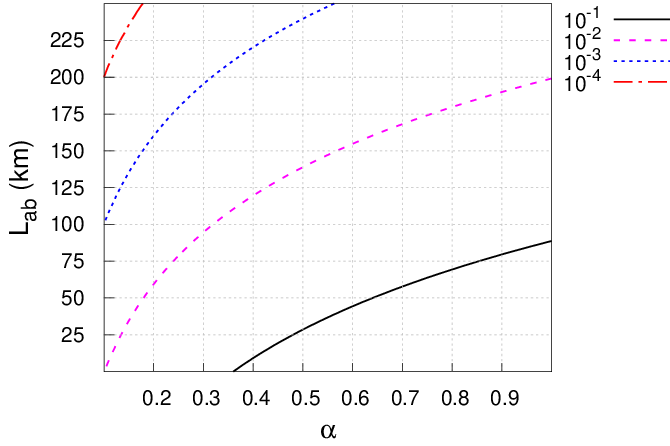}
    \caption{Contour plot for the probability ($\text{Pr}$) of obtaining the shared DV state with $L_\text{ab}$ and $\alpha$.
    Various curves show different contour values.
    We consider $\eta_0=1$.}
    \label{fig:probcont100}
\end{figure}

To begin with, in Fig. \ref{fig:probcont100}, we first plot the success probability of generating the DV entangled state \eqref{eq:probability} that manifests a non-monotonic character with $L_\text{ab}$ and $\alpha$, similar to earlier results observed in \cite{HOSEnt_Bose2024}.
For a fixed lab-separation, the probability of sharing the DV entangled pair first increases and then drops with increase in the coherent amplitude ($\alpha$).
This could be understood in terms of the interplay between probability of successful detection and loss-robustness of the transmitted signal.
As the $\alpha$ increases it enhances the chances of non-zero photon to detected by the click-event after passing through lossy optical fiber.
On the other hand, an increase in $\alpha$ also increases the mean photon-number of the signal which, in turn, makes the signal more vulnerable to transmission losses. 

\subsection{Bell violation and teleportation in presence of transmission losses only}
\label{subsec:bell_teleport_transloss}

\begin{figure}
    \centering
    \subfloat[\label{subfig:bell100}]{\includegraphics[width=\columnwidth]{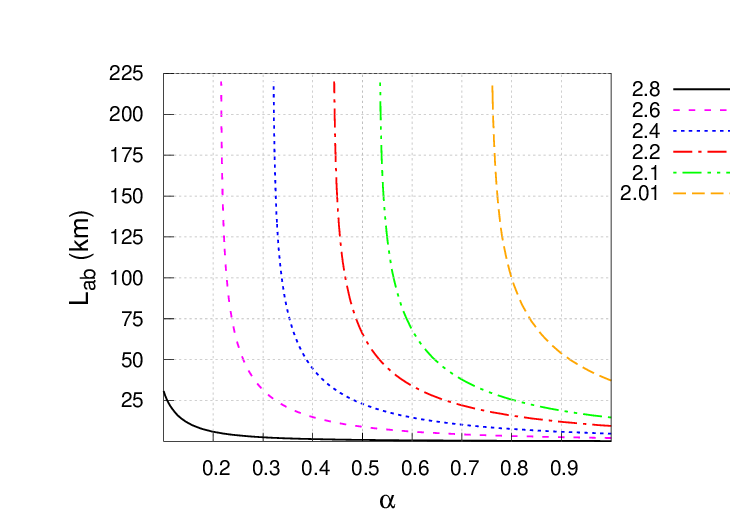}}
    \\
    \subfloat[\label{subfig:teleport100}] {\includegraphics[width=\columnwidth]{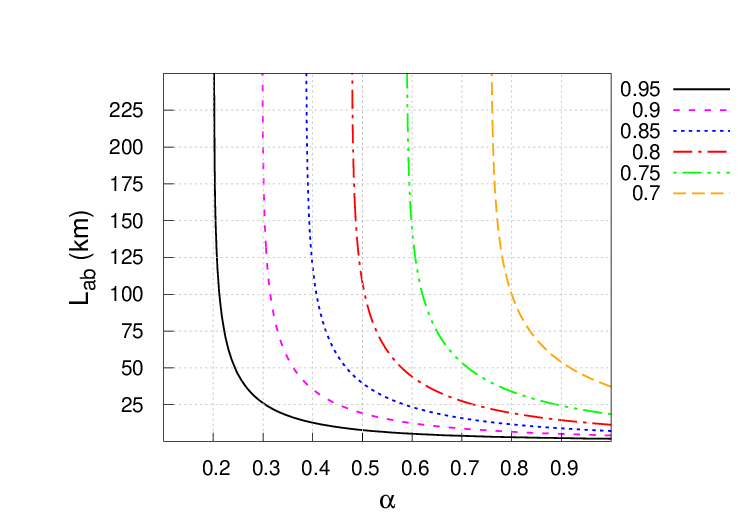}}
    \caption{Contour plot of \protect\subref{subfig:bell100} Bell-CHSH violation ($\mathcal{B}>2$) and \protect\subref{subfig:teleport100} quantum teleportation ($F_\text{av}>2/3$) vs lab separation ($L_\text{ab}$) and coherent amplitude ($\alpha$) with ideal detectors, i.e., $\eta_0=1$.} 
    \label{fig:bellteleport100}
\end{figure}

In Fig. \ref{fig:bellteleport100} we plot the Bell-violation ($\mathcal{B}>2$) and the fidelity of teleportation ($F_\text{av}>2/3$) for the shared DV state as a function of the lab-separation ($L_\text{ab}$) and the coherent amplitude ($\alpha$).
Here we have taken the detectors to be perfect, i.e., $\eta_0=1$. 
Transmission losses can be modeled by a lossy channel for which the transmission coefficient is given as $T = 10^{-0.2 L_\text{ab}/10}$ where $L_\text{ab}$ is the total lab separation in km.
(See {\bf Step 1} in the protocol section (\ref{sec:protocol}) for details).
It may be noted that, for the sake of simplicity, here we have displayed the results up to a distance ($\sim 250$ km).
However, from the Fig. \ref{fig:bellteleport100}, it is evident that our optical hybrid state based scheme can asymptotically lead to much larger distance. 

As it is evident from Figs. \ref{subfig:bell100} and \ref{subfig:teleport100}, the amount of Bell-CHSH violation and the fidelity of teleportation decrease with the increase in both the lab separation ($L_\text{ab}$) and the coherent amplitude ($\alpha$), even with ideal detectors.
This could be understood as follows.
The increase in lab separation introduces more transmission losses that lead to decrease in the correlation between the DV pair.
On the other hand, with increase in $\alpha$, the shared DV-state deviates from the ideal singlet state \eqref{eq:finalstate} further leading to a more mixed state between $\ket{\Psi^+}$ and $\ket{\Psi^-}$.

Besides, it is imperative to compare the Bell violation (Fig. \ref{subfig:bell100}), the fidelity of teleportation (Fig. \ref{subfig:teleport100}) and the success probability for the shared state (Fig. \ref{fig:probcont100}) in order to determine the optimal parameters that are experimentally accessible. 
At a given lab separation ($L_\text{ab}$), for very small $\alpha$ ($\sim 0.1$) the shared state closely approximates a perfect singlet state $\rho \sim \ket{\Psi^-}\bra{\Psi^-}$, which represents a maximally nonclocal state. 
As a result, it leads to high-quality teleportation with $F_\text{av} \sim 1$.
However, under the same conditions ($L_\text{ab}$ and $\alpha$), the success probability becomes negligible, i.e., $\text{Pr} \rightarrow 10^{-4}$.

Conversely, for a relatively larger value of $\alpha$ ($\gtrsim 0.5$) the shared state deviates significantly from an ideal singlet state ($\ket{\Psi^-}$) and thus the fidelity of teleportation drops closer to the classical limit $F_\text{av}=2/3$. 
Nonetheless, the success probability increases by $2 - 3$ orders of magnitude, reaching values on the order of $\text{Pr} \sim 10^{-1} - 10^{-2}$.
These observations indicate that, from a practical standpoint, it is essential to identify an optimal value of $\alpha$, depending on the physical constraints of the system.

\subsection{ Effect of detection inefficiency}
\label{subsec:detector_efficiency}

\begin{figure}
    \centering
    \subfloat[\label{bell95}]{\includegraphics[width=\columnwidth]{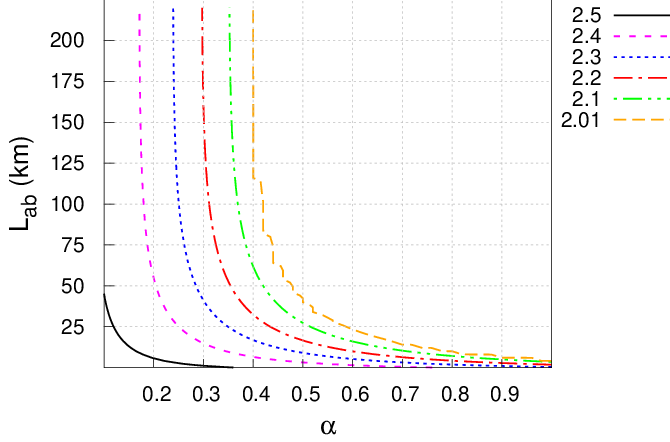}}
    \\
    \subfloat[\label{teleport95}] {\includegraphics[width=\columnwidth]{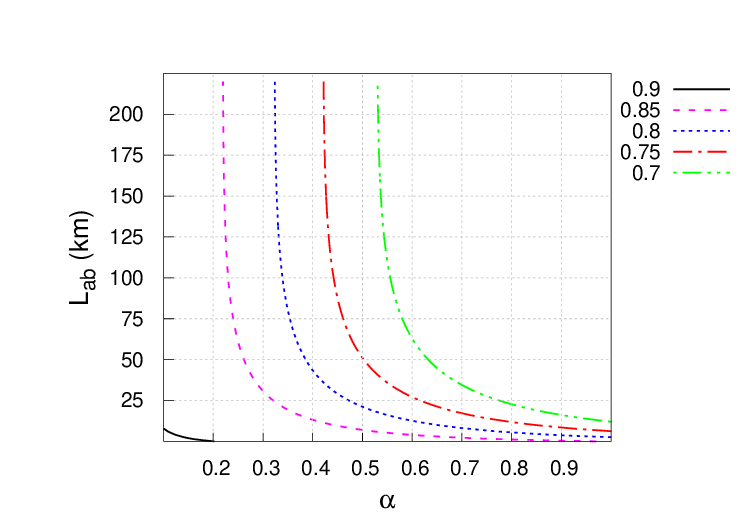}}
    \caption{Contour plot of \protect\subref{bell95} Bell-CHSH violation ($\mathcal{B}>2$) and \protect\subref{teleport95} quantum teleportation ($F_\text{av}>2/3$) vs lab separation ($L_\text{ab}$) and coherent amplitude ($\alpha$) with $5\%$ detection inefficiency, i.e., $\eta_0=0.95$.} 
    \label{fig:bellteleport_95}
\end{figure}

\begin{figure}
    \centering
    \subfloat[\label{bell90}]{\includegraphics[width=\columnwidth]{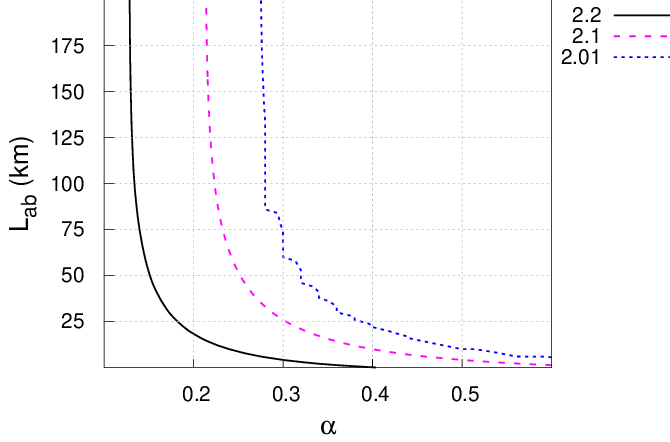}}
    \\
    \subfloat[\label{teleport90}] {\includegraphics[width=\columnwidth]{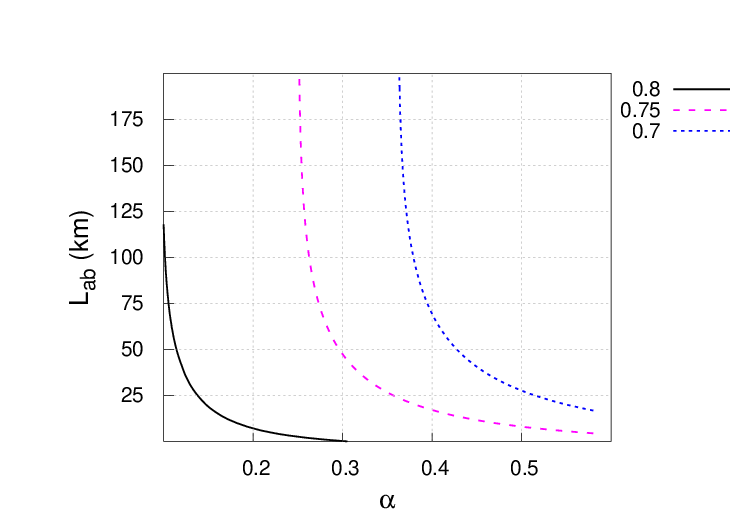}}
    \caption{Contour plot of \protect\subref{bell90} Bell-CHSH violation ($\mathcal{B}>2$) and \protect\subref{teleport90} quantum teleportation ($F_\text{av}>2/3$) vs lab separation ($L_\text{ab}$) and coherent amplitude ($\alpha$) with $10\%$ detection inefficiency, i.e., $\eta_0=0.90$.} 
    \label{fig:bellteleport_90}
\end{figure}
We now analyze the effect of inefficient detectors on the Bell nonlocality and the fidelity of teleportation for the shared DV state.
We elaborate our results on the Bell-CHSH violation ($\mathcal{B}>2$) and the quantum teleportation ($F_\text{av}>2/3$) in Figs. \ref{fig:bellteleport_95} and \ref{fig:bellteleport_90} with $95\%$ ($\eta_0=0.95$) and $90\%$ ($\eta_0=0.9$) detection efficiencies, respectively, in presence of transmission losses.

As it is evident from the Figs. \ref{fig:bellteleport_95} and \ref{fig:bellteleport_90}, the maximum Bell-CHSH violation as well as highest fidelity of teleportation get significantly reduced as the detection efficiency decreases.
The sharp drop in the performance of both noisy Bell measurement \eqref{eq:binary_measurement_rotated} and fidelity of teleportation \eqref{eq:avfid_teleport} is a reflection of the nonlinear nature of their dependence on the detector efficiency.
However, it may be noted that our scheme allows to achieve a distance ($\sim 200$ km) between the laboratories in presence of both transmission losses and inefficient detectors for measurable Bell-CHSH violation and teleportation fidelity of an unknown qubit.

\begin{figure}[htpb]
    \centering
    \includegraphics[width=\columnwidth]{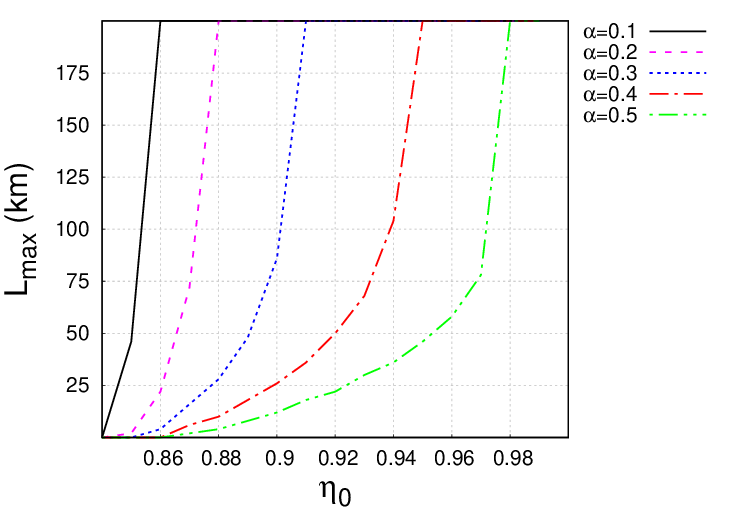}
    \caption{Maximum achievable distance vs detection efficiency for various values of $\alpha$.} 
    \label{fig:MaxLEtaCont}
\end{figure}

One may wonder how the maximum achievable lab separation, given $\mathcal{B}>2$, varies with the detection efficiency ($\eta_0$) and coherent amplitude ($\alpha$).
It may be noted that the expectation value of the joint measurement \eqref{eq:expression_jointmeasure} and thus the measurable quantity $\mathcal{B}$ scales as square of the detection efficiency ($\eta_0$).
This immediately implies that, even in the absence of any additional losses, the minimal detection efficiency required to observe Bell-violation is $\gtrsim 84\%$ ($\eta_0=0.84$).
As a consequence, in Fig. \ref{fig:MaxLEtaCont} we plot the maximum achievable lab separation ($L_\text{max}$) with varying detection efficiency ($0.84\leq \eta_0 \leq 1$). 
Various curves correspond to different choices of $\alpha$ as mentioned in the figure.
For the sake of simplicity, we have considered the lab separation up to $200$ km only; however, as it is evident from Fig. \ref{fig:MaxLEtaCont}, the maximum lab separation could be extended to larger distances based on the choice of $\alpha$ and $\eta_0$.

\section{Conclusions}
\label{sec:conclusion}

In this work, we have analyzed the efficacy of optical hybrid states in generating polarization-entangled DV states that exhibit Bell-CHSH correlations within a telecommunication framework. 
By leveraging the complementary advantages of both discrete-variable (DV) and continuous-variable (CV) systems, we have demonstrated that optical hybrid states enable significant Bell-CHSH violations over long distances, addressing a key challenge in fiber-optics-based quantum communication \cite{BellExp_Forgues2015, BellExp_Prabhakar2015, BellExp_Juan2016, BellExp_Martin2016, BellExp_Gonzalo2017, BellExp_Igor2018, BellExp_Oliver2018, BellExp_Zhang2018, BellExp_Zhong2019, BellFibOpt_Tim2022}.

Our numerical results indicate the feasibility of achieving high Bell-CHSH violation, supporting near-perfect quantum teleportation of an unknown qubit over distances up to $~250$ km. 
However, the performance of both the Bell-CHSH violation and the teleportation fidelity is constrained by detector inefficiencies.
Despite this limitation, our analysis shows that non-vanishing Bell-CHSH violation and reliable quantum teleportation remain achievable over distances of approximately $200$ km, even under transmission losses and with up to $10\%$ detection inefficiency. 
These results underscore the practical potential of optical hybrid states for long-distance quantum communication, presenting them as a compelling alternative to conventional DV-only and CV-only approaches.

It is worth noting that the issue of phase randomization in the transmitted signal—an inherent challenge in optical hybrid state based schemes, causes phase-mismatch between the incoming signals at Charlie's lab leading to destructive interference and thus destroys the correlation between Alice and Bob. 
This could be addressed through appropriate phase-locking mechanisms \cite{Phaselock_Minder2019, Phaselock_Wang2019, Phaselock_Pittaluga2021} where additional coherent state signals are sent for the phase reference.
Although, in a later approach, by employing tailored algorithms,   long-distance communication has been demonstrated without the need for phase-locking \cite{NoPhaselock_Pan2023}, applicability of such schemes in our protocol needs to verified.
Furthermore, the deterministic generation of high-purity optical hybrid states is reported within the $780-1064$ nm wavelength range \cite{HOSExp_Huang2019, HOSExp_Hacker2019, HOSExp_Giovanni2020}, When combined with frequency conversion techniques to telecom wavelengths ($\sim 1550$ nm) \cite{BellFibOpt_Tim2022} and time-bin encoding strategies \cite{HOSExp_Gouzien2020}, enhances the practical feasibility of our proposed scheme to integrate into existing fiber-optic infrastructure.

The current protocol is built on a generic structure for sharing entanglement over large distances based on the entanglement-swapping scheme as shown in \cite{HOSSwap_Lim2016}. 
We have analyzed the possibility of quantum teleportation over long distance as enabled by sharing Bell-CHSH nonlocal correlation between the distant labs. Accordingly, the measure of success is taken to be the maximum attainable distance that yields nonclassical fidelity of teleportation ($F>2/3$) under noisy transmission and imperfect detectors. 
Alternatively, in the context of communication one measure could be the shared secret key rate versus distance between the labs as explored in a previous work \cite{HOSEnt_Bose2024},  or by analyzing the feasibility of memoryless quantum repeater as outlined in \cite{RSBC_Li2023}.

In light of earlier studies on the distribution of Bell nonlocal correlations \cite{BellExp_Juan2016, BellExp_Martin2016, BellExp_Igor2018, BellExp_Oliver2018, BellExp_Zhong2019}, our analysis demonstrates the potential for significantly extending the lab separation to distances exceeding typical intercity scales. 
Additionally, our findings suggest a considerable improvement in the maximum achievable distance for faithful quantum teleportation \cite{BellFibOpt_Tim2022}.
Our results, particularly the viability of achieving Bell-CHSH violations over intercity distances, reinforce the promise of optical hybrid states as a robust alternative to DV-only and CV-only architectures, in line with previous works on entanglement distribution and hybrid quantum networking \cite{HOSSwap_Lim2016, HOSSwap_Parker2017, HOSEnt_Bose2024}.

Looking forward, this framework opens avenues for exploring loophole-free Bell tests \cite{Loopholefree_Xu2023} with hybrid states, potentially enabling device-independent quantum key distribution (DIQKD) \cite{DIQKD_Zhang2022, DIQKD_Liu2022, DIQKD_Victor2023, DIQKD_Lewis2024, DIQKD_Tan2024}. 
In this broader context, our results contribute to the ongoing efforts toward practical and scalable quantum information processing \cite{HOSQIP_Andersen2015}, offering resilience to teleportation noise \cite{HOSQT_Liu2024} and suitability for memoryless quantum communication \cite{RSBC_Li2023}.

\section{Acknowledgement}
This work was supported by the National Research Foundation of Korea (NRF) grant funded by the Korea government (MSIT) (Nos. RS-2024-00413957, RS-2024-00438415, and NRF-2023R1A2C1006115) and by the Institute of Information \& Communications Technology Planning \& Evaluation (IITP) grants funded by the Korea government (MSIT) (IITP-2025-RS-2020-II201606 and IITP-2025-RS-2024-00437191) through the Institute of Applied Physics at Seoul National University. 

Authors also declare no conflict of interest.


\begin{thebibliography}{108}%
\makeatletter
\providecommand \@ifxundefined [1]{%
 \@ifx{#1\undefined}
}%
\providecommand \@ifnum [1]{%
 \ifnum #1\expandafter \@firstoftwo
 \else \expandafter \@secondoftwo
 \fi
}%
\providecommand \@ifx [1]{%
 \ifx #1\expandafter \@firstoftwo
 \else \expandafter \@secondoftwo
 \fi
}%
\providecommand \natexlab [1]{#1}%
\providecommand \enquote  [1]{``#1''}%
\providecommand \bibnamefont  [1]{#1}%
\providecommand \bibfnamefont [1]{#1}%
\providecommand \citenamefont [1]{#1}%
\providecommand \href@noop [0]{\@secondoftwo}%
\providecommand \href [0]{\begingroup \@sanitize@url \@href}%
\providecommand \@href[1]{\@@startlink{#1}\@@href}%
\providecommand \@@href[1]{\endgroup#1\@@endlink}%
\providecommand \@sanitize@url [0]{\catcode `\\12\catcode `\$12\catcode
  `\&12\catcode `\#12\catcode `\^12\catcode `\_12\catcode `\%12\relax}%
\providecommand \@@startlink[1]{}%
\providecommand \@@endlink[0]{}%
\providecommand \url  [0]{\begingroup\@sanitize@url \@url }%
\providecommand \@url [1]{\endgroup\@href {#1}{\urlprefix }}%
\providecommand \urlprefix  [0]{URL }%
\providecommand \Eprint [0]{\href }%
\providecommand \doibase [0]{https://doi.org/}%
\providecommand \selectlanguage [0]{\@gobble}%
\providecommand \bibinfo  [0]{\@secondoftwo}%
\providecommand \bibfield  [0]{\@secondoftwo}%
\providecommand \translation [1]{[#1]}%
\providecommand \BibitemOpen [0]{}%
\providecommand \bibitemStop [0]{}%
\providecommand \bibitemNoStop [0]{.\EOS\space}%
\providecommand \EOS [0]{\spacefactor3000\relax}%
\providecommand \BibitemShut  [1]{\csname bibitem#1\endcsname}%
\let\auto@bib@innerbib\@empty
\bibitem [{\citenamefont {Bell}(1964)}]{Bell_Bell1964}%
  \BibitemOpen
  \bibfield  {author} {\bibinfo {author} {\bibfnamefont {J.~S.}\ \bibnamefont
  {Bell}},\ }\bibfield  {title} {\bibinfo {title} {On the einstein podolsky
  rosen paradox},\ }\href {https://doi.org/10.1103/PhysicsPhysiqueFizika.1.195}
  {\bibfield  {journal} {\bibinfo  {journal} {Physics Physique Fizika}\
  }\textbf {\bibinfo {volume} {1}},\ \bibinfo {pages} {195} (\bibinfo {year}
  {1964})}\BibitemShut {NoStop}%
\bibitem [{\citenamefont {Clauser}\ \emph {et~al.}(1969)\citenamefont
  {Clauser}, \citenamefont {Horne}, \citenamefont {Shimony},\ and\
  \citenamefont {Holt}}]{Bell_CHSH1969}%
  \BibitemOpen
  \bibfield  {author} {\bibinfo {author} {\bibfnamefont {J.~F.}\ \bibnamefont
  {Clauser}}, \bibinfo {author} {\bibfnamefont {M.~A.}\ \bibnamefont {Horne}},
  \bibinfo {author} {\bibfnamefont {A.}~\bibnamefont {Shimony}},\ and\ \bibinfo
  {author} {\bibfnamefont {R.~A.}\ \bibnamefont {Holt}},\ }\bibfield  {title}
  {\bibinfo {title} {Proposed experiment to test local hidden-variable
  theories},\ }\href {https://doi.org/10.1103/PhysRevLett.23.880} {\bibfield
  {journal} {\bibinfo  {journal} {Phys. Rev. Lett.}\ }\textbf {\bibinfo
  {volume} {23}},\ \bibinfo {pages} {880} (\bibinfo {year} {1969})}\BibitemShut
  {NoStop}%
\bibitem [{\citenamefont {Bell}\ and\ \citenamefont
  {Aspect}(2004)}]{BellBook_Aspect2004}%
  \BibitemOpen
  \bibfield  {author} {\bibinfo {author} {\bibfnamefont {J.~S.}\ \bibnamefont
  {Bell}}\ and\ \bibinfo {author} {\bibfnamefont {A.}~\bibnamefont {Aspect}},\
  }\href@noop {} {\emph {\bibinfo {title} {Speakable and Unspeakable in Quantum
  Mechanics: Collected Papers on Quantum Philosophy}}},\ \bibinfo {edition}
  {2nd}\ ed.\ (\bibinfo  {publisher} {Cambridge University Press},\ \bibinfo
  {year} {2004})\BibitemShut {NoStop}%
\bibitem [{\citenamefont {Zhang}\ \emph {et~al.}(2022)\citenamefont {Zhang},
  \citenamefont {van Leent}, \citenamefont {Redeker}, \citenamefont {Garthoff},
  \citenamefont {Schwonnek}, \citenamefont {Fertig}, \citenamefont {Eppelt},
  \citenamefont {Rosenfeld}, \citenamefont {Scarani}, \citenamefont {Lim},\
  and\ \citenamefont {Weinfurter}}]{DIQKD_Zhang2022}%
  \BibitemOpen
  \bibfield  {author} {\bibinfo {author} {\bibfnamefont {W.}~\bibnamefont
  {Zhang}}, \bibinfo {author} {\bibfnamefont {T.}~\bibnamefont {van Leent}},
  \bibinfo {author} {\bibfnamefont {K.}~\bibnamefont {Redeker}}, \bibinfo
  {author} {\bibfnamefont {R.}~\bibnamefont {Garthoff}}, \bibinfo {author}
  {\bibfnamefont {R.}~\bibnamefont {Schwonnek}}, \bibinfo {author}
  {\bibfnamefont {F.}~\bibnamefont {Fertig}}, \bibinfo {author} {\bibfnamefont
  {S.}~\bibnamefont {Eppelt}}, \bibinfo {author} {\bibfnamefont
  {W.}~\bibnamefont {Rosenfeld}}, \bibinfo {author} {\bibfnamefont
  {V.}~\bibnamefont {Scarani}}, \bibinfo {author} {\bibfnamefont {C.~C.-W.}\
  \bibnamefont {Lim}},\ and\ \bibinfo {author} {\bibfnamefont {H.}~\bibnamefont
  {Weinfurter}},\ }\bibfield  {title} {\bibinfo {title} {A device-independent
  quantum key distribution system for distant users},\ }\href
  {https://doi.org/10.1038/s41586-022-04891-y} {\bibfield  {journal} {\bibinfo
  {journal} {Nature}\ }\textbf {\bibinfo {volume} {607}},\ \bibinfo {pages}
  {687} (\bibinfo {year} {2022})}\BibitemShut {NoStop}%
\bibitem [{\citenamefont {Liu}\ \emph {et~al.}(2022)\citenamefont {Liu},
  \citenamefont {Zhang}, \citenamefont {Zhen}, \citenamefont {Li},
  \citenamefont {Liu}, \citenamefont {Fan}, \citenamefont {Xu}, \citenamefont
  {Zhang},\ and\ \citenamefont {Pan}}]{DIQKD_Liu2022}%
  \BibitemOpen
  \bibfield  {author} {\bibinfo {author} {\bibfnamefont {W.-Z.}\ \bibnamefont
  {Liu}}, \bibinfo {author} {\bibfnamefont {Y.-Z.}\ \bibnamefont {Zhang}},
  \bibinfo {author} {\bibfnamefont {Y.-Z.}\ \bibnamefont {Zhen}}, \bibinfo
  {author} {\bibfnamefont {M.-H.}\ \bibnamefont {Li}}, \bibinfo {author}
  {\bibfnamefont {Y.}~\bibnamefont {Liu}}, \bibinfo {author} {\bibfnamefont
  {J.}~\bibnamefont {Fan}}, \bibinfo {author} {\bibfnamefont {F.}~\bibnamefont
  {Xu}}, \bibinfo {author} {\bibfnamefont {Q.}~\bibnamefont {Zhang}},\ and\
  \bibinfo {author} {\bibfnamefont {J.-W.}\ \bibnamefont {Pan}},\ }\bibfield
  {title} {\bibinfo {title} {Toward a photonic demonstration of
  device-independent quantum key distribution},\ }\href
  {https://doi.org/10.1103/PhysRevLett.129.050502} {\bibfield  {journal}
  {\bibinfo  {journal} {Phys. Rev. Lett.}\ }\textbf {\bibinfo {volume} {129}},\
  \bibinfo {pages} {050502} (\bibinfo {year} {2022})}\BibitemShut {NoStop}%
\bibitem [{\citenamefont {Zapatero}\ \emph {et~al.}(2023)\citenamefont
  {Zapatero}, \citenamefont {van Leent}, \citenamefont {Arnon-Friedman},
  \citenamefont {Liu}, \citenamefont {Zhang}, \citenamefont {Weinfurter},\ and\
  \citenamefont {Curty}}]{DIQKD_Victor2023}%
  \BibitemOpen
  \bibfield  {author} {\bibinfo {author} {\bibfnamefont {V.}~\bibnamefont
  {Zapatero}}, \bibinfo {author} {\bibfnamefont {T.}~\bibnamefont {van Leent}},
  \bibinfo {author} {\bibfnamefont {R.}~\bibnamefont {Arnon-Friedman}},
  \bibinfo {author} {\bibfnamefont {W.-Z.}\ \bibnamefont {Liu}}, \bibinfo
  {author} {\bibfnamefont {Q.}~\bibnamefont {Zhang}}, \bibinfo {author}
  {\bibfnamefont {H.}~\bibnamefont {Weinfurter}},\ and\ \bibinfo {author}
  {\bibfnamefont {M.}~\bibnamefont {Curty}},\ }\bibfield  {title} {\bibinfo
  {title} {Advances in device-independent quantum key distribution},\ }\href
  {https://doi.org/10.1038/s41534-023-00684-x} {\bibfield  {journal} {\bibinfo
  {journal} {npj Quantum Information}\ ,\ \bibinfo {pages} {10}} (\bibinfo
  {year} {2023})}\BibitemShut {NoStop}%
\bibitem [{\citenamefont {Wooltorton}\ \emph {et~al.}(2024)\citenamefont
  {Wooltorton}, \citenamefont {Brown},\ and\ \citenamefont
  {Colbeck}}]{DIQKD_Lewis2024}%
  \BibitemOpen
  \bibfield  {author} {\bibinfo {author} {\bibfnamefont {L.}~\bibnamefont
  {Wooltorton}}, \bibinfo {author} {\bibfnamefont {P.}~\bibnamefont {Brown}},\
  and\ \bibinfo {author} {\bibfnamefont {R.}~\bibnamefont {Colbeck}},\
  }\bibfield  {title} {\bibinfo {title} {Device-independent quantum key
  distribution with arbitrarily small nonlocality},\ }\href
  {https://doi.org/10.1103/PhysRevLett.132.210802} {\bibfield  {journal}
  {\bibinfo  {journal} {Phys. Rev. Lett.}\ }\textbf {\bibinfo {volume} {132}},\
  \bibinfo {pages} {210802} (\bibinfo {year} {2024})}\BibitemShut {NoStop}%
\bibitem [{\citenamefont {Tan}\ and\ \citenamefont
  {Wolf}(2024)}]{DIQKD_Tan2024}%
  \BibitemOpen
  \bibfield  {author} {\bibinfo {author} {\bibfnamefont {E.~Y.-Z.}\
  \bibnamefont {Tan}}\ and\ \bibinfo {author} {\bibfnamefont {R.}~\bibnamefont
  {Wolf}},\ }\bibfield  {title} {\bibinfo {title} {Entropy bounds for
  device-independent quantum key distribution with local bell test},\ }\href
  {https://doi.org/10.1103/PhysRevLett.133.120803} {\bibfield  {journal}
  {\bibinfo  {journal} {Phys. Rev. Lett.}\ }\textbf {\bibinfo {volume} {133}},\
  \bibinfo {pages} {120803} (\bibinfo {year} {2024})}\BibitemShut {NoStop}%
\bibitem [{\citenamefont {Daniel}\ \emph {et~al.}(2022)\citenamefont {Daniel},
  \citenamefont {Zhu}, \citenamefont {Alderete}, \citenamefont {Buchemmavari},
  \citenamefont {Green}, \citenamefont {Nguyen}, \citenamefont {Thurtell},
  \citenamefont {Zhao}, \citenamefont {Linke},\ and\ \citenamefont
  {Miyake}}]{QCBell_Daniel2022}%
  \BibitemOpen
  \bibfield  {author} {\bibinfo {author} {\bibfnamefont {A.~K.}\ \bibnamefont
  {Daniel}}, \bibinfo {author} {\bibfnamefont {Y.}~\bibnamefont {Zhu}},
  \bibinfo {author} {\bibfnamefont {C.~H.}\ \bibnamefont {Alderete}}, \bibinfo
  {author} {\bibfnamefont {V.}~\bibnamefont {Buchemmavari}}, \bibinfo {author}
  {\bibfnamefont {A.~M.}\ \bibnamefont {Green}}, \bibinfo {author}
  {\bibfnamefont {N.~H.}\ \bibnamefont {Nguyen}}, \bibinfo {author}
  {\bibfnamefont {T.~G.}\ \bibnamefont {Thurtell}}, \bibinfo {author}
  {\bibfnamefont {A.}~\bibnamefont {Zhao}}, \bibinfo {author} {\bibfnamefont
  {N.~M.}\ \bibnamefont {Linke}},\ and\ \bibinfo {author} {\bibfnamefont
  {A.}~\bibnamefont {Miyake}},\ }\bibfield  {title} {\bibinfo {title} {Quantum
  computational advantage attested by nonlocal games with the cyclic cluster
  state},\ }\href {https://doi.org/10.1103/PhysRevResearch.4.033068} {\bibfield
   {journal} {\bibinfo  {journal} {Phys. Rev. Res.}\ }\textbf {\bibinfo
  {volume} {4}},\ \bibinfo {pages} {033068} (\bibinfo {year}
  {2022})}\BibitemShut {NoStop}%
\bibitem [{\citenamefont {Mackeprang}\ \emph {et~al.}(2023)\citenamefont
  {Mackeprang}, \citenamefont {Bhatti},\ and\ \citenamefont
  {Barz}}]{QCBell_Jelena2023}%
  \BibitemOpen
  \bibfield  {author} {\bibinfo {author} {\bibfnamefont {J.}~\bibnamefont
  {Mackeprang}}, \bibinfo {author} {\bibfnamefont {D.}~\bibnamefont {Bhatti}},\
  and\ \bibinfo {author} {\bibfnamefont {S.}~\bibnamefont {Barz}},\ }\bibfield
  {title} {\bibinfo {title} {Non-adaptive measurement-based quantum computation
  on ibm q},\ }\href {https://doi.org/10.1038/s41598-023-41025-4} {\bibfield
  {journal} {\bibinfo  {journal} {Scientific Reports}\ ,\ \bibinfo {pages}
  {15428}} (\bibinfo {year} {2023})}\BibitemShut {NoStop}%
\bibitem [{\citenamefont {Kalai}\ \emph {et~al.}(2023)\citenamefont {Kalai},
  \citenamefont {Lombardi}, \citenamefont {Vaikuntanathan},\ and\ \citenamefont
  {Yang}}]{QCBell_Kalai2023}%
  \BibitemOpen
  \bibfield  {author} {\bibinfo {author} {\bibfnamefont {Y.}~\bibnamefont
  {Kalai}}, \bibinfo {author} {\bibfnamefont {A.}~\bibnamefont {Lombardi}},
  \bibinfo {author} {\bibfnamefont {V.}~\bibnamefont {Vaikuntanathan}},\ and\
  \bibinfo {author} {\bibfnamefont {L.}~\bibnamefont {Yang}},\ }\bibfield
  {title} {\bibinfo {title} {Quantum advantage from any non-local game}\
  }(\bibinfo  {publisher} {Association for Computing Machinery},\ \bibinfo
  {year} {2023})\ p.\ \bibinfo {pages} {1617–1628}\BibitemShut {NoStop}%
\bibitem [{\citenamefont {Coladangelo}\ \emph {et~al.}(2017)\citenamefont
  {Coladangelo}, \citenamefont {Goh},\ and\ \citenamefont
  {Scarani}}]{BellSelfTEst_Andrea2017}%
  \BibitemOpen
  \bibfield  {author} {\bibinfo {author} {\bibfnamefont {A.}~\bibnamefont
  {Coladangelo}}, \bibinfo {author} {\bibfnamefont {K.~T.}\ \bibnamefont
  {Goh}},\ and\ \bibinfo {author} {\bibfnamefont {V.}~\bibnamefont {Scarani}},\
  }\bibfield  {title} {\bibinfo {title} {All pure bipartite entangled states
  can be self-tested},\ }\href {https://doi.org/10.1038/ncomms15485} {\bibfield
   {journal} {\bibinfo  {journal} {Nature Communications}\ }\textbf {\bibinfo
  {volume} {8}},\ \bibinfo {pages} {15485} (\bibinfo {year}
  {2017})}\BibitemShut {NoStop}%
\bibitem [{\citenamefont {Goswami}\ \emph {et~al.}(2018)\citenamefont
  {Goswami}, \citenamefont {Bhattacharya}, \citenamefont {Das}, \citenamefont
  {Sasmal}, \citenamefont {Jebaratnam},\ and\ \citenamefont
  {Majumdar}}]{BellSelfTest_Goswami2018}%
  \BibitemOpen
  \bibfield  {author} {\bibinfo {author} {\bibfnamefont {S.}~\bibnamefont
  {Goswami}}, \bibinfo {author} {\bibfnamefont {B.}~\bibnamefont
  {Bhattacharya}}, \bibinfo {author} {\bibfnamefont {D.}~\bibnamefont {Das}},
  \bibinfo {author} {\bibfnamefont {S.}~\bibnamefont {Sasmal}}, \bibinfo
  {author} {\bibfnamefont {C.}~\bibnamefont {Jebaratnam}},\ and\ \bibinfo
  {author} {\bibfnamefont {A.~S.}\ \bibnamefont {Majumdar}},\ }\bibfield
  {title} {\bibinfo {title} {One-sided device-independent self-testing of any
  pure two-qubit entangled state},\ }\href
  {https://doi.org/10.1103/PhysRevA.98.022311} {\bibfield  {journal} {\bibinfo
  {journal} {Phys. Rev. A}\ }\textbf {\bibinfo {volume} {98}},\ \bibinfo
  {pages} {022311} (\bibinfo {year} {2018})}\BibitemShut {NoStop}%
\bibitem [{\citenamefont {Zhang}\ \emph
  {et~al.}(2019{\natexlab{a}})\citenamefont {Zhang}, \citenamefont {Chen},
  \citenamefont {Peng}, \citenamefont {Ye}, \citenamefont {Yin}, \citenamefont
  {Xu}, \citenamefont {Xu}, \citenamefont {Li},\ and\ \citenamefont
  {Guo}}]{BellSelfTest_Zhang2019}%
  \BibitemOpen
  \bibfield  {author} {\bibinfo {author} {\bibfnamefont {W.-H.}\ \bibnamefont
  {Zhang}}, \bibinfo {author} {\bibfnamefont {G.}~\bibnamefont {Chen}},
  \bibinfo {author} {\bibfnamefont {X.-X.}\ \bibnamefont {Peng}}, \bibinfo
  {author} {\bibfnamefont {X.-J.}\ \bibnamefont {Ye}}, \bibinfo {author}
  {\bibfnamefont {P.}~\bibnamefont {Yin}}, \bibinfo {author} {\bibfnamefont
  {X.-Y.}\ \bibnamefont {Xu}}, \bibinfo {author} {\bibfnamefont {J.-S.}\
  \bibnamefont {Xu}}, \bibinfo {author} {\bibfnamefont {C.-F.}\ \bibnamefont
  {Li}},\ and\ \bibinfo {author} {\bibfnamefont {G.-C.}\ \bibnamefont {Guo}},\
  }\bibfield  {title} {\bibinfo {title} {Experimental realization of robust
  self-testing of bell state measurements},\ }\href
  {https://doi.org/10.1103/PhysRevLett.122.090402} {\bibfield  {journal}
  {\bibinfo  {journal} {Phys. Rev. Lett.}\ }\textbf {\bibinfo {volume} {122}},\
  \bibinfo {pages} {090402} (\bibinfo {year} {2019}{\natexlab{a}})}\BibitemShut
  {NoStop}%
\bibitem [{\citenamefont {Zhang}\ \emph
  {et~al.}(2019{\natexlab{b}})\citenamefont {Zhang}, \citenamefont {Chen},
  \citenamefont {Yin}, \citenamefont {Peng}, \citenamefont {Hu}, \citenamefont
  {Hou}, \citenamefont {Zhou}, \citenamefont {Yu}, \citenamefont {Ye},
  \citenamefont {Zhou}, \citenamefont {Xu}, \citenamefont {Tang}, \citenamefont
  {Xu}, \citenamefont {Han}, \citenamefont {Liu}, \citenamefont {Li},\ and\
  \citenamefont {Guo}}]{BellSelfTest_Zhang2019_2}%
  \BibitemOpen
  \bibfield  {author} {\bibinfo {author} {\bibfnamefont {W.-H.}\ \bibnamefont
  {Zhang}}, \bibinfo {author} {\bibfnamefont {G.}~\bibnamefont {Chen}},
  \bibinfo {author} {\bibfnamefont {P.}~\bibnamefont {Yin}}, \bibinfo {author}
  {\bibfnamefont {X.-X.}\ \bibnamefont {Peng}}, \bibinfo {author}
  {\bibfnamefont {X.-M.}\ \bibnamefont {Hu}}, \bibinfo {author} {\bibfnamefont
  {Z.-B.}\ \bibnamefont {Hou}}, \bibinfo {author} {\bibfnamefont {Z.-Y.}\
  \bibnamefont {Zhou}}, \bibinfo {author} {\bibfnamefont {S.}~\bibnamefont
  {Yu}}, \bibinfo {author} {\bibfnamefont {X.-J.}\ \bibnamefont {Ye}}, \bibinfo
  {author} {\bibfnamefont {Z.-Q.}\ \bibnamefont {Zhou}}, \bibinfo {author}
  {\bibfnamefont {X.-Y.}\ \bibnamefont {Xu}}, \bibinfo {author} {\bibfnamefont
  {J.-S.}\ \bibnamefont {Tang}}, \bibinfo {author} {\bibfnamefont {J.-S.}\
  \bibnamefont {Xu}}, \bibinfo {author} {\bibfnamefont {Y.-J.}\ \bibnamefont
  {Han}}, \bibinfo {author} {\bibfnamefont {B.-H.}\ \bibnamefont {Liu}},
  \bibinfo {author} {\bibfnamefont {C.-F.}\ \bibnamefont {Li}},\ and\ \bibinfo
  {author} {\bibfnamefont {G.-C.}\ \bibnamefont {Guo}},\ }\bibfield  {title}
  {\bibinfo {title} {Experimental demonstration of robust self-testing for
  bipartite entangled states},\ }\href
  {https://doi.org/10.1038/s41534-018-0120-0} {\bibfield  {journal} {\bibinfo
  {journal} {npj Quantum Information}\ }\textbf {\bibinfo {volume} {5}},\
  \bibinfo {pages} {4} (\bibinfo {year} {2019}{\natexlab{b}})}\BibitemShut
  {NoStop}%
\bibitem [{\citenamefont {Bierhorst}\ \emph {et~al.}(2018)\citenamefont
  {Bierhorst}, \citenamefont {Knill}, \citenamefont {Glancy}, \citenamefont
  {Zhang}, \citenamefont {Mink}, \citenamefont {Jordan}, \citenamefont
  {Rommal}, \citenamefont {Liu}, \citenamefont {Christensen}, \citenamefont
  {Nam}, \citenamefont {Stevens},\ and\ \citenamefont
  {Shalm}}]{BellDICert_Peter2016}%
  \BibitemOpen
  \bibfield  {author} {\bibinfo {author} {\bibfnamefont {P.}~\bibnamefont
  {Bierhorst}}, \bibinfo {author} {\bibfnamefont {E.}~\bibnamefont {Knill}},
  \bibinfo {author} {\bibfnamefont {S.}~\bibnamefont {Glancy}}, \bibinfo
  {author} {\bibfnamefont {Y.}~\bibnamefont {Zhang}}, \bibinfo {author}
  {\bibfnamefont {A.}~\bibnamefont {Mink}}, \bibinfo {author} {\bibfnamefont
  {S.}~\bibnamefont {Jordan}}, \bibinfo {author} {\bibfnamefont
  {A.}~\bibnamefont {Rommal}}, \bibinfo {author} {\bibfnamefont {Y.-K.}\
  \bibnamefont {Liu}}, \bibinfo {author} {\bibfnamefont {B.}~\bibnamefont
  {Christensen}}, \bibinfo {author} {\bibfnamefont {S.~W.}\ \bibnamefont
  {Nam}}, \bibinfo {author} {\bibfnamefont {M.~J.}\ \bibnamefont {Stevens}},\
  and\ \bibinfo {author} {\bibfnamefont {L.~K.}\ \bibnamefont {Shalm}},\
  }\bibfield  {title} {\bibinfo {title} {Experimentally generated randomness
  certified by the impossibility of superluminal signals},\ }\href
  {https://doi.org/10.1038/s41586-018-0019-0} {\bibfield  {journal} {\bibinfo
  {journal} {Nature}\ }\textbf {\bibinfo {volume} {556}},\ \bibinfo {pages}
  {223} (\bibinfo {year} {2018})}\BibitemShut {NoStop}%
\bibitem [{\citenamefont {Ac\'{\i}n}\ \emph {et~al.}(2016)\citenamefont
  {Ac\'{\i}n}, \citenamefont {Pironio}, \citenamefont {V\'ertesi},\ and\
  \citenamefont {Wittek}}]{BellDICert_Acin2016}%
  \BibitemOpen
  \bibfield  {author} {\bibinfo {author} {\bibfnamefont {A.}~\bibnamefont
  {Ac\'{\i}n}}, \bibinfo {author} {\bibfnamefont {S.}~\bibnamefont {Pironio}},
  \bibinfo {author} {\bibfnamefont {T.}~\bibnamefont {V\'ertesi}},\ and\
  \bibinfo {author} {\bibfnamefont {P.}~\bibnamefont {Wittek}},\ }\bibfield
  {title} {\bibinfo {title} {Optimal randomness certification from one
  entangled bit},\ }\href {https://doi.org/10.1103/PhysRevA.93.040102}
  {\bibfield  {journal} {\bibinfo  {journal} {Phys. Rev. A}\ }\textbf {\bibinfo
  {volume} {93}},\ \bibinfo {pages} {040102} (\bibinfo {year}
  {2016})}\BibitemShut {NoStop}%
\bibitem [{\citenamefont {Acín}\ and\ \citenamefont
  {Masanes}(2016)}]{BellDICert_Acin2016_2}%
  \BibitemOpen
  \bibfield  {author} {\bibinfo {author} {\bibfnamefont {A.}~\bibnamefont
  {Acín}}\ and\ \bibinfo {author} {\bibfnamefont {L.}~\bibnamefont
  {Masanes}},\ }\bibfield  {title} {\bibinfo {title} {Certified randomness in
  quantum physics},\ }\href {https://doi.org/10.1038/nature20119} {\bibfield
  {journal} {\bibinfo  {journal} {Nature}\ }\textbf {\bibinfo {volume} {540}},\
  \bibinfo {pages} {213} (\bibinfo {year} {2016})}\BibitemShut {NoStop}%
\bibitem [{\citenamefont {Bowles}\ \emph {et~al.}(2018)\citenamefont {Bowles},
  \citenamefont {\ifmmode \check{S}\else \v{S}\fi{}upi\ifmmode~\acute{c}\else
  \'{c}\fi{}}, \citenamefont {Cavalcanti},\ and\ \citenamefont
  {Ac\'{\i}n}}]{BellDICert_Joseph2018}%
  \BibitemOpen
  \bibfield  {author} {\bibinfo {author} {\bibfnamefont {J.}~\bibnamefont
  {Bowles}}, \bibinfo {author} {\bibfnamefont {I.}~\bibnamefont {\ifmmode
  \check{S}\else \v{S}\fi{}upi\ifmmode~\acute{c}\else \'{c}\fi{}}}, \bibinfo
  {author} {\bibfnamefont {D.}~\bibnamefont {Cavalcanti}},\ and\ \bibinfo
  {author} {\bibfnamefont {A.}~\bibnamefont {Ac\'{\i}n}},\ }\bibfield  {title}
  {\bibinfo {title} {Device-independent entanglement certification of all
  entangled states},\ }\href {https://doi.org/10.1103/PhysRevLett.121.180503}
  {\bibfield  {journal} {\bibinfo  {journal} {Phys. Rev. Lett.}\ }\textbf
  {\bibinfo {volume} {121}},\ \bibinfo {pages} {180503} (\bibinfo {year}
  {2018})}\BibitemShut {NoStop}%
\bibitem [{\citenamefont {Andersson}\ \emph {et~al.}(2018)\citenamefont
  {Andersson}, \citenamefont {Badzikag}, \citenamefont {Dumitru},\ and\
  \citenamefont {Cabello}}]{BellDICert_Ole2018}%
  \BibitemOpen
  \bibfield  {author} {\bibinfo {author} {\bibfnamefont {O.}~\bibnamefont
  {Andersson}}, \bibinfo {author} {\bibfnamefont {P.}~\bibnamefont {Badzikag}},
  \bibinfo {author} {\bibfnamefont {I.}~\bibnamefont {Dumitru}},\ and\ \bibinfo
  {author} {\bibfnamefont {A.}~\bibnamefont {Cabello}},\ }\bibfield  {title}
  {\bibinfo {title} {Device-independent certification of two bits of randomness
  from one entangled bit and gisin's elegant bell inequality},\ }\href
  {https://doi.org/10.1103/PhysRevA.97.012314} {\bibfield  {journal} {\bibinfo
  {journal} {Phys. Rev. A}\ }\textbf {\bibinfo {volume} {97}},\ \bibinfo
  {pages} {012314} (\bibinfo {year} {2018})}\BibitemShut {NoStop}%
\bibitem [{\citenamefont {Ghadimi}\ \emph {et~al.}(2018)\citenamefont
  {Ghadimi}, \citenamefont {Hall},\ and\ \citenamefont
  {Wiseman}}]{BellRev_Ghadimi2018}%
  \BibitemOpen
  \bibfield  {author} {\bibinfo {author} {\bibfnamefont {M.}~\bibnamefont
  {Ghadimi}}, \bibinfo {author} {\bibfnamefont {M.~J.~W.}\ \bibnamefont
  {Hall}},\ and\ \bibinfo {author} {\bibfnamefont {H.~M.}\ \bibnamefont
  {Wiseman}},\ }\bibfield  {title} {\bibinfo {title} {Nonlocality in bell’s
  theorem, in bohm’s theory, and in many interacting worlds theorising},\
  }\href {https://doi.org/10.3390/e20080567} {\bibfield  {journal} {\bibinfo
  {journal} {Entropy}\ }\textbf {\bibinfo {volume} {20}},\ \bibinfo {pages}
  {567} (\bibinfo {year} {2018})}\BibitemShut {NoStop}%
\bibitem [{\citenamefont {Scarani}(2019)}]{BellRevBook_Scarani2019}%
  \BibitemOpen
  \bibfield  {author} {\bibinfo {author} {\bibfnamefont {V.}~\bibnamefont
  {Scarani}},\ }\href {https://doi.org/10.1093/oso/9780198788416.001.0001}
  {\emph {\bibinfo {title} {Bell Nonlocality}}},\ \bibinfo {edition} {1st}\
  ed.\ (\bibinfo  {publisher} {Oxford University Press},\ \bibinfo {year}
  {2019})\BibitemShut {NoStop}%
\bibitem [{\citenamefont {Forgues}\ \emph {et~al.}(2015)\citenamefont
  {Forgues}, \citenamefont {Lupien},\ and\ \citenamefont
  {Reulet}}]{BellExp_Forgues2015}%
  \BibitemOpen
  \bibfield  {author} {\bibinfo {author} {\bibfnamefont {J.-C.}\ \bibnamefont
  {Forgues}}, \bibinfo {author} {\bibfnamefont {C.}~\bibnamefont {Lupien}},\
  and\ \bibinfo {author} {\bibfnamefont {B.}~\bibnamefont {Reulet}},\
  }\bibfield  {title} {\bibinfo {title} {Experimental violation of bell-like
  inequalities by electronic shot noise},\ }\href
  {https://doi.org/10.1103/PhysRevLett.114.130403} {\bibfield  {journal}
  {\bibinfo  {journal} {Phys. Rev. Lett.}\ }\textbf {\bibinfo {volume} {114}},\
  \bibinfo {pages} {130403} (\bibinfo {year} {2015})}\BibitemShut {NoStop}%
\bibitem [{\citenamefont {Prabhakar}\ \emph {et~al.}(2015)\citenamefont
  {Prabhakar}, \citenamefont {Reddy}, \citenamefont {Aadhi}, \citenamefont
  {Perumangatt}, \citenamefont {Samanta},\ and\ \citenamefont
  {Singh}}]{BellExp_Prabhakar2015}%
  \BibitemOpen
  \bibfield  {author} {\bibinfo {author} {\bibfnamefont {S.}~\bibnamefont
  {Prabhakar}}, \bibinfo {author} {\bibfnamefont {S.~G.}\ \bibnamefont
  {Reddy}}, \bibinfo {author} {\bibfnamefont {A.}~\bibnamefont {Aadhi}},
  \bibinfo {author} {\bibfnamefont {C.}~\bibnamefont {Perumangatt}}, \bibinfo
  {author} {\bibfnamefont {G.~K.}\ \bibnamefont {Samanta}},\ and\ \bibinfo
  {author} {\bibfnamefont {R.~P.}\ \bibnamefont {Singh}},\ }\bibfield  {title}
  {\bibinfo {title} {Violation of bell's inequality for phase-singular beams},\
  }\href {https://doi.org/10.1103/PhysRevA.92.023822} {\bibfield  {journal}
  {\bibinfo  {journal} {Phys. Rev. A}\ }\textbf {\bibinfo {volume} {92}},\
  \bibinfo {pages} {023822} (\bibinfo {year} {2015})}\BibitemShut {NoStop}%
\bibitem [{\citenamefont {Dehollain}\ \emph {et~al.}(2016)\citenamefont
  {Dehollain}, \citenamefont {Simmons}, \citenamefont {Muhonen}, \citenamefont
  {Kalra}, \citenamefont {Laucht}, \citenamefont {Hudson}, \citenamefont
  {Itoh}, \citenamefont {Jamieson}, \citenamefont {McCallum}, \citenamefont
  {Dzurak},\ and\ \citenamefont {Morello}}]{BellExp_Juan2016}%
  \BibitemOpen
  \bibfield  {author} {\bibinfo {author} {\bibfnamefont {J.~P.}\ \bibnamefont
  {Dehollain}}, \bibinfo {author} {\bibfnamefont {S.}~\bibnamefont {Simmons}},
  \bibinfo {author} {\bibfnamefont {J.~T.}\ \bibnamefont {Muhonen}}, \bibinfo
  {author} {\bibfnamefont {R.}~\bibnamefont {Kalra}}, \bibinfo {author}
  {\bibfnamefont {A.}~\bibnamefont {Laucht}}, \bibinfo {author} {\bibfnamefont
  {F.}~\bibnamefont {Hudson}}, \bibinfo {author} {\bibfnamefont {K.~M.}\
  \bibnamefont {Itoh}}, \bibinfo {author} {\bibfnamefont {D.~N.}\ \bibnamefont
  {Jamieson}}, \bibinfo {author} {\bibfnamefont {J.~C.}\ \bibnamefont
  {McCallum}}, \bibinfo {author} {\bibfnamefont {A.~S.}\ \bibnamefont
  {Dzurak}},\ and\ \bibinfo {author} {\bibfnamefont {A.}~\bibnamefont
  {Morello}},\ }\bibfield  {title} {\bibinfo {title} {Bell's inequality
  violation with spins in silicon},\ }\href
  {https://doi.org/10.1038/nnano.2015.262} {\bibfield  {journal} {\bibinfo
  {journal} {Nature Nanotechnology}\ }\textbf {\bibinfo {volume} {11}},\
  \bibinfo {pages} {242} (\bibinfo {year} {2016})}\BibitemShut {NoStop}%
\bibitem [{\citenamefont {Ringbauer}\ \emph {et~al.}(2016)\citenamefont
  {Ringbauer}, \citenamefont {Giarmatzi}, \citenamefont {Chaves}, \citenamefont
  {Costa}, \citenamefont {White},\ and\ \citenamefont
  {Fedrizzi}}]{BellExp_Martin2016}%
  \BibitemOpen
  \bibfield  {author} {\bibinfo {author} {\bibfnamefont {M.}~\bibnamefont
  {Ringbauer}}, \bibinfo {author} {\bibfnamefont {C.}~\bibnamefont
  {Giarmatzi}}, \bibinfo {author} {\bibfnamefont {R.}~\bibnamefont {Chaves}},
  \bibinfo {author} {\bibfnamefont {F.}~\bibnamefont {Costa}}, \bibinfo
  {author} {\bibfnamefont {A.~G.}\ \bibnamefont {White}},\ and\ \bibinfo
  {author} {\bibfnamefont {A.}~\bibnamefont {Fedrizzi}},\ }\bibfield  {title}
  {\bibinfo {title} {Experimental test of nonlocal causality},\ }\href
  {https://doi.org/10.1126/sciadv.1600162} {\bibfield  {journal} {\bibinfo
  {journal} {Science Advances}\ }\textbf {\bibinfo {volume} {2}},\ \bibinfo
  {pages} {e1600162} (\bibinfo {year} {2016})}\BibitemShut {NoStop}%
\bibitem [{Bel(2017)}]{BellExp_Gonzalo2017}%
  \BibitemOpen
  \bibfield  {title} {\bibinfo {title} {Experimental violation of local
  causality in a quantum network},\ }\href
  {https://doi.org/10.1038/ncomms14775} {\bibfield  {journal} {\bibinfo
  {journal} {Nature Communications}\ }\textbf {\bibinfo {volume} {8}},\
  \bibinfo {pages} {14775} (\bibinfo {year} {2017})}\BibitemShut {NoStop}%
\bibitem [{\citenamefont {Marinkovic}\ \emph {et~al.}(2018)\citenamefont
  {Marinkovic}, \citenamefont {Wallucks}, \citenamefont {Riedinger},
  \citenamefont {Hong}, \citenamefont {Aspelmeyer},\ and\ \citenamefont
  {Gr\"oblacher}}]{BellExp_Igor2018}%
  \BibitemOpen
  \bibfield  {author} {\bibinfo {author} {\bibfnamefont {I.}~\bibnamefont
  {Marinkovic}}, \bibinfo {author} {\bibfnamefont {A.}~\bibnamefont
  {Wallucks}}, \bibinfo {author} {\bibfnamefont {R.}~\bibnamefont {Riedinger}},
  \bibinfo {author} {\bibfnamefont {S.}~\bibnamefont {Hong}}, \bibinfo {author}
  {\bibfnamefont {M.}~\bibnamefont {Aspelmeyer}},\ and\ \bibinfo {author}
  {\bibfnamefont {S.}~\bibnamefont {Gr\"oblacher}},\ }\bibfield  {title}
  {\bibinfo {title} {Optomechanical bell test},\ }\href
  {https://doi.org/10.1103/PhysRevLett.121.220404} {\bibfield  {journal}
  {\bibinfo  {journal} {Phys. Rev. Lett.}\ }\textbf {\bibinfo {volume} {121}},\
  \bibinfo {pages} {220404} (\bibinfo {year} {2018})}\BibitemShut {NoStop}%
\bibitem [{\citenamefont {Thearle}\ \emph {et~al.}(2018)\citenamefont
  {Thearle}, \citenamefont {Janousek}, \citenamefont {Armstrong}, \citenamefont
  {Hosseini}, \citenamefont {Sch\"unemann~(Mraz)}, \citenamefont {Assad},
  \citenamefont {Symul}, \citenamefont {James}, \citenamefont {Huntington},
  \citenamefont {Ralph},\ and\ \citenamefont {Lam}}]{BellExp_Oliver2018}%
  \BibitemOpen
  \bibfield  {author} {\bibinfo {author} {\bibfnamefont {O.}~\bibnamefont
  {Thearle}}, \bibinfo {author} {\bibfnamefont {J.}~\bibnamefont {Janousek}},
  \bibinfo {author} {\bibfnamefont {S.}~\bibnamefont {Armstrong}}, \bibinfo
  {author} {\bibfnamefont {S.}~\bibnamefont {Hosseini}}, \bibinfo {author}
  {\bibfnamefont {M.}~\bibnamefont {Sch\"unemann~(Mraz)}}, \bibinfo {author}
  {\bibfnamefont {S.}~\bibnamefont {Assad}}, \bibinfo {author} {\bibfnamefont
  {T.}~\bibnamefont {Symul}}, \bibinfo {author} {\bibfnamefont {M.~R.}\
  \bibnamefont {James}}, \bibinfo {author} {\bibfnamefont {E.}~\bibnamefont
  {Huntington}}, \bibinfo {author} {\bibfnamefont {T.~C.}\ \bibnamefont
  {Ralph}},\ and\ \bibinfo {author} {\bibfnamefont {P.~K.}\ \bibnamefont
  {Lam}},\ }\bibfield  {title} {\bibinfo {title} {Violation of bell's
  inequality using continuous variable measurements},\ }\href
  {https://doi.org/10.1103/PhysRevLett.120.040406} {\bibfield  {journal}
  {\bibinfo  {journal} {Phys. Rev. Lett.}\ }\textbf {\bibinfo {volume} {120}},\
  \bibinfo {pages} {040406} (\bibinfo {year} {2018})}\BibitemShut {NoStop}%
\bibitem [{\citenamefont {Zhang}\ \emph {et~al.}(2018)\citenamefont {Zhang},
  \citenamefont {Qiu}, \citenamefont {Zhang},\ and\ \citenamefont
  {Chen}}]{BellExp_Zhang2018}%
  \BibitemOpen
  \bibfield  {author} {\bibinfo {author} {\bibfnamefont {D.}~\bibnamefont
  {Zhang}}, \bibinfo {author} {\bibfnamefont {X.}~\bibnamefont {Qiu}}, \bibinfo
  {author} {\bibfnamefont {W.}~\bibnamefont {Zhang}},\ and\ \bibinfo {author}
  {\bibfnamefont {L.}~\bibnamefont {Chen}},\ }\bibfield  {title} {\bibinfo
  {title} {Violation of a bell inequality in two-dimensional state spaces for
  radial quantum number},\ }\href {https://doi.org/10.1103/PhysRevA.98.042134}
  {\bibfield  {journal} {\bibinfo  {journal} {Phys. Rev. A}\ }\textbf {\bibinfo
  {volume} {98}},\ \bibinfo {pages} {042134} (\bibinfo {year}
  {2018})}\BibitemShut {NoStop}%
\bibitem [{\citenamefont {Zhong}\ \emph {et~al.}(2019)\citenamefont {Zhong},
  \citenamefont {Chang}, \citenamefont {Satzinger}, \citenamefont {Chou},
  \citenamefont {Bienfait}, \citenamefont {Conner}, \citenamefont {Dumur},
  \citenamefont {Grebel}, \citenamefont {Peairs}, \citenamefont {Povey},
  \citenamefont {Schuster},\ and\ \citenamefont {Cleland}}]{BellExp_Zhong2019}%
  \BibitemOpen
  \bibfield  {author} {\bibinfo {author} {\bibfnamefont {Y.~P.}\ \bibnamefont
  {Zhong}}, \bibinfo {author} {\bibfnamefont {H.-S.}\ \bibnamefont {Chang}},
  \bibinfo {author} {\bibfnamefont {K.~J.}\ \bibnamefont {Satzinger}}, \bibinfo
  {author} {\bibfnamefont {M.-H.}\ \bibnamefont {Chou}}, \bibinfo {author}
  {\bibfnamefont {A.}~\bibnamefont {Bienfait}}, \bibinfo {author}
  {\bibfnamefont {C.~R.}\ \bibnamefont {Conner}}, \bibinfo {author}
  {\bibfnamefont {E.}~\bibnamefont {Dumur}}, \bibinfo {author} {\bibfnamefont
  {J.}~\bibnamefont {Grebel}}, \bibinfo {author} {\bibfnamefont {G.~A.}\
  \bibnamefont {Peairs}}, \bibinfo {author} {\bibfnamefont {R.~G.}\
  \bibnamefont {Povey}}, \bibinfo {author} {\bibfnamefont {D.~I.}\ \bibnamefont
  {Schuster}},\ and\ \bibinfo {author} {\bibfnamefont {A.~N.}\ \bibnamefont
  {Cleland}},\ }\bibfield  {title} {\bibinfo {title} {Violating bell’s
  inequality with remotely connected superconducting qubits},\ }\href
  {https://doi.org/10.1038/s41567-019-0507-7} {\bibfield  {journal} {\bibinfo
  {journal} {Nature Physics}\ }\textbf {\bibinfo {volume} {15}},\ \bibinfo
  {pages} {741} (\bibinfo {year} {2019})}\BibitemShut {NoStop}%
\bibitem [{\citenamefont {van Leent}\ \emph {et~al.}(2022)\citenamefont {van
  Leent}, \citenamefont {Bock}, \citenamefont {Fertig}, \citenamefont
  {Garthoff}, \citenamefont {Eppelt}, \citenamefont {Zhou}, \citenamefont
  {Malik}, \citenamefont {Seubert}, \citenamefont {Bauer}, \citenamefont
  {Rosenfeld}, \citenamefont {Zhang}, \citenamefont {Becher},\ and\
  \citenamefont {Weinfurter}}]{BellFibOpt_Tim2022}%
  \BibitemOpen
  \bibfield  {author} {\bibinfo {author} {\bibfnamefont {T.}~\bibnamefont {van
  Leent}}, \bibinfo {author} {\bibfnamefont {M.}~\bibnamefont {Bock}}, \bibinfo
  {author} {\bibfnamefont {F.}~\bibnamefont {Fertig}}, \bibinfo {author}
  {\bibfnamefont {R.}~\bibnamefont {Garthoff}}, \bibinfo {author}
  {\bibfnamefont {S.}~\bibnamefont {Eppelt}}, \bibinfo {author} {\bibfnamefont
  {Y.}~\bibnamefont {Zhou}}, \bibinfo {author} {\bibfnamefont {P.}~\bibnamefont
  {Malik}}, \bibinfo {author} {\bibfnamefont {M.}~\bibnamefont {Seubert}},
  \bibinfo {author} {\bibfnamefont {T.}~\bibnamefont {Bauer}}, \bibinfo
  {author} {\bibfnamefont {W.}~\bibnamefont {Rosenfeld}}, \bibinfo {author}
  {\bibfnamefont {W.}~\bibnamefont {Zhang}}, \bibinfo {author} {\bibfnamefont
  {C.}~\bibnamefont {Becher}},\ and\ \bibinfo {author} {\bibfnamefont
  {H.}~\bibnamefont {Weinfurter}},\ }\bibfield  {title} {\bibinfo {title}
  {Entangling single atoms over 33km telecom fibre},\ }\href
  {https://doi.org/10.1038/s41586-022-04764-4} {\bibfield  {journal} {\bibinfo
  {journal} {Nature}\ }\textbf {\bibinfo {volume} {607}},\ \bibinfo {pages}
  {69} (\bibinfo {year} {2022})}\BibitemShut {NoStop}%
\bibitem [{\citenamefont {Wehner}\ \emph {et~al.}(2018)\citenamefont {Wehner},
  \citenamefont {Elkouss},\ and\ \citenamefont {Hanson}}]{QuantInt_Wehner2018}%
  \BibitemOpen
  \bibfield  {author} {\bibinfo {author} {\bibfnamefont {S.}~\bibnamefont
  {Wehner}}, \bibinfo {author} {\bibfnamefont {D.}~\bibnamefont {Elkouss}},\
  and\ \bibinfo {author} {\bibfnamefont {R.}~\bibnamefont {Hanson}},\
  }\bibfield  {title} {\bibinfo {title} {Quantum internet: A vision for the
  road ahead},\ }\href {https://doi.org/10.1126/science.aam9288} {\bibfield
  {journal} {\bibinfo  {journal} {Science}\ }\textbf {\bibinfo {volume}
  {362}},\ \bibinfo {pages} {eaam9288} (\bibinfo {year} {2018})}\BibitemShut
  {NoStop}%
\bibitem [{\citenamefont {Azuma}\ \emph {et~al.}(2023)\citenamefont {Azuma},
  \citenamefont {Economou}, \citenamefont {Elkouss}, \citenamefont {Hilaire},
  \citenamefont {Jiang}, \citenamefont {Lo},\ and\ \citenamefont
  {Tzitrin}}]{QuantInt_Koji2023}%
  \BibitemOpen
  \bibfield  {author} {\bibinfo {author} {\bibfnamefont {K.}~\bibnamefont
  {Azuma}}, \bibinfo {author} {\bibfnamefont {S.~E.}\ \bibnamefont {Economou}},
  \bibinfo {author} {\bibfnamefont {D.}~\bibnamefont {Elkouss}}, \bibinfo
  {author} {\bibfnamefont {P.}~\bibnamefont {Hilaire}}, \bibinfo {author}
  {\bibfnamefont {L.}~\bibnamefont {Jiang}}, \bibinfo {author} {\bibfnamefont
  {H.-K.}\ \bibnamefont {Lo}},\ and\ \bibinfo {author} {\bibfnamefont
  {I.}~\bibnamefont {Tzitrin}},\ }\bibfield  {title} {\bibinfo {title} {Quantum
  repeaters: From quantum networks to the quantum internet},\ }\href
  {https://doi.org/10.1103/RevModPhys.95.045006} {\bibfield  {journal}
  {\bibinfo  {journal} {Rev. Mod. Phys.}\ }\textbf {\bibinfo {volume} {95}},\
  \bibinfo {pages} {045006} (\bibinfo {year} {2023})}\BibitemShut {NoStop}%
\bibitem [{\citenamefont {Wiseman}\ \emph {et~al.}(2007)\citenamefont
  {Wiseman}, \citenamefont {Jones},\ and\ \citenamefont
  {Doherty}}]{BellGauss_Wiseman2007}%
  \BibitemOpen
  \bibfield  {author} {\bibinfo {author} {\bibfnamefont {H.~M.}\ \bibnamefont
  {Wiseman}}, \bibinfo {author} {\bibfnamefont {S.~J.}\ \bibnamefont {Jones}},\
  and\ \bibinfo {author} {\bibfnamefont {A.~C.}\ \bibnamefont {Doherty}},\
  }\bibfield  {title} {\bibinfo {title} {Steering, entanglement, nonlocality,
  and the einstein-podolsky-rosen paradox},\ }\href
  {https://doi.org/10.1103/PhysRevLett.98.140402} {\bibfield  {journal}
  {\bibinfo  {journal} {Phys. Rev. Lett.}\ }\textbf {\bibinfo {volume} {98}},\
  \bibinfo {pages} {140402} (\bibinfo {year} {2007})}\BibitemShut {NoStop}%
\bibitem [{\citenamefont {Buono}\ \emph {et~al.}(2014)\citenamefont {Buono},
  \citenamefont {Nocerino}, \citenamefont {Solimeno},\ and\ \citenamefont
  {Porzio}}]{BellGauss_Buono2014}%
  \BibitemOpen
  \bibfield  {author} {\bibinfo {author} {\bibfnamefont {D.}~\bibnamefont
  {Buono}}, \bibinfo {author} {\bibfnamefont {G.}~\bibnamefont {Nocerino}},
  \bibinfo {author} {\bibfnamefont {S.}~\bibnamefont {Solimeno}},\ and\
  \bibinfo {author} {\bibfnamefont {A.}~\bibnamefont {Porzio}},\ }\bibfield
  {title} {\bibinfo {title} {Different operational meanings of continuous
  variable gaussian entanglement criteria and bell inequalities},\ }\href
  {https://doi.org/10.1088/1054-660X/24/7/074008} {\bibfield  {journal}
  {\bibinfo  {journal} {Laser Physics}\ }\textbf {\bibinfo {volume} {24}},\
  \bibinfo {pages} {074008} (\bibinfo {year} {2014})}\BibitemShut {NoStop}%
\bibitem [{Qur(2020)}]{Qureshi2020}%
  \BibitemOpen
  \href {https://doi.org/10.1088/1361-6455/ab8b44} {\ \textbf {\bibinfo
  {volume} {53}},\ \bibinfo {pages} {135501} (\bibinfo {year}
  {2020})}\BibitemShut {NoStop}%
\bibitem [{\citenamefont {Lantz}\ \emph {et~al.}(2021)\citenamefont {Lantz},
  \citenamefont {Mabed},\ and\ \citenamefont {Devaux}}]{BellGauss_Lantz2021}%
  \BibitemOpen
  \bibfield  {author} {\bibinfo {author} {\bibfnamefont {E.}~\bibnamefont
  {Lantz}}, \bibinfo {author} {\bibfnamefont {M.}~\bibnamefont {Mabed}},\ and\
  \bibinfo {author} {\bibfnamefont {F.}~\bibnamefont {Devaux}},\ }\bibfield
  {title} {\bibinfo {title} {Violation of bell inequalities by stochastic
  simulations of gaussian states based on their positive wigner
  representation},\ }\href {https://doi.org/10.1088/1402-4896/abdd56}
  {\bibfield  {journal} {\bibinfo  {journal} {Physica Scripta}\ }\textbf
  {\bibinfo {volume} {96}},\ \bibinfo {pages} {045103} (\bibinfo {year}
  {2021})}\BibitemShut {NoStop}%
\bibitem [{\citenamefont {Jabbour}\ and\ \citenamefont
  {Brask}(2023)}]{BellGauss_Bohr2023}%
  \BibitemOpen
  \bibfield  {author} {\bibinfo {author} {\bibfnamefont {M.~G.}\ \bibnamefont
  {Jabbour}}\ and\ \bibinfo {author} {\bibfnamefont {J.~B.}\ \bibnamefont
  {Brask}},\ }\bibfield  {title} {\bibinfo {title} {Constructing local models
  for general measurements on bosonic gaussian states},\ }\href
  {https://doi.org/10.1103/PhysRevLett.131.110202} {\bibfield  {journal}
  {\bibinfo  {journal} {Phys. Rev. Lett.}\ }\textbf {\bibinfo {volume} {131}},\
  \bibinfo {pages} {110202} (\bibinfo {year} {2023})}\BibitemShut {NoStop}%
\bibitem [{\citenamefont {Xu}\ \emph {et~al.}(2015)\citenamefont {Xu},
  \citenamefont {Curty}, \citenamefont {Qi}, \citenamefont {Qian},\ and\
  \citenamefont {Lo}}]{DVCVCompare_Xu2015}%
  \BibitemOpen
  \bibfield  {author} {\bibinfo {author} {\bibfnamefont {F.}~\bibnamefont
  {Xu}}, \bibinfo {author} {\bibfnamefont {M.}~\bibnamefont {Curty}}, \bibinfo
  {author} {\bibfnamefont {B.}~\bibnamefont {Qi}}, \bibinfo {author}
  {\bibfnamefont {L.}~\bibnamefont {Qian}},\ and\ \bibinfo {author}
  {\bibfnamefont {H.-K.}\ \bibnamefont {Lo}},\ }\bibfield  {title} {\bibinfo
  {title} {Discrete and continuous variables for measurement-device-independent
  quantum cryptography},\ }\href {https://doi.org/10.1038/nphoton.2015.206}
  {\bibfield  {journal} {\bibinfo  {journal} {Nature Photonics}\ }\textbf
  {\bibinfo {volume} {9}},\ \bibinfo {pages} {772} (\bibinfo {year}
  {2015})}\BibitemShut {NoStop}%
\bibitem [{\citenamefont {Pirandola}\ \emph {et~al.}(2015)\citenamefont
  {Pirandola}, \citenamefont {Ottaviani}, \citenamefont {Spedalieri},
  \citenamefont {Weedbrook}, \citenamefont {Braunstein}, \citenamefont {Lloyd},
  \citenamefont {Gehring}, \citenamefont {Jacobsen},\ and\ \citenamefont
  {Andersen}}]{DVCVCompare_Pirandola2015}%
  \BibitemOpen
  \bibfield  {author} {\bibinfo {author} {\bibfnamefont {S.}~\bibnamefont
  {Pirandola}}, \bibinfo {author} {\bibfnamefont {C.}~\bibnamefont
  {Ottaviani}}, \bibinfo {author} {\bibfnamefont {G.}~\bibnamefont
  {Spedalieri}}, \bibinfo {author} {\bibfnamefont {C.}~\bibnamefont
  {Weedbrook}}, \bibinfo {author} {\bibfnamefont {S.~L.}\ \bibnamefont
  {Braunstein}}, \bibinfo {author} {\bibfnamefont {S.}~\bibnamefont {Lloyd}},
  \bibinfo {author} {\bibfnamefont {T.}~\bibnamefont {Gehring}}, \bibinfo
  {author} {\bibfnamefont {C.~S.}\ \bibnamefont {Jacobsen}},\ and\ \bibinfo
  {author} {\bibfnamefont {U.~L.}\ \bibnamefont {Andersen}},\ }\bibfield
  {title} {\bibinfo {title} {Reply to 'discrete and continuous variables for
  measurement-device-independent quantum cryptography'},\ }\href
  {https://doi.org/10.1038/nphoton.2015.207} {\bibfield  {journal} {\bibinfo
  {journal} {Nature Photonics}\ }\textbf {\bibinfo {volume} {9}},\ \bibinfo
  {pages} {773} (\bibinfo {year} {2015})}\BibitemShut {NoStop}%
\bibitem [{\citenamefont {Diamanti}\ \emph {et~al.}(2016)\citenamefont
  {Diamanti}, \citenamefont {Lo}, \citenamefont {Qi},\ and\ \citenamefont
  {Yuan}}]{DVCVCompare_Diamanti2016}%
  \BibitemOpen
  \bibfield  {author} {\bibinfo {author} {\bibfnamefont {E.}~\bibnamefont
  {Diamanti}}, \bibinfo {author} {\bibfnamefont {H.-K.}\ \bibnamefont {Lo}},
  \bibinfo {author} {\bibfnamefont {B.}~\bibnamefont {Qi}},\ and\ \bibinfo
  {author} {\bibfnamefont {Z.}~\bibnamefont {Yuan}},\ }\bibfield  {title}
  {\bibinfo {title} {Practical challenges in quantum key distribution},\ }\href
  {https://doi.org/10.1038/npjqi.2016.25} {\bibfield  {journal} {\bibinfo
  {journal} {npj Quantum Information}\ }\textbf {\bibinfo {volume} {2}},\
  \bibinfo {pages} {16025} (\bibinfo {year} {2016})}\BibitemShut {NoStop}%
\bibitem [{\citenamefont {Costanzo}\ \emph {et~al.}(2017)\citenamefont
  {Costanzo}, \citenamefont {Coelho}, \citenamefont {Biagi}, \citenamefont
  {Fiur\'a\ifmmode~\check{s}\else \v{s}\fi{}ek}, \citenamefont {Bellini},\ and\
  \citenamefont {Zavatta}}]{WCP_Costanzo2017}%
  \BibitemOpen
  \bibfield  {author} {\bibinfo {author} {\bibfnamefont {L.~S.}\ \bibnamefont
  {Costanzo}}, \bibinfo {author} {\bibfnamefont {A.~S.}\ \bibnamefont
  {Coelho}}, \bibinfo {author} {\bibfnamefont {N.}~\bibnamefont {Biagi}},
  \bibinfo {author} {\bibfnamefont {J.}~\bibnamefont
  {Fiur\'a\ifmmode~\check{s}\else \v{s}\fi{}ek}}, \bibinfo {author}
  {\bibfnamefont {M.}~\bibnamefont {Bellini}},\ and\ \bibinfo {author}
  {\bibfnamefont {A.}~\bibnamefont {Zavatta}},\ }\bibfield  {title} {\bibinfo
  {title} {Measurement-induced strong kerr nonlinearity for weak quantum states
  of light},\ }\href {https://doi.org/10.1103/PhysRevLett.119.013601}
  {\bibfield  {journal} {\bibinfo  {journal} {Phys. Rev. Lett.}\ }\textbf
  {\bibinfo {volume} {119}},\ \bibinfo {pages} {013601} (\bibinfo {year}
  {2017})}\BibitemShut {NoStop}%
\bibitem [{\citenamefont {Jain}\ \emph {et~al.}(2020)\citenamefont {Jain},
  \citenamefont {Sakhiya},\ and\ \citenamefont {Bahl}}]{WCP_Jain2020}%
  \BibitemOpen
  \bibfield  {author} {\bibinfo {author} {\bibfnamefont {A.}~\bibnamefont
  {Jain}}, \bibinfo {author} {\bibfnamefont {P.~V.}\ \bibnamefont {Sakhiya}},\
  and\ \bibinfo {author} {\bibfnamefont {R.~K.}\ \bibnamefont {Bahl}},\
  }\bibfield  {title} {\bibinfo {title} {Design and development of weak
  coherent pulse source for quantum key distribution system},\ }in\ \href
  {https://doi.org/10.1109/CONECCT50063.2020.9198351} {\emph {\bibinfo
  {booktitle} {2020 IEEE International Conference on Electronics, Computing and
  Communication Technologies (CONECCT)}}}\ (\bibinfo {year} {2020})\ pp.\
  \bibinfo {pages} {1--5}\BibitemShut {NoStop}%
\bibitem [{\citenamefont {Wiseman}\ \emph {et~al.}(2023)\citenamefont
  {Wiseman}, \citenamefont {Steinberg},\ and\ \citenamefont
  {Hallaji}}]{WCP_Wiseman2023}%
  \BibitemOpen
  \bibfield  {author} {\bibinfo {author} {\bibfnamefont {H.~M.}\ \bibnamefont
  {Wiseman}}, \bibinfo {author} {\bibfnamefont {A.~M.}\ \bibnamefont
  {Steinberg}},\ and\ \bibinfo {author} {\bibfnamefont {M.}~\bibnamefont
  {Hallaji}},\ }\bibfield  {title} {\bibinfo {title} {Obtaining a single-photon
  weak value from experiments using a strong (many-photon) coherent state},\
  }\href {https://doi.org/10.1116/5.0137579} {\bibfield  {journal} {\bibinfo
  {journal} {AVS Quantum Science}\ }\textbf {\bibinfo {volume} {5}},\ \bibinfo
  {pages} {024401} (\bibinfo {year} {2023})}\BibitemShut {NoStop}%
\bibitem [{\citenamefont {Duranti}\ \emph {et~al.}(2024)\citenamefont
  {Duranti}, \citenamefont {Wengerowsky}, \citenamefont {Feldmann},
  \citenamefont {Seri}, \citenamefont {Casabone},\ and\ \citenamefont
  {de~Riedmatten}}]{WCP_Duranti2024}%
  \BibitemOpen
  \bibfield  {author} {\bibinfo {author} {\bibfnamefont {S.}~\bibnamefont
  {Duranti}}, \bibinfo {author} {\bibfnamefont {S.}~\bibnamefont
  {Wengerowsky}}, \bibinfo {author} {\bibfnamefont {L.}~\bibnamefont
  {Feldmann}}, \bibinfo {author} {\bibfnamefont {A.}~\bibnamefont {Seri}},
  \bibinfo {author} {\bibfnamefont {B.}~\bibnamefont {Casabone}},\ and\
  \bibinfo {author} {\bibfnamefont {H.}~\bibnamefont {de~Riedmatten}},\
  }\bibfield  {title} {\bibinfo {title} {Efficient cavity-assisted storage of
  photonic qubits in a solid-state quantum memory},\ }\href
  {https://doi.org/10.1364/OE.512318} {\bibfield  {journal} {\bibinfo
  {journal} {Opt. Express}\ }\textbf {\bibinfo {volume} {32}},\ \bibinfo
  {pages} {26884} (\bibinfo {year} {2024})}\BibitemShut {NoStop}%
\bibitem [{\citenamefont {Jeong}(2005)}]{HOSDescrip_Jeong2005}%
  \BibitemOpen
  \bibfield  {author} {\bibinfo {author} {\bibfnamefont {H.}~\bibnamefont
  {Jeong}},\ }\bibfield  {title} {\bibinfo {title} {Using weak nonlinearity
  under decoherence for macroscopic entanglement generation and quantum
  computation},\ }\href {https://doi.org/10.1103/PhysRevA.72.034305} {\bibfield
   {journal} {\bibinfo  {journal} {Phys. Rev. A}\ }\textbf {\bibinfo {volume}
  {72}},\ \bibinfo {pages} {034305} (\bibinfo {year} {2005})}\BibitemShut
  {NoStop}%
\bibitem [{\citenamefont {Li}\ \emph {et~al.}(2006)\citenamefont {Li},
  \citenamefont {Jing},\ and\ \citenamefont {Zhan}}]{HOSDescript_Li2006}%
  \BibitemOpen
  \bibfield  {author} {\bibinfo {author} {\bibfnamefont {Y.}~\bibnamefont
  {Li}}, \bibinfo {author} {\bibfnamefont {H.}~\bibnamefont {Jing}},\ and\
  \bibinfo {author} {\bibfnamefont {M.-S.}\ \bibnamefont {Zhan}},\ }\bibfield
  {title} {\bibinfo {title} {Optical generation of a hybrid entangled state via
  an entangling single-photon-added coherent state},\ }\href
  {https://doi.org/10.1088/0953-4075/39/9/001} {\bibfield  {journal} {\bibinfo
  {journal} {Journal of Physics B: Atomic, Molecular and Optical Physics}\
  }\textbf {\bibinfo {volume} {39}},\ \bibinfo {pages} {2107} (\bibinfo {year}
  {2006})}\BibitemShut {NoStop}%
\bibitem [{\citenamefont {He}\ \emph {et~al.}(2011)\citenamefont {He},
  \citenamefont {Lin},\ and\ \citenamefont {Simon}}]{HOSDescrip_Bing2011}%
  \BibitemOpen
  \bibfield  {author} {\bibinfo {author} {\bibfnamefont {B.}~\bibnamefont
  {He}}, \bibinfo {author} {\bibfnamefont {Q.}~\bibnamefont {Lin}},\ and\
  \bibinfo {author} {\bibfnamefont {C.}~\bibnamefont {Simon}},\ }\bibfield
  {title} {\bibinfo {title} {Cross-kerr nonlinearity between continuous-mode
  coherent states and single photons},\ }\href
  {https://doi.org/10.1103/PhysRevA.83.053826} {\bibfield  {journal} {\bibinfo
  {journal} {Phys. Rev. A}\ }\textbf {\bibinfo {volume} {83}},\ \bibinfo
  {pages} {053826} (\bibinfo {year} {2011})}\BibitemShut {NoStop}%
\bibitem [{\citenamefont {Le}\ \emph {et~al.}(2021)\citenamefont {Le},
  \citenamefont {Asavanant},\ and\ \citenamefont {An}}]{HOSDescript_Le2021}%
  \BibitemOpen
  \bibfield  {author} {\bibinfo {author} {\bibfnamefont {D.~T.}\ \bibnamefont
  {Le}}, \bibinfo {author} {\bibfnamefont {W.}~\bibnamefont {Asavanant}},\ and\
  \bibinfo {author} {\bibfnamefont {N.~B.}\ \bibnamefont {An}},\ }\bibfield
  {title} {\bibinfo {title} {Heralded preparation of polarization entanglement
  via quantum scissors},\ }\href {https://doi.org/10.1103/PhysRevA.104.012612}
  {\bibfield  {journal} {\bibinfo  {journal} {Phys. Rev. A}\ }\textbf {\bibinfo
  {volume} {104}},\ \bibinfo {pages} {012612} (\bibinfo {year}
  {2021})}\BibitemShut {NoStop}%
\bibitem [{\citenamefont {Kwon}\ and\ \citenamefont
  {Jeong}(2013)}]{HOSBell_Kwon2013}%
  \BibitemOpen
  \bibfield  {author} {\bibinfo {author} {\bibfnamefont {H.}~\bibnamefont
  {Kwon}}\ and\ \bibinfo {author} {\bibfnamefont {H.}~\bibnamefont {Jeong}},\
  }\bibfield  {title} {\bibinfo {title} {Violation of the
  bell--clauser-horne-shimony-holt inequality using imperfect photodetectors
  with optical hybrid states},\ }\href
  {https://doi.org/10.1103/PhysRevA.88.052127} {\bibfield  {journal} {\bibinfo
  {journal} {Phys. Rev. A}\ }\textbf {\bibinfo {volume} {88}},\ \bibinfo
  {pages} {052127} (\bibinfo {year} {2013})}\BibitemShut {NoStop}%
\bibitem [{\citenamefont {Kwon}\ and\ \citenamefont
  {Jeong}(2015)}]{HOSBell_Kwon2014}%
  \BibitemOpen
  \bibfield  {author} {\bibinfo {author} {\bibfnamefont {H.}~\bibnamefont
  {Kwon}}\ and\ \bibinfo {author} {\bibfnamefont {H.}~\bibnamefont {Jeong}},\
  }\bibfield  {title} {\bibinfo {title} {Generation of hybrid entanglement
  between a single-photon polarization qubit and a coherent state},\ }\href
  {https://doi.org/10.1103/PhysRevA.91.012340} {\bibfield  {journal} {\bibinfo
  {journal} {Phys. Rev. A}\ }\textbf {\bibinfo {volume} {91}},\ \bibinfo
  {pages} {012340} (\bibinfo {year} {2015})}\BibitemShut {NoStop}%
\bibitem [{\citenamefont {Moradi}\ \emph {et~al.}(2024)\citenamefont {Moradi},
  \citenamefont {Carreño}, \citenamefont {Buraczewski}, \citenamefont
  {McDermott}, \citenamefont {Asenbeck}, \citenamefont {Laurat},\ and\
  \citenamefont {Stobińska}}]{HOSBell_Moradi2024}%
  \BibitemOpen
  \bibfield  {author} {\bibinfo {author} {\bibfnamefont {M.}~\bibnamefont
  {Moradi}}, \bibinfo {author} {\bibfnamefont {J.~C.~L.}\ \bibnamefont
  {Carreño}}, \bibinfo {author} {\bibfnamefont {A.}~\bibnamefont
  {Buraczewski}}, \bibinfo {author} {\bibfnamefont {T.}~\bibnamefont
  {McDermott}}, \bibinfo {author} {\bibfnamefont {B.~E.}\ \bibnamefont
  {Asenbeck}}, \bibinfo {author} {\bibfnamefont {J.}~\bibnamefont {Laurat}},\
  and\ \bibinfo {author} {\bibfnamefont {M.}~\bibnamefont {Stobińska}},\
  }\bibfield  {title} {\bibinfo {title} {Chsh bell tests for optical hybrid
  entanglement},\ }\href {https://doi.org/10.1088/1367-2630/ad2d40} {\bibfield
  {journal} {\bibinfo  {journal} {New Journal of Physics}\ }\textbf {\bibinfo
  {volume} {26}},\ \bibinfo {pages} {033019} (\bibinfo {year}
  {2024})}\BibitemShut {NoStop}%
\bibitem [{\citenamefont {Cavalcanti}\ \emph {et~al.}(2011)\citenamefont
  {Cavalcanti}, \citenamefont {Brunner}, \citenamefont {Skrzypczyk},
  \citenamefont {Salles},\ and\ \citenamefont
  {Scarani}}]{HOSBellMeas_Daniel2011}%
  \BibitemOpen
  \bibfield  {author} {\bibinfo {author} {\bibfnamefont {D.}~\bibnamefont
  {Cavalcanti}}, \bibinfo {author} {\bibfnamefont {N.}~\bibnamefont {Brunner}},
  \bibinfo {author} {\bibfnamefont {P.}~\bibnamefont {Skrzypczyk}}, \bibinfo
  {author} {\bibfnamefont {A.}~\bibnamefont {Salles}},\ and\ \bibinfo {author}
  {\bibfnamefont {V.}~\bibnamefont {Scarani}},\ }\bibfield  {title} {\bibinfo
  {title} {Large violation of bell inequalities using both particle andwave
  measurements},\ }\href {https://doi.org/10.1103/PhysRevA.84.022105}
  {\bibfield  {journal} {\bibinfo  {journal} {Phys. Rev. A}\ }\textbf {\bibinfo
  {volume} {84}},\ \bibinfo {pages} {022105} (\bibinfo {year}
  {2011})}\BibitemShut {NoStop}%
\bibitem [{\citenamefont {Park}\ \emph {et~al.}(2012)\citenamefont {Park},
  \citenamefont {Lee},\ and\ \citenamefont {Jeong}}]{HOSTeleport_Park2012}%
  \BibitemOpen
  \bibfield  {author} {\bibinfo {author} {\bibfnamefont {K.}~\bibnamefont
  {Park}}, \bibinfo {author} {\bibfnamefont {S.-W.}\ \bibnamefont {Lee}},\ and\
  \bibinfo {author} {\bibfnamefont {H.}~\bibnamefont {Jeong}},\ }\bibfield
  {title} {\bibinfo {title} {Quantum teleportation between particlelike and
  fieldlike qubits using hybrid entanglement under decoherence effects},\
  }\href {https://doi.org/10.1103/PhysRevA.86.062301} {\bibfield  {journal}
  {\bibinfo  {journal} {Phys. Rev. A}\ }\textbf {\bibinfo {volume} {86}},\
  \bibinfo {pages} {062301} (\bibinfo {year} {2012})}\BibitemShut {NoStop}%
\bibitem [{\citenamefont {Lee}\ and\ \citenamefont
  {Jeong}(2013)}]{HOSTeleport_Lee2013}%
  \BibitemOpen
  \bibfield  {author} {\bibinfo {author} {\bibfnamefont {S.-W.}\ \bibnamefont
  {Lee}}\ and\ \bibinfo {author} {\bibfnamefont {H.}~\bibnamefont {Jeong}},\
  }\bibfield  {title} {\bibinfo {title} {Near-deterministic quantum
  teleportation and resource-efficient quantum computation using linear optics
  and hybrid qubits},\ }\href {https://doi.org/10.1103/PhysRevA.87.022326}
  {\bibfield  {journal} {\bibinfo  {journal} {Phys. Rev. A}\ }\textbf {\bibinfo
  {volume} {87}},\ \bibinfo {pages} {022326} (\bibinfo {year}
  {2013})}\BibitemShut {NoStop}%
\bibitem [{\citenamefont {Kim}\ \emph {et~al.}(2016)\citenamefont {Kim},
  \citenamefont {Lee},\ and\ \citenamefont {Jeong}}]{HOSTeleport_Kim2016}%
  \BibitemOpen
  \bibfield  {author} {\bibinfo {author} {\bibfnamefont {H.}~\bibnamefont
  {Kim}}, \bibinfo {author} {\bibfnamefont {S.-W.}\ \bibnamefont {Lee}},\ and\
  \bibinfo {author} {\bibfnamefont {H.}~\bibnamefont {Jeong}},\ }\bibfield
  {title} {\bibinfo {title} {Two different types of optical hybrid qubits for
  teleportation in a lossy environment},\ }\href
  {https://doi.org/10.1007/s11128-016-1408-7} {\bibfield  {journal} {\bibinfo
  {journal} {Quantum Information Processing}\ }\textbf {\bibinfo {volume}
  {15}},\ \bibinfo {pages} {4729} (\bibinfo {year} {2016})}\BibitemShut
  {NoStop}%
\bibitem [{\citenamefont {Ulanov}\ \emph {et~al.}(2017)\citenamefont {Ulanov},
  \citenamefont {Sychev}, \citenamefont {Pushkina}, \citenamefont {Fedorov},\
  and\ \citenamefont {Lvovsky}}]{HOSTelport_Ulanov2017}%
  \BibitemOpen
  \bibfield  {author} {\bibinfo {author} {\bibfnamefont {A.~E.}\ \bibnamefont
  {Ulanov}}, \bibinfo {author} {\bibfnamefont {D.}~\bibnamefont {Sychev}},
  \bibinfo {author} {\bibfnamefont {A.~A.}\ \bibnamefont {Pushkina}}, \bibinfo
  {author} {\bibfnamefont {I.~A.}\ \bibnamefont {Fedorov}},\ and\ \bibinfo
  {author} {\bibfnamefont {A.~I.}\ \bibnamefont {Lvovsky}},\ }\bibfield
  {title} {\bibinfo {title} {Quantum teleportation between discrete and
  continuous encodings of an optical qubit},\ }\href
  {https://doi.org/10.1103/PhysRevLett.118.160501} {\bibfield  {journal}
  {\bibinfo  {journal} {Phys. Rev. Lett.}\ }\textbf {\bibinfo {volume} {118}},\
  \bibinfo {pages} {160501} (\bibinfo {year} {2017})}\BibitemShut {NoStop}%
\bibitem [{\citenamefont {Sychev}\ \emph {et~al.}(2018)\citenamefont {Sychev},
  \citenamefont {Ulanov}, \citenamefont {Tiunov}, \citenamefont {Pushkina},
  \citenamefont {Kuzhamuratov}, \citenamefont {Novikov},\ and\ \citenamefont
  {Lvovsky}}]{HOSTelport_Sychev2018}%
  \BibitemOpen
  \bibfield  {author} {\bibinfo {author} {\bibfnamefont {D.~V.}\ \bibnamefont
  {Sychev}}, \bibinfo {author} {\bibfnamefont {A.~E.}\ \bibnamefont {Ulanov}},
  \bibinfo {author} {\bibfnamefont {E.~S.}\ \bibnamefont {Tiunov}}, \bibinfo
  {author} {\bibfnamefont {A.~A.}\ \bibnamefont {Pushkina}}, \bibinfo {author}
  {\bibfnamefont {A.}~\bibnamefont {Kuzhamuratov}}, \bibinfo {author}
  {\bibfnamefont {V.}~\bibnamefont {Novikov}},\ and\ \bibinfo {author}
  {\bibfnamefont {A.~I.}\ \bibnamefont {Lvovsky}},\ }\bibfield  {title}
  {\bibinfo {title} {Entanglement and teleportation between polarization and
  wave-like encodings of an optical qubit},\ }\href
  {https://doi.org/10.1038/s41467-018-06055-x} {\bibfield  {journal} {\bibinfo
  {journal} {Nature Communications}\ }\textbf {\bibinfo {volume} {9}},\
  \bibinfo {pages} {3672} (\bibinfo {year} {2018})}\BibitemShut {NoStop}%
\bibitem [{\citenamefont {Dom\'{i}nguez-Serna}\ \emph
  {et~al.}(2020)\citenamefont {Dom\'{i}nguez-Serna}, \citenamefont {Rojas},\
  and\ \citenamefont {Garay-Palmett}}]{HOSTelport_Francisco2020}%
  \BibitemOpen
  \bibfield  {author} {\bibinfo {author} {\bibfnamefont {F.~A.}\ \bibnamefont
  {Dom\'{i}nguez-Serna}}, \bibinfo {author} {\bibfnamefont {F.}~\bibnamefont
  {Rojas}},\ and\ \bibinfo {author} {\bibfnamefont {K.}~\bibnamefont
  {Garay-Palmett}},\ }\bibfield  {title} {\bibinfo {title} {Quantum
  teleportation with hybrid entangled resources prepared from heralded quantum
  states},\ }\href {https://doi.org/10.1364/JOSAB.377687} {\bibfield  {journal}
  {\bibinfo  {journal} {J. Opt. Soc. Am. B}\ }\textbf {\bibinfo {volume}
  {37}},\ \bibinfo {pages} {695} (\bibinfo {year} {2020})}\BibitemShut
  {NoStop}%
\bibitem [{\citenamefont {Bose}\ and\ \citenamefont
  {Jeong}(2022)}]{HOSTeleport_Bose2022}%
  \BibitemOpen
  \bibfield  {author} {\bibinfo {author} {\bibfnamefont {S.}~\bibnamefont
  {Bose}}\ and\ \bibinfo {author} {\bibfnamefont {H.}~\bibnamefont {Jeong}},\
  }\bibfield  {title} {\bibinfo {title} {Quantum teleportation of hybrid qubits
  and single-photon qubits using gaussian resources},\ }\href
  {https://doi.org/10.1103/PhysRevA.105.032434} {\bibfield  {journal} {\bibinfo
   {journal} {Phys. Rev. A}\ }\textbf {\bibinfo {volume} {105}},\ \bibinfo
  {pages} {032434} (\bibinfo {year} {2022})}\BibitemShut {NoStop}%
\bibitem [{\citenamefont {He}\ and\ \citenamefont
  {Malaney}(2022)}]{HOSTeleport_He2022}%
  \BibitemOpen
  \bibfield  {author} {\bibinfo {author} {\bibfnamefont {M.}~\bibnamefont
  {He}}\ and\ \bibinfo {author} {\bibfnamefont {R.}~\bibnamefont {Malaney}},\
  }\bibfield  {title} {\bibinfo {title} {Teleportation of hybrid entangled
  states with continuous-variable entanglement},\ }\href
  {https://doi.org/10.1038/s41598-022-21283-4} {\bibfield  {journal} {\bibinfo
  {journal} {Scientific Reports}\ }\textbf {\bibinfo {volume} {12}},\ \bibinfo
  {pages} {17169} (\bibinfo {year} {2022})}\BibitemShut {NoStop}%
\bibitem [{\citenamefont {Kirdi}\ \emph {et~al.}(2023)\citenamefont {Kirdi},
  \citenamefont {Slaoui}, \citenamefont {Ikken}, \citenamefont {Daoud},\ and\
  \citenamefont {Laamara}}]{HOSTeleport_Kirdi2023}%
  \BibitemOpen
  \bibfield  {author} {\bibinfo {author} {\bibfnamefont {M.~E.}\ \bibnamefont
  {Kirdi}}, \bibinfo {author} {\bibfnamefont {A.}~\bibnamefont {Slaoui}},
  \bibinfo {author} {\bibfnamefont {N.}~\bibnamefont {Ikken}}, \bibinfo
  {author} {\bibfnamefont {M.}~\bibnamefont {Daoud}},\ and\ \bibinfo {author}
  {\bibfnamefont {R.~A.}\ \bibnamefont {Laamara}},\ }\bibfield  {title}
  {\bibinfo {title} {Controlled quantum teleportation between discrete and
  continuous physical systems},\ }\href
  {https://doi.org/10.1088/1402-4896/acacd2} {\bibfield  {journal} {\bibinfo
  {journal} {Physica Scripta}\ }\textbf {\bibinfo {volume} {98}},\ \bibinfo
  {pages} {025101} (\bibinfo {year} {2023})}\BibitemShut {NoStop}%
\bibitem [{\citenamefont {Liu}\ \emph {et~al.}(2024)\citenamefont {Liu},
  \citenamefont {Siltanen}, \citenamefont {Kuusela}, \citenamefont {Miao},
  \citenamefont {Ning}, \citenamefont {Li}, \citenamefont {Guo},\ and\
  \citenamefont {Piilo}}]{HOSQT_Liu2024}%
  \BibitemOpen
  \bibfield  {author} {\bibinfo {author} {\bibfnamefont {Z.-D.}\ \bibnamefont
  {Liu}}, \bibinfo {author} {\bibfnamefont {O.}~\bibnamefont {Siltanen}},
  \bibinfo {author} {\bibfnamefont {T.}~\bibnamefont {Kuusela}}, \bibinfo
  {author} {\bibfnamefont {R.-H.}\ \bibnamefont {Miao}}, \bibinfo {author}
  {\bibfnamefont {C.-X.}\ \bibnamefont {Ning}}, \bibinfo {author}
  {\bibfnamefont {C.-F.}\ \bibnamefont {Li}}, \bibinfo {author} {\bibfnamefont
  {G.-C.}\ \bibnamefont {Guo}},\ and\ \bibinfo {author} {\bibfnamefont
  {J.}~\bibnamefont {Piilo}},\ }\bibfield  {title} {\bibinfo {title}
  {Overcoming noise in quantum teleportation with multipartite hybrid
  entanglement},\ }\href {https://doi.org/10.1126/sciadv.adj3435} {\bibfield
  {journal} {\bibinfo  {journal} {Science Advances}\ }\textbf {\bibinfo
  {volume} {10}},\ \bibinfo {pages} {eadj3435} (\bibinfo {year}
  {2024})}\BibitemShut {NoStop}%
\bibitem [{\citenamefont {Omkar}\ \emph {et~al.}(2020)\citenamefont {Omkar},
  \citenamefont {Teo},\ and\ \citenamefont {Jeong}}]{HOSQComp_Omkar2020}%
  \BibitemOpen
  \bibfield  {author} {\bibinfo {author} {\bibfnamefont {S.}~\bibnamefont
  {Omkar}}, \bibinfo {author} {\bibfnamefont {Y.~S.}\ \bibnamefont {Teo}},\
  and\ \bibinfo {author} {\bibfnamefont {H.}~\bibnamefont {Jeong}},\ }\bibfield
   {title} {\bibinfo {title} {Resource-efficient topological fault-tolerant
  quantum computation with hybrid entanglement of light},\ }\href
  {https://doi.org/10.1103/PhysRevLett.125.060501} {\bibfield  {journal}
  {\bibinfo  {journal} {Phys. Rev. Lett.}\ }\textbf {\bibinfo {volume} {125}},\
  \bibinfo {pages} {060501} (\bibinfo {year} {2020})}\BibitemShut {NoStop}%
\bibitem [{\citenamefont {Omkar}\ \emph {et~al.}(2021)\citenamefont {Omkar},
  \citenamefont {Teo}, \citenamefont {Lee},\ and\ \citenamefont
  {Jeong}}]{HOSQComp_Omkar2021}%
  \BibitemOpen
  \bibfield  {author} {\bibinfo {author} {\bibfnamefont {S.}~\bibnamefont
  {Omkar}}, \bibinfo {author} {\bibfnamefont {Y.~S.}\ \bibnamefont {Teo}},
  \bibinfo {author} {\bibfnamefont {S.-W.}\ \bibnamefont {Lee}},\ and\ \bibinfo
  {author} {\bibfnamefont {H.}~\bibnamefont {Jeong}},\ }\bibfield  {title}
  {\bibinfo {title} {Highly photon-loss-tolerant quantum computing using hybrid
  qubits},\ }\href {https://doi.org/10.1103/PhysRevA.103.032602} {\bibfield
  {journal} {\bibinfo  {journal} {Phys. Rev. A}\ }\textbf {\bibinfo {volume}
  {103}},\ \bibinfo {pages} {032602} (\bibinfo {year} {2021})}\BibitemShut
  {NoStop}%
\bibitem [{\citenamefont {Darras}\ \emph {et~al.}(2023)\citenamefont {Darras},
  \citenamefont {Asenbeck}, \citenamefont {Guccione}, \citenamefont
  {Cavaillès}, \citenamefont {Le~Jeannic},\ and\ \citenamefont
  {Laurat}}]{HOSQComp_Tom2023}%
  \BibitemOpen
  \bibfield  {author} {\bibinfo {author} {\bibfnamefont {T.}~\bibnamefont
  {Darras}}, \bibinfo {author} {\bibfnamefont {B.~E.}\ \bibnamefont
  {Asenbeck}}, \bibinfo {author} {\bibfnamefont {G.}~\bibnamefont {Guccione}},
  \bibinfo {author} {\bibfnamefont {A.}~\bibnamefont {Cavaillès}}, \bibinfo
  {author} {\bibfnamefont {H.}~\bibnamefont {Le~Jeannic}},\ and\ \bibinfo
  {author} {\bibfnamefont {J.}~\bibnamefont {Laurat}},\ }\bibfield  {title}
  {\bibinfo {title} {A quantum-bit encoding converter},\ }\href
  {https://doi.org/10.1038/s41566-022-01117-5} {\bibfield  {journal} {\bibinfo
  {journal} {Nature Photonics}\ }\textbf {\bibinfo {volume} {17}},\ \bibinfo
  {pages} {165} (\bibinfo {year} {2023})}\BibitemShut {NoStop}%
\bibitem [{\citenamefont {Lee}\ \emph {et~al.}(2024)\citenamefont {Lee},
  \citenamefont {Kang}, \citenamefont {Lee}, \citenamefont {Jeong},
  \citenamefont {Jiang},\ and\ \citenamefont {Lee}}]{HOSQComp_Lee2024}%
  \BibitemOpen
  \bibfield  {author} {\bibinfo {author} {\bibfnamefont {J.}~\bibnamefont
  {Lee}}, \bibinfo {author} {\bibfnamefont {N.}~\bibnamefont {Kang}}, \bibinfo
  {author} {\bibfnamefont {S.-H.}\ \bibnamefont {Lee}}, \bibinfo {author}
  {\bibfnamefont {H.}~\bibnamefont {Jeong}}, \bibinfo {author} {\bibfnamefont
  {L.}~\bibnamefont {Jiang}},\ and\ \bibinfo {author} {\bibfnamefont {S.-W.}\
  \bibnamefont {Lee}},\ }\bibfield  {title} {\bibinfo {title} {Fault-tolerant
  quantum computation by hybrid qubits with bosonic cat code and single
  photons},\ }\href {https://doi.org/10.1103/PRXQuantum.5.030322} {\bibfield
  {journal} {\bibinfo  {journal} {PRX Quantum}\ }\textbf {\bibinfo {volume}
  {5}},\ \bibinfo {pages} {030322} (\bibinfo {year} {2024})}\BibitemShut
  {NoStop}%
\bibitem [{\citenamefont {Lim}\ \emph {et~al.}(2016)\citenamefont {Lim},
  \citenamefont {Joo}, \citenamefont {Spiller},\ and\ \citenamefont
  {Jeong}}]{HOSSwap_Lim2016}%
  \BibitemOpen
  \bibfield  {author} {\bibinfo {author} {\bibfnamefont {Y.}~\bibnamefont
  {Lim}}, \bibinfo {author} {\bibfnamefont {J.}~\bibnamefont {Joo}}, \bibinfo
  {author} {\bibfnamefont {T.~P.}\ \bibnamefont {Spiller}},\ and\ \bibinfo
  {author} {\bibfnamefont {H.}~\bibnamefont {Jeong}},\ }\bibfield  {title}
  {\bibinfo {title} {Loss-resilient photonic entanglement swapping using
  optical hybrid states},\ }\href {https://doi.org/10.1103/PhysRevA.94.062337}
  {\bibfield  {journal} {\bibinfo  {journal} {Phys. Rev. A}\ }\textbf {\bibinfo
  {volume} {94}},\ \bibinfo {pages} {062337} (\bibinfo {year}
  {2016})}\BibitemShut {NoStop}%
\bibitem [{\citenamefont {Parker}\ \emph {et~al.}(2017)\citenamefont {Parker},
  \citenamefont {Joo}, \citenamefont {Razavi},\ and\ \citenamefont
  {Spiller}}]{HOSSwap_Parker2017}%
  \BibitemOpen
  \bibfield  {author} {\bibinfo {author} {\bibfnamefont {R.~C.}\ \bibnamefont
  {Parker}}, \bibinfo {author} {\bibfnamefont {J.}~\bibnamefont {Joo}},
  \bibinfo {author} {\bibfnamefont {M.}~\bibnamefont {Razavi}},\ and\ \bibinfo
  {author} {\bibfnamefont {T.~P.}\ \bibnamefont {Spiller}},\ }\bibfield
  {title} {\bibinfo {title} {Hybrid photonic loss resilient entanglement
  swapping},\ }\href {https://doi.org/10.1088/2040-8986/aa858a} {\bibfield
  {journal} {\bibinfo  {journal} {Journal of Optics}\ }\textbf {\bibinfo
  {volume} {19}},\ \bibinfo {pages} {104004} (\bibinfo {year}
  {2017})}\BibitemShut {NoStop}%
\bibitem [{\citenamefont {Bose}\ \emph {et~al.}(2024)\citenamefont {Bose},
  \citenamefont {Singh}, \citenamefont {Cabello},\ and\ \citenamefont
  {Jeong}}]{HOSEnt_Bose2024}%
  \BibitemOpen
  \bibfield  {author} {\bibinfo {author} {\bibfnamefont {S.}~\bibnamefont
  {Bose}}, \bibinfo {author} {\bibfnamefont {J.}~\bibnamefont {Singh}},
  \bibinfo {author} {\bibfnamefont {A.}~\bibnamefont {Cabello}},\ and\ \bibinfo
  {author} {\bibfnamefont {H.}~\bibnamefont {Jeong}},\ }\bibfield  {title}
  {\bibinfo {title} {Long-distance entanglement sharing using hybrid states of
  discrete and continuous variables},\ }\href
  {https://doi.org/10.1103/PhysRevApplied.21.064013} {\bibfield  {journal}
  {\bibinfo  {journal} {Phys. Rev. Appl.}\ }\textbf {\bibinfo {volume} {21}},\
  \bibinfo {pages} {064013} (\bibinfo {year} {2024})}\BibitemShut {NoStop}%
\bibitem [{\citenamefont {Jeong}\ \emph {et~al.}(2014)\citenamefont {Jeong},
  \citenamefont {Zavatta}, \citenamefont {Kang}, \citenamefont {Lee},
  \citenamefont {Costanzo}, \citenamefont {Grandi}, \citenamefont {Ralph},\
  and\ \citenamefont {Bellini}}]{HOSExp_Jeong2014}%
  \BibitemOpen
  \bibfield  {author} {\bibinfo {author} {\bibfnamefont {H.}~\bibnamefont
  {Jeong}}, \bibinfo {author} {\bibfnamefont {A.}~\bibnamefont {Zavatta}},
  \bibinfo {author} {\bibfnamefont {M.}~\bibnamefont {Kang}}, \bibinfo {author}
  {\bibfnamefont {S.-W.}\ \bibnamefont {Lee}}, \bibinfo {author} {\bibfnamefont
  {L.~S.}\ \bibnamefont {Costanzo}}, \bibinfo {author} {\bibfnamefont
  {S.}~\bibnamefont {Grandi}}, \bibinfo {author} {\bibfnamefont {T.~C.}\
  \bibnamefont {Ralph}},\ and\ \bibinfo {author} {\bibfnamefont
  {M.}~\bibnamefont {Bellini}},\ }\bibfield  {title} {\bibinfo {title}
  {Generation of hybrid entanglement of light},\ }\href
  {https://doi.org/10.1038/nphoton.2014.136} {\bibfield  {journal} {\bibinfo
  {journal} {Nature Photonics}\ }\textbf {\bibinfo {volume} {8}},\ \bibinfo
  {pages} {564} (\bibinfo {year} {2014})}\BibitemShut {NoStop}%
\bibitem [{\citenamefont {Morin}\ \emph {et~al.}(2014)\citenamefont {Morin},
  \citenamefont {Huang}, \citenamefont {Liu}, \citenamefont {Le~Jeannic},
  \citenamefont {Fabre},\ and\ \citenamefont {Laurat}}]{HOSEXxp_Morin2014}%
  \BibitemOpen
  \bibfield  {author} {\bibinfo {author} {\bibfnamefont {O.}~\bibnamefont
  {Morin}}, \bibinfo {author} {\bibfnamefont {K.}~\bibnamefont {Huang}},
  \bibinfo {author} {\bibfnamefont {J.}~\bibnamefont {Liu}}, \bibinfo {author}
  {\bibfnamefont {H.}~\bibnamefont {Le~Jeannic}}, \bibinfo {author}
  {\bibfnamefont {C.}~\bibnamefont {Fabre}},\ and\ \bibinfo {author}
  {\bibfnamefont {J.}~\bibnamefont {Laurat}},\ }\bibfield  {title} {\bibinfo
  {title} {Remote creation of hybrid entanglement between particle-like and
  wave-like optical qubits},\ }\href {https://doi.org/10.1038/nphoton.2014.137}
  {\bibfield  {journal} {\bibinfo  {journal} {Nature Photonics}\ }\textbf
  {\bibinfo {volume} {8}},\ \bibinfo {pages} {570} (\bibinfo {year}
  {2014})}\BibitemShut {NoStop}%
\bibitem [{\citenamefont {Jeannic}\ \emph {et~al.}(2018)\citenamefont
  {Jeannic}, \citenamefont {Cavaill\`{e}s}, \citenamefont {Raskop},
  \citenamefont {Huang},\ and\ \citenamefont {Laurat}}]{HOSExp_Jeannic2018}%
  \BibitemOpen
  \bibfield  {author} {\bibinfo {author} {\bibfnamefont {H.~L.}\ \bibnamefont
  {Jeannic}}, \bibinfo {author} {\bibfnamefont {A.}~\bibnamefont
  {Cavaill\`{e}s}}, \bibinfo {author} {\bibfnamefont {J.}~\bibnamefont
  {Raskop}}, \bibinfo {author} {\bibfnamefont {K.}~\bibnamefont {Huang}},\ and\
  \bibinfo {author} {\bibfnamefont {J.}~\bibnamefont {Laurat}},\ }\bibfield
  {title} {\bibinfo {title} {Remote preparation of continuous-variable qubits
  using loss-tolerant hybrid entanglement of light},\ }\href
  {https://doi.org/10.1364/OPTICA.5.001012} {\bibfield  {journal} {\bibinfo
  {journal} {Optica}\ }\textbf {\bibinfo {volume} {5}},\ \bibinfo {pages}
  {1012} (\bibinfo {year} {2018})}\BibitemShut {NoStop}%
\bibitem [{\citenamefont {Huang}\ \emph {et~al.}(2019)\citenamefont {Huang},
  \citenamefont {Jeannic}, \citenamefont {Morin}, \citenamefont {Darras},
  \citenamefont {Guccione}, \citenamefont {Cavaillès},\ and\ \citenamefont
  {Laurat}}]{HOSExp_Huang2019}%
  \BibitemOpen
  \bibfield  {author} {\bibinfo {author} {\bibfnamefont {K.}~\bibnamefont
  {Huang}}, \bibinfo {author} {\bibfnamefont {H.~L.}\ \bibnamefont {Jeannic}},
  \bibinfo {author} {\bibfnamefont {O.}~\bibnamefont {Morin}}, \bibinfo
  {author} {\bibfnamefont {T.}~\bibnamefont {Darras}}, \bibinfo {author}
  {\bibfnamefont {G.}~\bibnamefont {Guccione}}, \bibinfo {author}
  {\bibfnamefont {A.}~\bibnamefont {Cavaillès}},\ and\ \bibinfo {author}
  {\bibfnamefont {J.}~\bibnamefont {Laurat}},\ }\bibfield  {title} {\bibinfo
  {title} {Engineering optical hybrid entanglement between discrete- and
  continuous-variable states},\ }\href
  {https://doi.org/10.1088/1367-2630/ab34e7} {\bibfield  {journal} {\bibinfo
  {journal} {New Journal of Physics}\ }\textbf {\bibinfo {volume} {21}},\
  \bibinfo {pages} {083033} (\bibinfo {year} {2019})}\BibitemShut {NoStop}%
\bibitem [{\citenamefont {Hacker}\ \emph {et~al.}(2019)\citenamefont {Hacker},
  \citenamefont {Welte}, \citenamefont {Daiss}, \citenamefont {Shaukat},
  \citenamefont {Ritter}, \citenamefont {Li},\ and\ \citenamefont
  {Rempe}}]{HOSExp_Hacker2019}%
  \BibitemOpen
  \bibfield  {author} {\bibinfo {author} {\bibfnamefont {B.}~\bibnamefont
  {Hacker}}, \bibinfo {author} {\bibfnamefont {S.}~\bibnamefont {Welte}},
  \bibinfo {author} {\bibfnamefont {S.}~\bibnamefont {Daiss}}, \bibinfo
  {author} {\bibfnamefont {A.}~\bibnamefont {Shaukat}}, \bibinfo {author}
  {\bibfnamefont {S.}~\bibnamefont {Ritter}}, \bibinfo {author} {\bibfnamefont
  {L.}~\bibnamefont {Li}},\ and\ \bibinfo {author} {\bibfnamefont
  {G.}~\bibnamefont {Rempe}},\ }\bibfield  {title} {\bibinfo {title}
  {Deterministic creation of entangled atom–light schrödinger-cat states},\
  }\href {https://doi.org/10.1038/s41566-018-0339-5} {\bibfield  {journal}
  {\bibinfo  {journal} {Nature Photonics}\ }\textbf {\bibinfo {volume} {13}},\
  \bibinfo {pages} {110} (\bibinfo {year} {2019})}\BibitemShut {NoStop}%
\bibitem [{\citenamefont {Guccione}\ \emph {et~al.}(2020)\citenamefont
  {Guccione}, \citenamefont {Darras}, \citenamefont {Jeannic}, \citenamefont
  {Verma}, \citenamefont {Nam}, \citenamefont {Cavaillès},\ and\ \citenamefont
  {Laurat}}]{HOSExp_Giovanni2020}%
  \BibitemOpen
  \bibfield  {author} {\bibinfo {author} {\bibfnamefont {G.}~\bibnamefont
  {Guccione}}, \bibinfo {author} {\bibfnamefont {T.}~\bibnamefont {Darras}},
  \bibinfo {author} {\bibfnamefont {H.~L.}\ \bibnamefont {Jeannic}}, \bibinfo
  {author} {\bibfnamefont {V.~B.}\ \bibnamefont {Verma}}, \bibinfo {author}
  {\bibfnamefont {S.~W.}\ \bibnamefont {Nam}}, \bibinfo {author} {\bibfnamefont
  {A.}~\bibnamefont {Cavaillès}},\ and\ \bibinfo {author} {\bibfnamefont
  {J.}~\bibnamefont {Laurat}},\ }\bibfield  {title} {\bibinfo {title}
  {Connecting heterogeneous quantum networks by hybrid entanglement swapping},\
  }\href {https://doi.org/10.1126/sciadv.aba4508} {\bibfield  {journal}
  {\bibinfo  {journal} {Science Advances}\ }\textbf {\bibinfo {volume} {6}},\
  \bibinfo {pages} {eaba4508} (\bibinfo {year} {2020})}\BibitemShut {NoStop}%
\bibitem [{\citenamefont {Gouzien}\ \emph {et~al.}(2020)\citenamefont
  {Gouzien}, \citenamefont {Brunel}, \citenamefont {Tanzilli},\ and\
  \citenamefont {D'Auria}}]{HOSExp_Gouzien2020}%
  \BibitemOpen
  \bibfield  {author} {\bibinfo {author} {\bibfnamefont {E.}~\bibnamefont
  {Gouzien}}, \bibinfo {author} {\bibfnamefont {F.}~\bibnamefont {Brunel}},
  \bibinfo {author} {\bibfnamefont {S.}~\bibnamefont {Tanzilli}},\ and\
  \bibinfo {author} {\bibfnamefont {V.}~\bibnamefont {D'Auria}},\ }\bibfield
  {title} {\bibinfo {title} {Scheme for the generation of hybrid entanglement
  between time-bin and wavelike encodings},\ }\href
  {https://doi.org/10.1103/PhysRevA.102.012603} {\bibfield  {journal} {\bibinfo
   {journal} {Phys. Rev. A}\ }\textbf {\bibinfo {volume} {102}},\ \bibinfo
  {pages} {012603} (\bibinfo {year} {2020})}\BibitemShut {NoStop}%
\bibitem [{\citenamefont {Wen}\ \emph {et~al.}(2021)\citenamefont {Wen},
  \citenamefont {Novikova}, \citenamefont {Qian}, \citenamefont {Zhang},\ and\
  \citenamefont {Du}}]{HOSExp_Wen2021}%
  \BibitemOpen
  \bibfield  {author} {\bibinfo {author} {\bibfnamefont {J.}~\bibnamefont
  {Wen}}, \bibinfo {author} {\bibfnamefont {I.}~\bibnamefont {Novikova}},
  \bibinfo {author} {\bibfnamefont {C.}~\bibnamefont {Qian}}, \bibinfo {author}
  {\bibfnamefont {C.}~\bibnamefont {Zhang}},\ and\ \bibinfo {author}
  {\bibfnamefont {S.}~\bibnamefont {Du}},\ }\bibfield  {title} {\bibinfo
  {title} {Hybrid entanglement between optical discrete polarizations and
  continuous quadrature variables},\ }\href
  {https://doi.org/10.3390/photonics8120552} {\bibfield  {journal} {\bibinfo
  {journal} {Photonics}\ }\textbf {\bibinfo {volume} {8}},\ \bibinfo {pages}
  {552} (\bibinfo {year} {2021})}\BibitemShut {NoStop}%
\bibitem [{\citenamefont {Li}\ \emph {et~al.}(2021)\citenamefont {Li},
  \citenamefont {He}, \citenamefont {Deng}, \citenamefont {Xue}, \citenamefont
  {Xu},\ and\ \citenamefont {Wang}}]{HOSExp_Li2021}%
  \BibitemOpen
  \bibfield  {author} {\bibinfo {author} {\bibfnamefont {S.}~\bibnamefont
  {Li}}, \bibinfo {author} {\bibfnamefont {Y.}~\bibnamefont {He}}, \bibinfo
  {author} {\bibfnamefont {Q.}~\bibnamefont {Deng}}, \bibinfo {author}
  {\bibfnamefont {J.}~\bibnamefont {Xue}}, \bibinfo {author} {\bibfnamefont
  {Z.}~\bibnamefont {Xu}},\ and\ \bibinfo {author} {\bibfnamefont
  {H.}~\bibnamefont {Wang}},\ }\bibfield  {title} {\bibinfo {title}
  {Improvement of hybrid entanglement by dual-way photon polarization
  measurement},\ }\href {https://doi.org/10.1007/s11128-021-03221-x} {\bibfield
   {journal} {\bibinfo  {journal} {Quantum Information Processing}\ }\textbf
  {\bibinfo {volume} {20}},\ \bibinfo {pages} {295} (\bibinfo {year}
  {2021})}\BibitemShut {NoStop}%
\bibitem [{\citenamefont {Zopf}\ \emph {et~al.}(2019)\citenamefont {Zopf},
  \citenamefont {Keil}, \citenamefont {Chen}, \citenamefont {Yang},
  \citenamefont {Chen}, \citenamefont {Ding},\ and\ \citenamefont
  {Schmidt}}]{BellSwap_Zopf2019}%
  \BibitemOpen
  \bibfield  {author} {\bibinfo {author} {\bibfnamefont {M.}~\bibnamefont
  {Zopf}}, \bibinfo {author} {\bibfnamefont {R.}~\bibnamefont {Keil}}, \bibinfo
  {author} {\bibfnamefont {Y.}~\bibnamefont {Chen}}, \bibinfo {author}
  {\bibfnamefont {J.}~\bibnamefont {Yang}}, \bibinfo {author} {\bibfnamefont
  {D.}~\bibnamefont {Chen}}, \bibinfo {author} {\bibfnamefont {F.}~\bibnamefont
  {Ding}},\ and\ \bibinfo {author} {\bibfnamefont {O.~G.}\ \bibnamefont
  {Schmidt}},\ }\bibfield  {title} {\bibinfo {title} {Entanglement swapping
  with semiconductor-generated photons violates bell's inequality},\ }\href
  {https://doi.org/10.1103/PhysRevLett.123.160502} {\bibfield  {journal}
  {\bibinfo  {journal} {Phys. Rev. Lett.}\ }\textbf {\bibinfo {volume} {123}},\
  \bibinfo {pages} {160502} (\bibinfo {year} {2019})}\BibitemShut {NoStop}%
\bibitem [{\citenamefont {Tsujimoto}\ \emph {et~al.}(2020)\citenamefont
  {Tsujimoto}, \citenamefont {You}, \citenamefont {Wakui}, \citenamefont
  {Fujiwara}, \citenamefont {Hayasaka}, \citenamefont {Miki}, \citenamefont
  {Terai}, \citenamefont {Sasaki}, \citenamefont {Dowling},\ and\ \citenamefont
  {Takeoka}}]{BellSwap_Tsujimoto2020}%
  \BibitemOpen
  \bibfield  {author} {\bibinfo {author} {\bibfnamefont {Y.}~\bibnamefont
  {Tsujimoto}}, \bibinfo {author} {\bibfnamefont {C.}~\bibnamefont {You}},
  \bibinfo {author} {\bibfnamefont {K.}~\bibnamefont {Wakui}}, \bibinfo
  {author} {\bibfnamefont {M.}~\bibnamefont {Fujiwara}}, \bibinfo {author}
  {\bibfnamefont {K.}~\bibnamefont {Hayasaka}}, \bibinfo {author}
  {\bibfnamefont {S.}~\bibnamefont {Miki}}, \bibinfo {author} {\bibfnamefont
  {H.}~\bibnamefont {Terai}}, \bibinfo {author} {\bibfnamefont
  {M.}~\bibnamefont {Sasaki}}, \bibinfo {author} {\bibfnamefont {J.~P.}\
  \bibnamefont {Dowling}},\ and\ \bibinfo {author} {\bibfnamefont
  {M.}~\bibnamefont {Takeoka}},\ }\bibfield  {title} {\bibinfo {title}
  {Heralded amplification of nonlocality via entanglement swapping},\ }\href
  {https://doi.org/10.1088/1367-2630/ab61da} {\bibfield  {journal} {\bibinfo
  {journal} {New Journal of Physics}\ }\textbf {\bibinfo {volume} {22}},\
  \bibinfo {pages} {023008} (\bibinfo {year} {2020})}\BibitemShut {NoStop}%
\bibitem [{\citenamefont {Huang}\ \emph {et~al.}(2022)\citenamefont {Huang},
  \citenamefont {Hu}, \citenamefont {Guo}, \citenamefont {Zhang}, \citenamefont
  {Liu}, \citenamefont {Huang}, \citenamefont {Li}, \citenamefont {Guo},
  \citenamefont {Gisin}, \citenamefont {Branciard},\ and\ \citenamefont
  {Tavakoli}}]{BellSwap_Huang2022}%
  \BibitemOpen
  \bibfield  {author} {\bibinfo {author} {\bibfnamefont {C.-X.}\ \bibnamefont
  {Huang}}, \bibinfo {author} {\bibfnamefont {X.-M.}\ \bibnamefont {Hu}},
  \bibinfo {author} {\bibfnamefont {Y.}~\bibnamefont {Guo}}, \bibinfo {author}
  {\bibfnamefont {C.}~\bibnamefont {Zhang}}, \bibinfo {author} {\bibfnamefont
  {B.-H.}\ \bibnamefont {Liu}}, \bibinfo {author} {\bibfnamefont {Y.-F.}\
  \bibnamefont {Huang}}, \bibinfo {author} {\bibfnamefont {C.-F.}\ \bibnamefont
  {Li}}, \bibinfo {author} {\bibfnamefont {G.-C.}\ \bibnamefont {Guo}},
  \bibinfo {author} {\bibfnamefont {N.}~\bibnamefont {Gisin}}, \bibinfo
  {author} {\bibfnamefont {C.}~\bibnamefont {Branciard}},\ and\ \bibinfo
  {author} {\bibfnamefont {A.}~\bibnamefont {Tavakoli}},\ }\bibfield  {title}
  {\bibinfo {title} {Entanglement swapping and quantum correlations via
  symmetric joint measurements},\ }\href
  {https://doi.org/10.1103/PhysRevLett.129.030502} {\bibfield  {journal}
  {\bibinfo  {journal} {Phys. Rev. Lett.}\ }\textbf {\bibinfo {volume} {129}},\
  \bibinfo {pages} {030502} (\bibinfo {year} {2022})}\BibitemShut {NoStop}%
\bibitem [{\citenamefont {Bjerrum}\ \emph {et~al.}(2023)\citenamefont
  {Bjerrum}, \citenamefont {Brask}, \citenamefont {Neergaard-Nielsen},\ and\
  \citenamefont {Andersen}}]{BellSwap_Anders2023}%
  \BibitemOpen
  \bibfield  {author} {\bibinfo {author} {\bibfnamefont {A.~J.~E.}\
  \bibnamefont {Bjerrum}}, \bibinfo {author} {\bibfnamefont {J.~B.}\
  \bibnamefont {Brask}}, \bibinfo {author} {\bibfnamefont {J.~S.}\ \bibnamefont
  {Neergaard-Nielsen}},\ and\ \bibinfo {author} {\bibfnamefont {U.~L.}\
  \bibnamefont {Andersen}},\ }\bibfield  {title} {\bibinfo {title} {Proposal
  for a long-distance nonlocality test with entanglement swapping and
  displacement-based measurements},\ }\href
  {https://doi.org/10.1103/PhysRevA.107.052611} {\bibfield  {journal} {\bibinfo
   {journal} {Phys. Rev. A}\ }\textbf {\bibinfo {volume} {107}},\ \bibinfo
  {pages} {052611} (\bibinfo {year} {2023})}\BibitemShut {NoStop}%
\bibitem [{\citenamefont {Sheng}\ \emph {et~al.}(2013)\citenamefont {Sheng},
  \citenamefont {Zhou},\ and\ \citenamefont {Long}}]{HOSEntPurify_Sheng2013}%
  \BibitemOpen
  \bibfield  {author} {\bibinfo {author} {\bibfnamefont {Y.-B.}\ \bibnamefont
  {Sheng}}, \bibinfo {author} {\bibfnamefont {L.}~\bibnamefont {Zhou}},\ and\
  \bibinfo {author} {\bibfnamefont {G.-L.}\ \bibnamefont {Long}},\ }\bibfield
  {title} {\bibinfo {title} {Hybrid entanglement purification for quantum
  repeaters},\ }\href {https://doi.org/10.1103/PhysRevA.88.022302} {\bibfield
  {journal} {\bibinfo  {journal} {Phys. Rev. A}\ }\textbf {\bibinfo {volume}
  {88}},\ \bibinfo {pages} {022302} (\bibinfo {year} {2013})}\BibitemShut
  {NoStop}%
\bibitem [{\citenamefont {Sheng}\ and\ \citenamefont
  {Zhou}(2015)}]{BellPolarization_Sheng2015}%
  \BibitemOpen
  \bibfield  {author} {\bibinfo {author} {\bibfnamefont {Y.-B.}\ \bibnamefont
  {Sheng}}\ and\ \bibinfo {author} {\bibfnamefont {L.}~\bibnamefont {Zhou}},\
  }\bibfield  {title} {\bibinfo {title} {Two-step complete polarization logic
  bell-state analysis},\ }\href {https://doi.org/10.1038/srep13453} {\bibfield
  {journal} {\bibinfo  {journal} {Scientific Reports}\ }\textbf {\bibinfo
  {volume} {5}},\ \bibinfo {pages} {13453} (\bibinfo {year}
  {2015})}\BibitemShut {NoStop}%
\bibitem [{\citenamefont {Bayerbach}\ \emph {et~al.}(2023)\citenamefont
  {Bayerbach}, \citenamefont {D’Aurelio}, \citenamefont {van Loock},\ and\
  \citenamefont {Barz}}]{BellPolarization_Matthias2023}%
  \BibitemOpen
  \bibfield  {author} {\bibinfo {author} {\bibfnamefont {M.~J.}\ \bibnamefont
  {Bayerbach}}, \bibinfo {author} {\bibfnamefont {S.~E.}\ \bibnamefont
  {D’Aurelio}}, \bibinfo {author} {\bibfnamefont {P.}~\bibnamefont {van
  Loock}},\ and\ \bibinfo {author} {\bibfnamefont {S.}~\bibnamefont {Barz}},\
  }\bibfield  {title} {\bibinfo {title} {Bell-state measurement exceeding 50\%
  success probability with linear optics},\ }\href
  {https://doi.org/10.1126/sciadv.adf4080} {\bibfield  {journal} {\bibinfo
  {journal} {Science Advances}\ }\textbf {\bibinfo {volume} {9}},\ \bibinfo
  {pages} {eadf4080} (\bibinfo {year} {2023})}\BibitemShut {NoStop}%
\bibitem [{\citenamefont {Mishra}\ and\ \citenamefont
  {Singh}(2024)}]{BellPolarization_Mishra2024}%
  \BibitemOpen
  \bibfield  {author} {\bibinfo {author} {\bibfnamefont {S.}~\bibnamefont
  {Mishra}}\ and\ \bibinfo {author} {\bibfnamefont {R.}~\bibnamefont {Singh}},\
  }\bibfield  {title} {\bibinfo {title} {Transformation of bell states using
  linear optics},\ }\href
  {https://doi.org/https://doi.org/10.1016/j.physo.2023.100199} {\bibfield
  {journal} {\bibinfo  {journal} {Physics Open}\ }\textbf {\bibinfo {volume}
  {18}},\ \bibinfo {pages} {100199} (\bibinfo {year} {2024})}\BibitemShut
  {NoStop}%
\bibitem [{\citenamefont {Arensk\"otter}\ \emph {et~al.}(2024)\citenamefont
  {Arensk\"otter}, \citenamefont {Kucera}, \citenamefont {Elshehy},
  \citenamefont {Bergerhoff}, \citenamefont {Kreis}, \citenamefont {Brunel},\
  and\ \citenamefont {Eschner}}]{PolarizationQuantNet_Elena2024}%
  \BibitemOpen
  \bibfield  {author} {\bibinfo {author} {\bibfnamefont {E.}~\bibnamefont
  {Arensk\"otter}}, \bibinfo {author} {\bibfnamefont {S.}~\bibnamefont
  {Kucera}}, \bibinfo {author} {\bibfnamefont {O.}~\bibnamefont {Elshehy}},
  \bibinfo {author} {\bibfnamefont {M.}~\bibnamefont {Bergerhoff}}, \bibinfo
  {author} {\bibfnamefont {M.}~\bibnamefont {Kreis}}, \bibinfo {author}
  {\bibfnamefont {L.}~\bibnamefont {Brunel}},\ and\ \bibinfo {author}
  {\bibfnamefont {J.}~\bibnamefont {Eschner}},\ }\bibfield  {title} {\bibinfo
  {title} {Full bell-basis measurement of an atom-photon 2-qubit state and its
  application for quantum networks},\ }\href
  {https://doi.org/10.1103/PhysRevResearch.6.023061} {\bibfield  {journal}
  {\bibinfo  {journal} {Phys. Rev. Res.}\ }\textbf {\bibinfo {volume} {6}},\
  \bibinfo {pages} {023061} (\bibinfo {year} {2024})}\BibitemShut {NoStop}%
\bibitem [{\citenamefont {Pennacchietti}\ \emph {et~al.}(2024)\citenamefont
  {Pennacchietti}, \citenamefont {Cunard}, \citenamefont {Nahar}, \citenamefont
  {Zeeshan}, \citenamefont {Gangopadhyay}, \citenamefont {Poole}, \citenamefont
  {Dalacu}, \citenamefont {Fognini}, \citenamefont {Jöns}, \citenamefont
  {Zwiller}, \citenamefont {Jennewein}, \citenamefont {Lütkenhaus},\ and\
  \citenamefont {Reimer}}]{PolarizationQKD_Matteo2024}%
  \BibitemOpen
  \bibfield  {author} {\bibinfo {author} {\bibfnamefont {M.}~\bibnamefont
  {Pennacchietti}}, \bibinfo {author} {\bibfnamefont {B.}~\bibnamefont
  {Cunard}}, \bibinfo {author} {\bibfnamefont {S.}~\bibnamefont {Nahar}},
  \bibinfo {author} {\bibfnamefont {M.}~\bibnamefont {Zeeshan}}, \bibinfo
  {author} {\bibfnamefont {S.}~\bibnamefont {Gangopadhyay}}, \bibinfo {author}
  {\bibfnamefont {P.~J.}\ \bibnamefont {Poole}}, \bibinfo {author}
  {\bibfnamefont {D.}~\bibnamefont {Dalacu}}, \bibinfo {author} {\bibfnamefont
  {A.}~\bibnamefont {Fognini}}, \bibinfo {author} {\bibfnamefont {K.~D.}\
  \bibnamefont {Jöns}}, \bibinfo {author} {\bibfnamefont {V.}~\bibnamefont
  {Zwiller}}, \bibinfo {author} {\bibfnamefont {T.}~\bibnamefont {Jennewein}},
  \bibinfo {author} {\bibfnamefont {N.}~\bibnamefont {Lütkenhaus}},\ and\
  \bibinfo {author} {\bibfnamefont {M.~E.}\ \bibnamefont {Reimer}},\ }\bibfield
   {title} {\bibinfo {title} {Oscillating photonic bell state from a
  semiconductor quantum dot for quantum key distribution},\ }\href
  {https://doi.org/10.1038/s42005-024-01547-3} {\bibfield  {journal} {\bibinfo
  {journal} {Communications Physics}\ }\textbf {\bibinfo {volume} {7}},\
  \bibinfo {pages} {62} (\bibinfo {year} {2024})}\BibitemShut {NoStop}%
\bibitem [{\citenamefont {Bennett}\ \emph {et~al.}(1993)\citenamefont
  {Bennett}, \citenamefont {Brassard}, \citenamefont {Cr\'epeau}, \citenamefont
  {Jozsa}, \citenamefont {Peres},\ and\ \citenamefont
  {Wootters}}]{Teleport_Bennett1993}%
  \BibitemOpen
  \bibfield  {author} {\bibinfo {author} {\bibfnamefont {C.~H.}\ \bibnamefont
  {Bennett}}, \bibinfo {author} {\bibfnamefont {G.}~\bibnamefont {Brassard}},
  \bibinfo {author} {\bibfnamefont {C.}~\bibnamefont {Cr\'epeau}}, \bibinfo
  {author} {\bibfnamefont {R.}~\bibnamefont {Jozsa}}, \bibinfo {author}
  {\bibfnamefont {A.}~\bibnamefont {Peres}},\ and\ \bibinfo {author}
  {\bibfnamefont {W.~K.}\ \bibnamefont {Wootters}},\ }\bibfield  {title}
  {\bibinfo {title} {Teleporting an unknown quantum state via dual classical
  and einstein-podolsky-rosen channels},\ }\href
  {https://doi.org/10.1103/PhysRevLett.70.1895} {\bibfield  {journal} {\bibinfo
   {journal} {Phys. Rev. Lett.}\ }\textbf {\bibinfo {volume} {70}},\ \bibinfo
  {pages} {1895} (\bibinfo {year} {1993})}\BibitemShut {NoStop}%
\bibitem [{\citenamefont {Caleffi}\ \emph {et~al.}(2024)\citenamefont
  {Caleffi}, \citenamefont {Amoretti}, \citenamefont {Ferrari}, \citenamefont
  {Illiano}, \citenamefont {Manzalini},\ and\ \citenamefont
  {Cacciapuoti}}]{QCDist_Caleffi2024}%
  \BibitemOpen
  \bibfield  {author} {\bibinfo {author} {\bibfnamefont {M.}~\bibnamefont
  {Caleffi}}, \bibinfo {author} {\bibfnamefont {M.}~\bibnamefont {Amoretti}},
  \bibinfo {author} {\bibfnamefont {D.}~\bibnamefont {Ferrari}}, \bibinfo
  {author} {\bibfnamefont {J.}~\bibnamefont {Illiano}}, \bibinfo {author}
  {\bibfnamefont {A.}~\bibnamefont {Manzalini}},\ and\ \bibinfo {author}
  {\bibfnamefont {A.~S.}\ \bibnamefont {Cacciapuoti}},\ }\bibfield  {title}
  {\bibinfo {title} {Distributed quantum computing: A survey},\ }\href
  {https://doi.org/10.1016/j.comnet.2024.110672} {\bibfield  {journal}
  {\bibinfo  {journal} {Computer Networks}\ }\textbf {\bibinfo {volume}
  {254}},\ \bibinfo {pages} {110672} (\bibinfo {year} {2024})}\BibitemShut
  {NoStop}%
\bibitem [{\citenamefont {Ganguly}\ \emph {et~al.}(2011)\citenamefont
  {Ganguly}, \citenamefont {Adhikari}, \citenamefont {Majumdar},\ and\
  \citenamefont {Chatterjee}}]{EntWit_Ganguly2011}%
  \BibitemOpen
  \bibfield  {author} {\bibinfo {author} {\bibfnamefont {N.}~\bibnamefont
  {Ganguly}}, \bibinfo {author} {\bibfnamefont {S.}~\bibnamefont {Adhikari}},
  \bibinfo {author} {\bibfnamefont {A.~S.}\ \bibnamefont {Majumdar}},\ and\
  \bibinfo {author} {\bibfnamefont {J.}~\bibnamefont {Chatterjee}},\ }\bibfield
   {title} {\bibinfo {title} {Entanglement witness operator for quantum
  teleportation},\ }\href {https://doi.org/10.1103/PhysRevLett.107.270501}
  {\bibfield  {journal} {\bibinfo  {journal} {Phys. Rev. Lett.}\ }\textbf
  {\bibinfo {volume} {107}},\ \bibinfo {pages} {270501} (\bibinfo {year}
  {2011})}\BibitemShut {NoStop}%
\bibitem [{\citenamefont {Adhikari}\ \emph {et~al.}(2010)\citenamefont
  {Adhikari}, \citenamefont {Majumdar}, \citenamefont {Roy}, \citenamefont
  {Ghosh},\ and\ \citenamefont {Nayak}}]{QT_Adhikari2010}%
  \BibitemOpen
  \bibfield  {author} {\bibinfo {author} {\bibfnamefont {S.}~\bibnamefont
  {Adhikari}}, \bibinfo {author} {\bibfnamefont {A.~S.}\ \bibnamefont
  {Majumdar}}, \bibinfo {author} {\bibfnamefont {S.}~\bibnamefont {Roy}},
  \bibinfo {author} {\bibfnamefont {B.}~\bibnamefont {Ghosh}},\ and\ \bibinfo
  {author} {\bibfnamefont {N.}~\bibnamefont {Nayak}},\ }\bibfield  {title}
  {\bibinfo {title} {Teleportation via maximally and non-maximally entangled
  mixed states},\ }\href {https://doi.org/10.26421/QIC10.5-6-3} {\bibfield
  {journal} {\bibinfo  {journal} {Quant. Inf. Compt.}\ }\textbf {\bibinfo
  {volume} {10}} (\bibinfo {year} {2010})}\BibitemShut {NoStop}%
\bibitem [{\citenamefont {Gisin}(1996)}]{BellTeleport_Gisin1996}%
  \BibitemOpen
  \bibfield  {author} {\bibinfo {author} {\bibfnamefont {N.}~\bibnamefont
  {Gisin}},\ }\bibfield  {title} {\bibinfo {title} {Nonlocality criteria for
  quantum teleportation},\ }\href
  {https://doi.org/https://doi.org/10.1016/S0375-9601(96)80002-8} {\bibfield
  {journal} {\bibinfo  {journal} {Physics Letters A}\ }\textbf {\bibinfo
  {volume} {210}},\ \bibinfo {pages} {157} (\bibinfo {year}
  {1996})}\BibitemShut {NoStop}%
\bibitem [{\citenamefont {Horodecki}\ \emph {et~al.}(1996)\citenamefont
  {Horodecki}, \citenamefont {Horodecki},\ and\ \citenamefont
  {Horodecki}}]{BellTeleport_Horodecki1996}%
  \BibitemOpen
  \bibfield  {author} {\bibinfo {author} {\bibfnamefont {R.}~\bibnamefont
  {Horodecki}}, \bibinfo {author} {\bibfnamefont {M.}~\bibnamefont
  {Horodecki}},\ and\ \bibinfo {author} {\bibfnamefont {P.}~\bibnamefont
  {Horodecki}},\ }\bibfield  {title} {\bibinfo {title} {Teleportation, bell's
  inequalities and inseparability},\ }\href
  {https://doi.org/https://doi.org/10.1016/0375-9601(96)00639-1} {\bibfield
  {journal} {\bibinfo  {journal} {Physics Letters A}\ }\textbf {\bibinfo
  {volume} {222}},\ \bibinfo {pages} {21} (\bibinfo {year} {1996})}\BibitemShut
  {NoStop}%
\bibitem [{\citenamefont {Takesue}\ \emph {et~al.}(2015)\citenamefont
  {Takesue}, \citenamefont {Dyer}, \citenamefont {Stevens}, \citenamefont
  {Verma}, \citenamefont {Mirin},\ and\ \citenamefont
  {Nam}}]{FibOptTp_Hiroki2015}%
  \BibitemOpen
  \bibfield  {author} {\bibinfo {author} {\bibfnamefont {H.}~\bibnamefont
  {Takesue}}, \bibinfo {author} {\bibfnamefont {S.~D.}\ \bibnamefont {Dyer}},
  \bibinfo {author} {\bibfnamefont {M.~J.}\ \bibnamefont {Stevens}}, \bibinfo
  {author} {\bibfnamefont {V.}~\bibnamefont {Verma}}, \bibinfo {author}
  {\bibfnamefont {R.~P.}\ \bibnamefont {Mirin}},\ and\ \bibinfo {author}
  {\bibfnamefont {S.~W.}\ \bibnamefont {Nam}},\ }\bibfield  {title} {\bibinfo
  {title} {Quantum teleportation over 100 km of fiber using highly efficient
  superconducting nanowire single-photon detectors},\ }\href
  {https://doi.org/10.1364/OPTICA.2.000832} {\bibfield  {journal} {\bibinfo
  {journal} {Optica}\ }\textbf {\bibinfo {volume} {2}},\ \bibinfo {pages} {832}
  (\bibinfo {year} {2015})}\BibitemShut {NoStop}%
\bibitem [{\citenamefont {Huo}\ \emph {et~al.}(2018)\citenamefont {Huo},
  \citenamefont {Qin}, \citenamefont {Cheng}, \citenamefont {Yan},
  \citenamefont {Qin}, \citenamefont {Su}, \citenamefont {Jia}, \citenamefont
  {Xie},\ and\ \citenamefont {Peng}}]{FibOptTp_Huo2018}%
  \BibitemOpen
  \bibfield  {author} {\bibinfo {author} {\bibfnamefont {M.}~\bibnamefont
  {Huo}}, \bibinfo {author} {\bibfnamefont {J.}~\bibnamefont {Qin}}, \bibinfo
  {author} {\bibfnamefont {J.}~\bibnamefont {Cheng}}, \bibinfo {author}
  {\bibfnamefont {Z.}~\bibnamefont {Yan}}, \bibinfo {author} {\bibfnamefont
  {Z.}~\bibnamefont {Qin}}, \bibinfo {author} {\bibfnamefont {X.}~\bibnamefont
  {Su}}, \bibinfo {author} {\bibfnamefont {X.}~\bibnamefont {Jia}}, \bibinfo
  {author} {\bibfnamefont {C.}~\bibnamefont {Xie}},\ and\ \bibinfo {author}
  {\bibfnamefont {K.}~\bibnamefont {Peng}},\ }\bibfield  {title} {\bibinfo
  {title} {Deterministic quantum teleportation through fiber channels},\ }\href
  {https://doi.org/10.1126/sciadv.aas9401} {\bibfield  {journal} {\bibinfo
  {journal} {Science Advances}\ }\textbf {\bibinfo {volume} {4}},\ \bibinfo
  {pages} {eaas9401} (\bibinfo {year} {2018})}\BibitemShut {NoStop}%
\bibitem [{\citenamefont {Zhao}\ \emph {et~al.}(2022)\citenamefont {Zhao},
  \citenamefont {Feng}, \citenamefont {Sun}, \citenamefont {Li},\ and\
  \citenamefont {Zhang}}]{FibOptTp_Zhao2022}%
  \BibitemOpen
  \bibfield  {author} {\bibinfo {author} {\bibfnamefont {H.}~\bibnamefont
  {Zhao}}, \bibinfo {author} {\bibfnamefont {J.}~\bibnamefont {Feng}}, \bibinfo
  {author} {\bibfnamefont {J.}~\bibnamefont {Sun}}, \bibinfo {author}
  {\bibfnamefont {Y.}~\bibnamefont {Li}},\ and\ \bibinfo {author}
  {\bibfnamefont {K.}~\bibnamefont {Zhang}},\ }\bibfield  {title} {\bibinfo
  {title} {Real time deterministic quantum teleportation over 10 km of single
  optical fiber channel},\ }\href {https://doi.org/10.1364/OE.447603}
  {\bibfield  {journal} {\bibinfo  {journal} {Opt. Express}\ }\textbf {\bibinfo
  {volume} {30}},\ \bibinfo {pages} {3770} (\bibinfo {year}
  {2022})}\BibitemShut {NoStop}%
\bibitem [{\citenamefont {Shen}\ \emph {et~al.}(2023)\citenamefont {Shen},
  \citenamefont {Yuan}, \citenamefont {Zhang}, \citenamefont {Yu},
  \citenamefont {Zhang}, \citenamefont {Yang}, \citenamefont {Li},
  \citenamefont {Wang}, \citenamefont {Wang}, \citenamefont {Deng},
  \citenamefont {Song}, \citenamefont {You}, \citenamefont {Fan}, \citenamefont
  {Guo},\ and\ \citenamefont {Zhou}}]{FibOptTp_Shen2023}%
  \BibitemOpen
  \bibfield  {author} {\bibinfo {author} {\bibfnamefont {S.}~\bibnamefont
  {Shen}}, \bibinfo {author} {\bibfnamefont {C.}~\bibnamefont {Yuan}}, \bibinfo
  {author} {\bibfnamefont {Z.}~\bibnamefont {Zhang}}, \bibinfo {author}
  {\bibfnamefont {H.}~\bibnamefont {Yu}}, \bibinfo {author} {\bibfnamefont
  {R.}~\bibnamefont {Zhang}}, \bibinfo {author} {\bibfnamefont
  {C.}~\bibnamefont {Yang}}, \bibinfo {author} {\bibfnamefont {H.}~\bibnamefont
  {Li}}, \bibinfo {author} {\bibfnamefont {Z.}~\bibnamefont {Wang}}, \bibinfo
  {author} {\bibfnamefont {Y.}~\bibnamefont {Wang}}, \bibinfo {author}
  {\bibfnamefont {G.}~\bibnamefont {Deng}}, \bibinfo {author} {\bibfnamefont
  {H.}~\bibnamefont {Song}}, \bibinfo {author} {\bibfnamefont {L.}~\bibnamefont
  {You}}, \bibinfo {author} {\bibfnamefont {Y.}~\bibnamefont {Fan}}, \bibinfo
  {author} {\bibfnamefont {G.}~\bibnamefont {Guo}},\ and\ \bibinfo {author}
  {\bibfnamefont {Q.}~\bibnamefont {Zhou}},\ }\bibfield  {title} {\bibinfo
  {title} {Hertz-rate metropolitan quantum teleportation},\ }\href
  {https://doi.org/10.1038/s41377-023-01158-7} {\bibfield  {journal} {\bibinfo
  {journal} {Light: Science $\&$ Applications}\ }\textbf {\bibinfo {volume}
  {12}},\ \bibinfo {pages} {115} (\bibinfo {year} {2023})}\BibitemShut
  {NoStop}%
\bibitem [{\citenamefont {Lago-Rivera}\ \emph {et~al.}(2023)\citenamefont
  {Lago-Rivera}, \citenamefont {Rakonjac}, \citenamefont {Grandi},\ and\
  \citenamefont {Riedmatten}}]{FibOptTp_Rivera2023}%
  \BibitemOpen
  \bibfield  {author} {\bibinfo {author} {\bibfnamefont {D.}~\bibnamefont
  {Lago-Rivera}}, \bibinfo {author} {\bibfnamefont {J.~V.}\ \bibnamefont
  {Rakonjac}}, \bibinfo {author} {\bibfnamefont {S.}~\bibnamefont {Grandi}},\
  and\ \bibinfo {author} {\bibfnamefont {H.~d.}\ \bibnamefont {Riedmatten}},\
  }\bibfield  {title} {\bibinfo {title} {Long distance multiplexed quantum
  teleportation from a telecom photon to a solid-state qubit},\ }\href
  {https://doi.org/10.1038/s41467-023-37518-5} {\bibfield  {journal} {\bibinfo
  {journal} {Nature Communications}\ }\textbf {\bibinfo {volume} {14}},\
  \bibinfo {pages} {1889} (\bibinfo {year} {2023})}\BibitemShut {NoStop}%
\bibitem [{\citenamefont {Minder}\ \emph {et~al.}(2019)\citenamefont {Minder},
  \citenamefont {Pittaluga}, \citenamefont {Roberts}, \citenamefont
  {Lucamarini}, \citenamefont {Dynes}, \citenamefont {Yuan},\ and\
  \citenamefont {Shields}}]{Phaselock_Minder2019}%
  \BibitemOpen
  \bibfield  {author} {\bibinfo {author} {\bibfnamefont {M.}~\bibnamefont
  {Minder}}, \bibinfo {author} {\bibfnamefont {M.}~\bibnamefont {Pittaluga}},
  \bibinfo {author} {\bibfnamefont {G.~L.}\ \bibnamefont {Roberts}}, \bibinfo
  {author} {\bibfnamefont {M.}~\bibnamefont {Lucamarini}}, \bibinfo {author}
  {\bibfnamefont {J.~F.}\ \bibnamefont {Dynes}}, \bibinfo {author}
  {\bibfnamefont {Z.~L.}\ \bibnamefont {Yuan}},\ and\ \bibinfo {author}
  {\bibfnamefont {A.~J.}\ \bibnamefont {Shields}},\ }\bibfield  {title}
  {\bibinfo {title} {Experimental quantum key distribution beyond the
  repeaterless secret key capacity},\ }\href
  {https://doi.org/10.1038/s41566-019-0377-7} {\bibfield  {journal} {\bibinfo
  {journal} {Nature Photonics}\ }\textbf {\bibinfo {volume} {13}},\ \bibinfo
  {pages} {334} (\bibinfo {year} {2019})}\BibitemShut {NoStop}%
\bibitem [{\citenamefont {Wang}\ \emph {et~al.}(2019)\citenamefont {Wang},
  \citenamefont {He}, \citenamefont {Yin}, \citenamefont {Lu}, \citenamefont
  {Cui}, \citenamefont {Chen}, \citenamefont {Zhou}, \citenamefont {Guo},\ and\
  \citenamefont {Han}}]{Phaselock_Wang2019}%
  \BibitemOpen
  \bibfield  {author} {\bibinfo {author} {\bibfnamefont {S.}~\bibnamefont
  {Wang}}, \bibinfo {author} {\bibfnamefont {D.-Y.}\ \bibnamefont {He}},
  \bibinfo {author} {\bibfnamefont {Z.-Q.}\ \bibnamefont {Yin}}, \bibinfo
  {author} {\bibfnamefont {F.-Y.}\ \bibnamefont {Lu}}, \bibinfo {author}
  {\bibfnamefont {C.-H.}\ \bibnamefont {Cui}}, \bibinfo {author} {\bibfnamefont
  {W.}~\bibnamefont {Chen}}, \bibinfo {author} {\bibfnamefont {Z.}~\bibnamefont
  {Zhou}}, \bibinfo {author} {\bibfnamefont {G.-C.}\ \bibnamefont {Guo}},\ and\
  \bibinfo {author} {\bibfnamefont {Z.-F.}\ \bibnamefont {Han}},\ }\bibfield
  {title} {\bibinfo {title} {Beating the fundamental rate-distance limit in a
  proof-of-principle quantum key distribution system},\ }\href
  {https://doi.org/10.1103/PhysRevX.9.021046} {\bibfield  {journal} {\bibinfo
  {journal} {Phys. Rev. X}\ }\textbf {\bibinfo {volume} {9}},\ \bibinfo {pages}
  {021046} (\bibinfo {year} {2019})}\BibitemShut {NoStop}%
\bibitem [{\citenamefont {Pittaluga}\ \emph {et~al.}(2021)\citenamefont
  {Pittaluga}, \citenamefont {Minder}, \citenamefont {Lucamarini},
  \citenamefont {Sanzaro}, \citenamefont {Woodward}, \citenamefont {Li},
  \citenamefont {Yuan},\ and\ \citenamefont
  {Shields}}]{Phaselock_Pittaluga2021}%
  \BibitemOpen
  \bibfield  {author} {\bibinfo {author} {\bibfnamefont {M.}~\bibnamefont
  {Pittaluga}}, \bibinfo {author} {\bibfnamefont {M.}~\bibnamefont {Minder}},
  \bibinfo {author} {\bibfnamefont {M.}~\bibnamefont {Lucamarini}}, \bibinfo
  {author} {\bibfnamefont {M.}~\bibnamefont {Sanzaro}}, \bibinfo {author}
  {\bibfnamefont {R.~I.}\ \bibnamefont {Woodward}}, \bibinfo {author}
  {\bibfnamefont {M.-J.}\ \bibnamefont {Li}}, \bibinfo {author} {\bibfnamefont
  {Z.}~\bibnamefont {Yuan}},\ and\ \bibinfo {author} {\bibfnamefont {A.~J.}\
  \bibnamefont {Shields}},\ }\bibfield  {title} {\bibinfo {title} {600-km
  repeater-like quantum communications with dual-band stabilization},\ }\href
  {https://doi.org/10.1038/s41566-021-00811-0} {\bibfield  {journal} {\bibinfo
  {journal} {Nature Photonics}\ }\textbf {\bibinfo {volume} {15}},\ \bibinfo
  {pages} {530} (\bibinfo {year} {2021})}\BibitemShut {NoStop}%
\bibitem [{\citenamefont {Li}\ \emph {et~al.}(2023)\citenamefont {Li},
  \citenamefont {Zhang}, \citenamefont {Lu}, \citenamefont {Li}, \citenamefont
  {Jiang}, \citenamefont {Liu}, \citenamefont {Huang}, \citenamefont {Li},
  \citenamefont {Wang}, \citenamefont {Wang}, \citenamefont {Zhang},
  \citenamefont {You}, \citenamefont {Xu},\ and\ \citenamefont
  {Pan}}]{NoPhaselock_Pan2023}%
  \BibitemOpen
  \bibfield  {author} {\bibinfo {author} {\bibfnamefont {W.}~\bibnamefont
  {Li}}, \bibinfo {author} {\bibfnamefont {L.}~\bibnamefont {Zhang}}, \bibinfo
  {author} {\bibfnamefont {Y.}~\bibnamefont {Lu}}, \bibinfo {author}
  {\bibfnamefont {Z.-P.}\ \bibnamefont {Li}}, \bibinfo {author} {\bibfnamefont
  {C.}~\bibnamefont {Jiang}}, \bibinfo {author} {\bibfnamefont
  {Y.}~\bibnamefont {Liu}}, \bibinfo {author} {\bibfnamefont {J.}~\bibnamefont
  {Huang}}, \bibinfo {author} {\bibfnamefont {H.}~\bibnamefont {Li}}, \bibinfo
  {author} {\bibfnamefont {Z.}~\bibnamefont {Wang}}, \bibinfo {author}
  {\bibfnamefont {X.-B.}\ \bibnamefont {Wang}}, \bibinfo {author}
  {\bibfnamefont {Q.}~\bibnamefont {Zhang}}, \bibinfo {author} {\bibfnamefont
  {L.}~\bibnamefont {You}}, \bibinfo {author} {\bibfnamefont {F.}~\bibnamefont
  {Xu}},\ and\ \bibinfo {author} {\bibfnamefont {J.-W.}\ \bibnamefont {Pan}},\
  }\bibfield  {title} {\bibinfo {title} {Twin-field quantum key distribution
  without phase locking},\ }\href
  {https://doi.org/10.1103/PhysRevLett.130.250802} {\bibfield  {journal}
  {\bibinfo  {journal} {Phys. Rev. Lett.}\ }\textbf {\bibinfo {volume} {130}},\
  \bibinfo {pages} {250802} (\bibinfo {year} {2023})}\BibitemShut {NoStop}%
\bibitem [{\citenamefont {Li}\ and\ \citenamefont {van
  Loock}(2023)}]{RSBC_Li2023}%
  \BibitemOpen
  \bibfield  {author} {\bibinfo {author} {\bibfnamefont {P.-Z.}\ \bibnamefont
  {Li}}\ and\ \bibinfo {author} {\bibfnamefont {P.}~\bibnamefont {van Loock}},\
  }\bibfield  {title} {\bibinfo {title} {Memoryless quantum repeaters based on
  cavity-qed and coherent states},\ }\href
  {https://doi.org/https://doi.org/10.1002/qute.202200151} {\bibfield
  {journal} {\bibinfo  {journal} {Advanced Quantum Technologies}\ }\textbf
  {\bibinfo {volume} {6}},\ \bibinfo {pages} {2200151} (\bibinfo {year}
  {2023})}\BibitemShut {NoStop}%
\bibitem [{\citenamefont {Xu}\ \emph {et~al.}(2023)\citenamefont {Xu},
  \citenamefont {Steinberg}, \citenamefont {Singh}, \citenamefont
  {L{\'{o}}pez-Tarrida}, \citenamefont {Portillo},\ and\ \citenamefont
  {Cabello}}]{Loopholefree_Xu2023}%
  \BibitemOpen
  \bibfield  {author} {\bibinfo {author} {\bibfnamefont {Z.-P.}\ \bibnamefont
  {Xu}}, \bibinfo {author} {\bibfnamefont {J.}~\bibnamefont {Steinberg}},
  \bibinfo {author} {\bibfnamefont {J.}~\bibnamefont {Singh}}, \bibinfo
  {author} {\bibfnamefont {A.~J.}\ \bibnamefont {L{\'{o}}pez-Tarrida}},
  \bibinfo {author} {\bibfnamefont {J.~R.}\ \bibnamefont {Portillo}},\ and\
  \bibinfo {author} {\bibfnamefont {A.}~\bibnamefont {Cabello}},\ }\bibfield
  {title} {\bibinfo {title} {Graph-theoretic approach to {B}ell experiments
  with low detection efficiency},\ }\href
  {https://doi.org/10.22331/q-2023-02-16-922} {\bibfield  {journal} {\bibinfo
  {journal} {{Quantum}}\ }\textbf {\bibinfo {volume} {7}},\ \bibinfo {pages}
  {922} (\bibinfo {year} {2023})}\BibitemShut {NoStop}%
\bibitem [{\citenamefont {Andersen}\ \emph {et~al.}(2015)\citenamefont
  {Andersen}, \citenamefont {Neergaard-Nielsen}, \citenamefont {van Loock},\
  and\ \citenamefont {Furusawa}}]{HOSQIP_Andersen2015}%
  \BibitemOpen
  \bibfield  {author} {\bibinfo {author} {\bibfnamefont {U.~L.}\ \bibnamefont
  {Andersen}}, \bibinfo {author} {\bibfnamefont {J.~S.}\ \bibnamefont
  {Neergaard-Nielsen}}, \bibinfo {author} {\bibfnamefont {P.}~\bibnamefont {van
  Loock}},\ and\ \bibinfo {author} {\bibfnamefont {A.}~\bibnamefont
  {Furusawa}},\ }\bibfield  {title} {\bibinfo {title} {Hybrid discrete- and
  continuous-variable quantum information},\ }\href
  {https://doi.org/10.1038/nphys3410} {\bibfield  {journal} {\bibinfo
  {journal} {Nature Physics}\ }\textbf {\bibinfo {volume} {11}},\ \bibinfo
  {pages} {713} (\bibinfo {year} {2015})}\BibitemShut {NoStop}%
\end{thebibliography}

\clearpage
\appendix
\section{Hybrid entangled state after passing through loss-only channel}
\label{append_sec:hes_lossonly_channel}

We consider a hybrid entangled  (HE)-state in modes $a$ and $b$ given as \cite{HOSTelport_Sychev2018}
\begin{align}
    \ket{\Psi_{ab}} &= \frac{1}{\sqrt{2}}\argp{ \ket H_a\ket{\alpha}_b + \ket V_a\ket{-\alpha}_b }.
    \label{append_eq:hes_def}
\end{align}
Here, the single-photon-polarization-state $\ket H(\ket V)$ represents the discrete-variable (DV) part and the coherent state $\ket\alpha (\ket{-\alpha})$ forms the continuous-variable (CV) part.
Let us now consider that only the CV part passes through the loss-only channel that can be modelled as mixing the signal with an ancilla vacuum through a BS with transmittance $T$, and then tracing out the ancilla.
Considering the action of such a BS on an incoming coherent state, $U_\text{bs}(T):\ket{\alpha,0} \rightarrow \ket{\sqrt{T}\alpha,\sqrt{1-T}\alpha}$, the HE-state after passing through the channel becomes
\begin{widetext}
\begin{align}
    \rho_{ab}^\text{ch,cv} &= \rm{Tr}_c \argc{ U_\text{bs}^{bc}(T) \argp{ \rho_{ab}\otimes \ket 0_c\bra 0 } \args{ U_\text{bs}^{bc} (T) }^\dagger }
    \nonumber 
    \\
    &= \frac{1}{2} \left[
    \rho_a^{HH} \otimes \ket{\sqrt{T}\alpha}_b\bra{\sqrt{T}\alpha}
    +
    \rho_a^{VV} \otimes \ket{-\sqrt{T}\alpha}_{b}\bra{-\sqrt{T}\alpha}
    +
    e^{-2(1-T)\alpha^2} \left(
    \rho_a^{HV} \otimes \ket{\sqrt{T}\alpha}_{b}\bra{-\sqrt{T}\alpha}
    \right.
    \right.
    \nonumber 
    \\
    &~~+ \left.
    \left.
    \rho_a^{VH} \otimes \ket{-\sqrt{T}\alpha}_{b}\bra{\sqrt{T}\alpha}
    \right)
    \right],
\end{align}
\end{widetext}
where $\rho^{x,y} = \ket{x}\bra{y}$ ($x,y=H,V$).

\section{Obtaining the shared DV-state after noisy transmission followed by on-off measurement at Charlie's lab}
\label{append_sec:dvstate_alice_bob}

\subsection{$4$-mode state transmission through noisy channel and mixing}
\label{append_subsec:hes_channel_charlie}
Let us consider that both Alice and Bob prepare their individual HE-states in modes $\argc{a_1,a_2}$ and $\argc{b_1,b_2}$, respectively.
\begin{align}
    \rho_{a_1a_2} &= \ket{\Psi_{a_1a_2}}\bra{\Psi_{a_1a_2}} ~~~~\rm{and}
    \nonumber 
    \\
    \rho_{b_1b_2} &= \ket{\Psi_{b_1b_2}}\bra{\Psi_{b_1b_2}},
    \label{append_eq:hes_alice_bob}
\end{align}

For the sake of convenience we use a compact notation for the DV and the CV parts as $\rho^{xy,uv} = \rho^{xy}_{a_1} \otimes \rho^{uv}_{b_1}$ ($x,y,u,v=H,V$) and $\ket{\alpha,\beta} = \ket{\alpha}_{a_2}\ket{\beta}_{b_2}$ ($\alpha,\beta \in \mathcal{C}^2$).

After Alice and Bob send their coherent states through the loss-only channel to Charlie, the $4$-mode state is given by
\begin{widetext}
\begin{align}
    \rho_\text{charlie} &= \rho_{a_1a_2}^\text{ch} \otimes \rho_{b_1b_2}^\text{ch}
    \nonumber 
    \\
    &= \frac{1}{4} \left(\left\lbrace 
    \args{ 
    \rho^{HH,HH} \otimes \ket{\sqrt{T}\alpha,\sqrt{T}\alpha}\bra{\sqrt{T}\alpha,\sqrt{T}\alpha}
    +
    \rho^{HH,VV} \otimes \ket{\sqrt{T}\alpha,-\sqrt{T}\alpha}\bra{\sqrt{T}\alpha,-\sqrt{T}\alpha}
    } 
    \right.\right.
    \nonumber 
    \\
    &~~
    + e^{-2(1-T)\alpha^2}\left.\args{ 
    \rho^{HH,HV} \otimes \ket{\sqrt{T}\alpha,\sqrt{T}\alpha}\bra{\sqrt{T}\alpha,-\sqrt{T}\alpha}
    +
    \rho^{HH,VH} \otimes \ket{\sqrt{T}\alpha,-\sqrt{T}\alpha}\bra{\sqrt{T}\alpha,\sqrt{T}\alpha} 
    }
    \right\rbrace
    \nonumber 
    \\
    &~~+ \left\lbrace
    \args{ 
    \rho^{VV,HH} \otimes \ket{-\sqrt{T}\alpha,\sqrt{T}\alpha}\bra{-\sqrt{T}\alpha,\sqrt{T}\alpha}
    +
    \rho^{VV,VV} \otimes \ket{-\sqrt{T}\alpha,-\sqrt{T}\alpha}\bra{-\sqrt{T}\alpha,-\sqrt{T}\alpha}
    } 
    \right.
    \nonumber 
    \\
    &~~
    + e^{-2(1-T)\alpha^2}\left.\args{
    \rho^{VV,HV} \otimes \ket{-\sqrt{T}\alpha,\sqrt{T}\alpha}\bra{-\sqrt{T}\alpha,-\sqrt{T}\alpha}
    +
    \rho^{VV,VH} \otimes \ket{-\sqrt{T}\alpha,-\sqrt{T}\alpha}\bra{-\sqrt{T}\alpha,\sqrt{T}\alpha}
    }
    \right\rbrace
    \nonumber 
    \\
        &~~+ e^{-2(1-T)\alpha^2}\left\lbrace
        \args{ 
        \rho^{HV,HH} \otimes \ket{\sqrt{T}\alpha,\sqrt{T}\alpha}\bra{-\sqrt{T}\alpha,\sqrt{T}\alpha}
        +
        \rho^{HV,VV} \otimes \ket{\sqrt{T}\alpha,-\sqrt{T}\alpha}\bra{-\sqrt{T}\alpha,-\sqrt{T}\alpha} 
        } 
        \right.
        \nonumber 
        \\
        &~~
        + e^{-2(1-T)\alpha^2}\left.\args{ 
        \rho^{HV,HV} \otimes \ket{\sqrt{T}\alpha,\sqrt{T}\alpha}\bra{-\sqrt{T}\alpha,-\sqrt{T}\alpha}
        +
        \rho^{HV,VH} \otimes \ket{\sqrt{T}\alpha,-\sqrt{T}\alpha}\bra{-\sqrt{T}\alpha,\sqrt{T}\alpha}
        }
        \right\rbrace
        \nonumber 
        \\
        &~~+ e^{-2(1-T)\alpha^2}\left\lbrace
        \args{ 
        \rho^{VH,HH} \otimes \ket{-\sqrt{T}\alpha,\sqrt{T}\alpha}\bra{\sqrt{T}\alpha,\sqrt{T}\alpha}
        +
        \rho^{VH,VV} \otimes \ket{-\sqrt{T}\alpha,-\sqrt{T}\alpha}\bra{\sqrt{T}\alpha,-\sqrt{T}\alpha} 
        } 
        \right.
        \nonumber 
        \\
        &~~+ e^{-2(1-T)\alpha^2}\left.\left.\args{ 
        \rho^{VH,HV} \otimes \ket{-\sqrt{T}\alpha,\sqrt{T}\alpha}\bra{\sqrt{T}\alpha,-\sqrt{T}\alpha}
        +
        \rho^{VH,VH} \otimes \ket{-\sqrt{T}\alpha,-\sqrt{T}\alpha}\bra{\sqrt{T}\alpha,\sqrt{T}\alpha}
        }
        \right\rbrace \right).
\end{align}
    
Upon mixing at the $50:50$ BS ($T=1/2$) by Charlie, the state becomes
\begin{align}
        \rho_\text{charlie}^\text{mix} &= U_\text{bs}^{a_2b_2}(1/2) \rho_\text{charlie} \args{ U_\text{bs}^{a_2b_2}(1/2) }^\dagger
        \nonumber 
        \\
        &= \frac{1}{4} \left(\left\lbrace 
        \args{ 
        \rho^{HH,HH} \otimes \ket{\sqrt{2T}\alpha,0}\bra{\sqrt{2T}\alpha,0}
        +
        \rho^{HH,VV} \otimes \ket{0,-\sqrt{2T}\alpha}\bra{0,-\sqrt{2T}\alpha}
        } 
        \right.\right.
        \nonumber 
        \\
        &~~
        + e^{-2(1-T)\alpha^2}\left.\args{ 
        \rho^{HH,HV} \otimes \ket{\sqrt{2T}\alpha,0}\bra{0,\sqrt{2T}\alpha}
        +
        \rho^{HH,VH} \otimes \ket{0,\sqrt{2T}\alpha}\bra{\sqrt{2T}\alpha,0} 
        }
        \right\rbrace
        \nonumber 
        \\
        &~~+ \left\lbrace
        \args{ 
        \rho^{VV,HH} \otimes \ket{0,-\sqrt{2T}\alpha}\bra{0,-\sqrt{2T}\alpha}
        +
        \rho^{VV,VV} \otimes \ket{-\sqrt{2T}\alpha,0}\bra{-\sqrt{2T}\alpha,0}
        } 
        \right.
        \nonumber 
        \\
        &~~
        + e^{-2(1-T)\alpha^2}\left.\args{
        \rho^{VV,HV} \otimes \ket{0,-\sqrt{2T}\alpha}\bra{-\sqrt{2T}\alpha,0}
        +
        \rho^{VV,VH} \otimes \ket{-\sqrt{2T}\alpha,0}\bra{0,-\sqrt{2T}\alpha}
        }
        \right\rbrace
        \nonumber 
        \\
        &~~+ e^{-2(1-T)\alpha^2}\left\lbrace
        \args{ 
        \rho^{HV,HH} \otimes \ket{\sqrt{2T}\alpha,0}\bra{0,-\sqrt{2T}\alpha}
        +
        \rho^{HV,VV} \otimes \ket{0,-\sqrt{2T}\alpha}\bra{-\sqrt{2T}\alpha,0} 
        } 
        \right.
        \nonumber 
        \\
        &~~
        + e^{-2(1-T)\alpha^2}\left.\args{ 
        \rho^{HV,HV} \otimes \ket{\sqrt{2T}\alpha,0}\bra{-\sqrt{2T}\alpha,0}
        +
        \rho^{HV,VH} \otimes \ket{0,\sqrt{2T}\alpha}\bra{0,-\sqrt{2T}\alpha}
        }
        \right\rbrace
        \nonumber 
        \\
        &~~+ e^{-2(1-T)\alpha^2}\left\lbrace
        \args{ 
        \rho^{VH,HH} \otimes \ket{0,-\sqrt{2T}\alpha}\bra{\sqrt{2T}\alpha,0}
        +
        \rho^{VH,VV} \otimes \ket{-\sqrt{2T}\alpha,0}\bra{0,\sqrt{2T}\alpha} 
        } 
        \right.
        \nonumber 
        \\
        &~~+ e^{-2(1-T)\alpha^2}\left.\left.\args{ 
        \rho^{VH,HV} \otimes \ket{0,-\sqrt{2T}\alpha}\bra{0,\sqrt{2T}\alpha}
        +
        \rho^{VH,VH} \otimes \ket{-\sqrt{2T}\alpha,0}\bra{\sqrt{2T}\alpha,0}
    }
    \right\rbrace \right).
\end{align}
\end{widetext}

\subsection{Post-measurement shared DV-state}
\label{append_subsec:dvstate_charlie_measurement}

After mixing the incoming signal Charlie checks whether the single-photon-detector (SPD) D$_2$ clicks.
An inefficient SPD is described the set of operators $\argc{\Pi_1,\Pi_{\neg 1}=\mathbf{I} - \Pi_1}$ such that \cite{HOSTeleport_Park2012}
\begin{equation}
    \Pi_m = \eta_0^m \sum_k {}^{k+m}C_m (1-\eta_0)^k \ket{k+m}\bra{k+m}
    \label{append_eq:noisy_detector_m}
\end{equation}
represents the noisy $m$-photon detection, where $\eta_0$ is the efficiency of the SPD and $\ket k$ represents the $k$-photon-number state.
${}^{k+m}C_m = \frac{(k+m)!}{k!m!}$ represents the binomial coefficient.
 Charlie's measurement is described by the operator $\mathcal{M}_{a_2b_2} = \Pi_{\neg 1,1}$, such that $\Pi_{\alpha,\beta} = \Pi_\alpha \otimes \Pi_\beta$ ($\alpha,\beta = 1,\neg 1$) where the first and the second operator operates on modes $a_2$ and $b_2$, respectively.

The shared DV-state between Alice and Bob, after Charlie's measurement, becomes $\rho_{a_1b_1} = \frac{1}{P_\text{dv}} \rho_{a_1b_1}^{\neg 1,1}$ where $P_\text{dv} = \text{Tr}_{a_1b_1}\argp{\rho_{a_1b_1}^{\neg 1,1}}$ is the probability of obtaining the state $\rho_{a_1b_1}$ and $\rho_{a_1b_1}^{\neg 1,1} = \text{Tr}_{a_2b_2}\argp{\Pi_{\neg 1,1}\rho_\text{charlie}^\text{mix}}$.
Now, applying the following results
\begin{align}
    \text{Tr}\argp{\Pi_1 \ket\alpha\bra\beta} &= \eta_0 \sum_k {}^{k+1}C_1 (1-\eta_0)^k \braket{\beta}{k+1}\braket{k+1}{\alpha}
    \nonumber 
    \\
    &= \eta_0 e^{-\frac{\alpha^2+\beta^2}{2}} \alpha\beta \sum_k \frac{\args{\beta\alpha(1-\eta_1)}^k}{k!}
    \nonumber 
    \\
    &= \eta_0 \alpha\beta \braket{\beta}{\alpha} e^{-\eta_0\beta\alpha} 
    \nonumber 
    \\
    \text{Tr}\argp{\Pi_{\neg 1} \ket\alpha\bra\beta} &= \text{Tr}\args{(\mathbf{I} - \Pi_1) \ket\alpha\bra\beta}
    \nonumber 
    \\
    &= \braket{\beta}{\alpha} \argp{ 1 - \eta_0\alpha\beta e^{-\eta_0\beta\alpha}},
\end{align}
leads to the projected DV state onto the modes $a_1$ and $b_1$ as
\begin{widetext}
\begin{align}
        \rho_{a_1b_1}^{\neg 1,1} &= \text{Tr}_{a_2b_2}\argp{\Pi_{\neg 1,1}\rho_\text{charlie}^\text{mix}}
        \nonumber 
        \\
        &= \frac{1}{4} \left[
        \rho^{HH,VV} \eta_0 (2T\alpha^2) e^{-2T\eta_0\alpha^2}
        +
        \rho^{VV,HH} \eta_0 (2T\alpha^2) e^{-2T\eta_0\alpha^2}
        +
        e^{-2(1-T)\alpha^2} e^{-2(1-T)\alpha^2} \rho^{HV,VH} \eta_0 (-2T\alpha^2) e^{-4T\alpha^2} e^{2T\eta_0\alpha^2}
        \right. 
        \nonumber 
        \\
        &~~\left. +
        e^{-2(1-T)\alpha^2} e^{-2(1-T)\alpha^2} \rho^{VH,HV} \eta_0 (-2T\alpha^2) e^{-4T\alpha^2} e^{2T\eta_0\alpha^2}
        \right]
        \nonumber 
        \\
        &= \frac{T\eta_0\alpha^2}{2} e^{-2T\eta_0\alpha^2}\args{
        \argp{ \rho^{HH,VV} + \rho^{VV,HH} } 
        -
        e^{-4(1-T\eta_0)\alpha^2}\argp{ \rho^{HV,VH} + \rho^{VH,HV} } 
        }
        \nonumber 
        \\
        &= \frac{T\eta_0\alpha^2}{2} e^{-2T\eta_0\alpha^2}\args{
        \Big( \ket{H,V}\bra{H,V} + \ket{V,H}\bra{V,H} \Big) 
        -
        e^{-4(1-T\eta_0)\alpha^2}\Big( \ket{H,V}\bra{V,H} + \ket{V,H}\bra{H,V} \Big) 
        }.
        \label{append_eq:dvstate_preli}
    \end{align}
    with probability 
    \begin{equation}
        P^{\neg 1,1} = \text{Tr} \argp{ \rho_{a_1b_1}^{\neg 1,1} }
        = T\eta_0\alpha^2 e^{-2T\eta_0\alpha^2}
        \label{append_eq:dvstate_probability}
    \end{equation}

    Thus the final normalized shared DV state is given by
    \begin{align}
        \rho_{a_1b_1} &= \frac{1}{P^{\neg 1,1}} \rho_{a_1b_1}^{\neg 1,1}
        \nonumber 
        \\
        &= \frac{1}{2} \args{ 
        \argp{ \ket{H,V}\bra{H,V} + \ket{V,H}\bra{V,H} }
        - e^{-4(1-T\eta_0)\alpha^2}\argp{ \ket{H,V}\bra{V,H} + \ket{V,H}\bra{H,V} }  
        }.
        \label{append_eq:dvstate_final}
\end{align}
\end{widetext}

If one defines the $4$ Bell-states in polarization basis as
\begin{align}
    \ket{\Psi^\pm} &= \frac{1}{\sqrt{2}}\argp{ \ket{H,V} \pm \ket{V,H}}
    \nonumber 
    \\
    \ket{\Phi^\pm} &= \frac{1}{\sqrt{2}}\argp{ \ket{H,H} \pm \ket{V,V}}.
    \label{append_eq:bellstate_polarization}
\end{align}
one can further recast \eqref{append_eq:dvstate_final} in a compact form as
\begin{align}
    \rho_{a_1b_1} &= \ket{\Psi^-}\bra{\Psi^-} + R\argp{ \ket{\Psi^+}\bra{\Psi^+} - \ket{\Psi^-}\bra{\Psi^-} }
    \nonumber 
    \\
    &= \argp{ 1-R }\ket{\Psi^-}\bra{\Psi^-} + R\ket{\Psi^+}\bra{\Psi^+},
    \label{append_eq:dvstate_final_bellbasis}
\end{align}
where $R=\frac{1-e^{-4(1-T\eta_0)\alpha^2}}{2}$.

\section{Bell function for the shared DV-state between Alice and Bob}
\label{append_sec:bell_function_alice_bob}

The PRBOs are given as 
\begin{align}
    \hat{O}(\zeta,\theta) &= U(\zeta,\theta) \Pi(\eta_0) U^\dagger(\zeta,\theta)
    \nonumber 
    \\
    &= \eta_0
    \begin{pmatrix}
        \ket H &\ket V
    \end{pmatrix}
    \begin{bmatrix}
        N_{hh}(\zeta,\theta) &-N_{hv}(\zeta,\theta)\\
        -N_{hv}^*(\zeta,\theta) &N_{vv}(\zeta,\theta)
    \end{bmatrix}
    \begin{pmatrix}
        \bra H \\
        \bra V
    \end{pmatrix},
    \label{append_eq:binary_measurement_rotated}
\end{align}
where $N_{hh}(\zeta,\theta) = -\eta_0(1-2\zeta)$, $N_{vv}(\zeta,\theta) = \eta_0(1-2\zeta)$ and $N_{hv}(\zeta,\theta) = 2e^{\mathsf{i}\theta}\eta_0\sqrt{\zeta(1-\zeta)}$.
This leads to the joint binary-outcome measurement, described by $\hat{O}_{a_1,b_1}(\zeta,\theta,\xi,\phi) = \hat{O}_{a_1}(\zeta,\theta) \otimes \hat{O}_{b_1}(\xi,\phi)$, as
\begin{widetext}
\begin{align}
        \hat{O}(\zeta,\theta,\xi,\phi) &= \hat{O}(\zeta,\theta) \otimes \hat{O}(\xi,\phi)
        \nonumber 
        \\
        &= \argc{
        \Big[\{N_{hh}(\zeta,\theta) \ket H\bra H + N_{vv}(\zeta,\theta) \ket V\bra V\Big] - \Big[N_{hv}(\zeta,\theta)\ket H\bra V + N_{hv}^*(\zeta,\theta)\ket V\bra H\Big]
        } 
        \otimes 
        \nonumber 
        \\
        &~~\argc{
        \Big[ 
        N_{hh}(\xi,\phi) \ket H\bra H + N_{vv}(\xi,\phi) \ket V\bra V \Big] - \Big[ N_{hv}(\xi,\phi)\ket H\bra V + N_{hv}^*(\xi,\phi)\ket V\bra H       
        \Big]
        }
        \nonumber 
        \\
        &= \Big\{ \Big[ N_{hh}(\zeta,\theta)N_{vv}(\xi,\phi) \ket{H,V}\bra{H,V} + N_{vv}(\zeta,\theta)N_{hh}(\xi,\phi) \ket{V,H}\bra{V,H} \Big]
        \nonumber 
        \\
        &~~ + \Big[ N_{hv}(\zeta,\theta)N_{hv}^*(\xi,\phi) \ket{H,V}\bra{V,H} 
        + N_{hv}^*(\zeta,\theta)N_{hv}(\xi,\phi) \ket{V,H}\bra{H,V} \Big] \Big\}
        + \mathcal{O}_{hv},
        \label{append_eq:bell_operator_rotated}
\end{align}
\end{widetext}
where $\mathcal{O}_{h,v}$ contains all other terms where the same polarization state occurs in either "ket" or "bra" vectors.
The expectation value of the joint measurement, $\hat{O}_{a_1,b_1}(\zeta,\theta,\xi,\phi)$ is hence given by
\begin{widetext}
\begin{align}
        &\mathcal{E}\argp{\zeta,\theta,\xi,\phi} = \left\langle \hat{O}_{a_1,b_1}(\zeta,\theta,\xi,\phi) \right\rangle_{\rho_{a_1b_1}}
        = \text{Tr}\args{ \rho_{a_1b_1} \hat{O}_{a_1,b_1}(\zeta,\theta,\xi,\phi) }
        \nonumber 
        \\
        &= \frac{1}{2} \argc{
        \Big[ N_{hh}(\zeta,\theta)N_{vv}(\xi,\phi) + N_{vv}(\zeta,\theta)N_{hh}(\xi,\phi) \Big]
        - e^{-4(1-T\eta_0)\alpha^2}\Big[ N_{hv}(\zeta,\theta)N_{hv}^*(\xi,\phi) + N_{hv}^*(\zeta,\theta)N_{hv}(\xi,\phi) \big]
        }
        \nonumber 
        \\
        &= -\eta_0^2\args{
        (1-2\zeta)(1-2\xi) + 4e^{-4(1-T\eta_0)\alpha^2}\sqrt{\zeta(1-\zeta)\xi(1-\xi)}\cos{2(\theta-\phi)}
        }.
        \label{append_eq:expectation_joint_measure}
\end{align}
\end{widetext}

\section{Fidelity of Teleportation for an input polarization qubit}
\label{append_sec:telfid_qubit_pure}
Let us now consider the teleportion of an unknown input pure-state $\ket{\psi_\text{in}} = \sqrt{p} \ket H + \sqrt{1-p}e^{\mathsf{i}\theta} \ket V$.
In the present case, the total tripartite state is given by
\begin{align}
    \rho_\text{tot} &= \ket{\psi_\text{in}}\bra{\psi_\text{in}} \otimes \rho_{a_1b_1}
    \nonumber 
    \\
    &= \left( 1-R \right)\ket{\Psi^{(1)}}\bra{\Psi^{(1)}} +R\ket{\Psi^{(2)}}\bra{\Psi^{(2)}},
    \label{append_eq:totalstate}
\end{align}
where
\begin{align}
        \ket{\Psi^{(1)}} &= \ket{\psi_\text{in}}\ket{\Psi^-}
        \nonumber 
        \\
        &= \frac{1}{2}\left(
        -\ket{\Psi^+}_{\text{in},a_1} \otimes \sigma_z\ket{\psi_\text{in}}_{b_1}
        -\ket{\Psi^-}_{\text{in},a_1} \otimes \ket{\psi_\text{in}}_{b_1}
        \right.
        \nonumber 
        \\
        &~~\left.
        +\ket{\Phi^+}_{\text{in},a_1} \otimes \sigma_x\sigma_z\ket{\psi_\text{in}}_{b_1}
        +\ket{\Phi^-}_{\text{in},a_1} \otimes \sigma_x\ket{\psi_\text{in}}_{b_1}
        \right)
        \nonumber 
        \\
        \ket{\Psi^{(2)}} &= \ket{\psi_\text{in}}\ket{\Psi^+}
        \nonumber 
        \\
        &= \frac{1}{2}\left(
        +\ket{\Psi^+}_{\text{in},a_1} \otimes \ket{\psi_\text{in}}_{b_1}
        +\ket{\Psi^-}_{\text{in},a_1} \otimes \sigma_z\ket{\psi_\text{in}}_{b_1}
        \right.
        \nonumber 
        \\
        &~~\left.
        +\ket{\Phi^+}_{\text{in},a_1} \otimes \sigma_x\ket{\psi_\text{in}}_{b_1}
        +\ket{\Phi^-}_{\text{in},a_1} \otimes \sigma_x\sigma_z\ket{\psi_\text{in}}_{b_1}
        \right)
        \label{append_eq:psi1psi2}
\end{align}

The Bell-state measurement $\Pi_{\psi/\phi}^\pm(\eta_0)$ yields the state in mode $b_1$ as $\rho_{b_1,\psi/\phi}^\pm(\eta_0) = \text{Tr}_{\text{in},a_1}\args{ \Pi_{\psi/\phi}^\pm(\eta_0) \rho_\text{tot}}$ with probability $P_{\psi/\phi}^\pm(\eta_0) = \text{Tr}_{\text{in},a_1,b_1}\args{ \Pi_{\psi/\phi}^\pm(\eta_0) \rho_\text{tot}} = \text{Tr}\args{ \rho_{b_1,\psi/\phi}^\pm(\eta_0) }$. 
This leads to the normalised state corresponding to the Bell-state measurement $\Pi_{\psi/\phi}^\pm(\eta_0)$ as $\rho_{\psi/\phi}^\pm (\eta_0) = \frac{\rho_{b_1,\psi/\phi}^\pm(\eta_0)}{P_{\psi/\phi}^\pm(\eta_0)}$.

In a straightforward calculation it can be shown that
\begin{align}
    \rho_{b_1,\psi}^+ &= \frac{\eta_0^2}{4}\args{
    \argp{1-R} \sigma_z\ket{\psi_\text{in}} \bra{\psi_\text{in}}\sigma_z + R \ket{\psi_\text{in}}\bra{\psi_\text{in}}
    }
    \nonumber 
    \\
    \rho_{b_1,\psi}^- &= \frac{\eta_0^2}{4}\args{
    \argp{1-R} \ket{\psi_\text{in}} \bra{\psi_\text{in}} + R \sigma_z\ket{\psi_\text{in}}\bra{\psi_\text{in}}\sigma_z
    }
    \nonumber 
    \\
    \rho_{b_1,\phi}^+ &= \frac{\eta_0^2}{4}\left[ 
    \argp{1-R} \sigma_x\sigma_z\ket{\psi_\text{in}} \bra{\psi_\text{in}}\sigma_z\sigma_x 
    + R \sigma_x\ket{\psi_\text{in}}\bra{\psi_\text{in}}\sigma_x
    \right]
    \nonumber 
    \\
    \rho_{b_1,\psi}^- &= \frac{\eta_0^2}{4}\args{
    \argp{1-R} \ket{\psi_\text{in}} \bra{\psi_\text{in}} + R \sigma_x\ket{\psi_\text{in}}\bra{\psi_\text{in}}\sigma_x
    }
    \label{append_eq:outstate_bellstate_measurment}
\end{align}
with the corresponding probability $P_{\psi/\phi}^\pm = \frac{\eta_0^2}{4}$.
The corresponding unitary operations are given as - $\Pi_\psi^+:\sigma_z$, $\Pi_\psi^-:\mathbf{I}$, $\Pi_\phi^+:-\mathsf{i}\sigma_y = \sigma_z\sigma_x$ and $\Pi_\phi^-:\sigma_x$.
It can be easily seen from \eqref{append_eq:outstate_bellstate_measurment} that with these choices of the unitary rotations, in the limit $R \rightarrow 0$ ($\eta_0\rightarrow 1, T\rightarrow 1$), we recover the ideal case.

After applying the suitable unitary rotation, the fidelity of teleportation for the four Bell-state measurements is given by
\begin{align}
    F_{\psi/\phi}^\pm(\eta_0) &= \argp{ 1 - R } + R \left| \bra{\psi_\text{in}} \sigma_z \ket{\psi_\text{in}} \right|^2
    \nonumber 
    \\
    &= \argp{ 1 - R } + R\argp{ 2p -1 }^2  
    \label{append_eq:telfid_bellstate_measurement}
\end{align}

Consequently, after summing over the Bell-state measurements and averaging over the probability amplitude $p$, we get
\begin{equation}
    F_\text{av} = \eta_0^2 \argp{ 1 - \frac{2R}{3}} = \eta_0^2 \frac{2 + e^{-4(1-T\eta_0)\alpha^2}}{3}.  
    \label{append_eq:avfid_teleport_input}
\end{equation}

It is evident from \eqref{append_eq:avfid_teleport_input} that with ideal detector ($\eta_0=1$) and no-loss ($T=1$) we obtain the perfect teleportation fidelity, i.e., $F_\text{av} = 1$.
\end{document}